\documentclass[
superscriptaddress , 
journal  =  jpccck ,
manuscript = article ,
layout = traditional 
]{achemso}

\usepackage[version=3]{mhchem} 
\usepackage[english]{babel}
\usepackage{natbib}
\usepackage{braket}
\usepackage{gensymb}
\usepackage{graphicx}
\usepackage{courier}
\usepackage{amsmath}
\usepackage{amssymb}
\usepackage{textcomp}
\usepackage{xspace}
\usepackage{multirow}
\usepackage{hyperref}

\usepackage{multicol}

\SectionNumbersOn

\author{Andrea Zen}
\altaffiliation{These two authors contributed equally}
\affiliation{ Thomas Young Centre and London Centre for Nanotechnology, 17--19 Gordon Street, London WC1H 0AH, United Kingdom }
\alsoaffiliation{ Department of Physics and Astronomy, University College London, Gower Street, London WC1E 6BT, United Kingdom }
\email{a.zen@ucl.ac.uk}

\author{Lo\"\i c M. Roch}
\altaffiliation{These two authors contributed equally}
\affiliation{ Thomas Young Centre and London Centre for Nanotechnology, 17--19 Gordon Street, London WC1H 0AH, United Kingdom }
\alsoaffiliation{ Department of Physics and Astronomy, University College London, Gower Street, London WC1E 6BT, United Kingdom }

\author{Stephen J. Cox}
\affiliation{ Thomas Young Centre and London Centre for Nanotechnology, 17--19 Gordon Street, London WC1H 0AH, United Kingdom }
\alsoaffiliation{ Department of Chemistry, University College London, 20 Gordon Street, London WC1H 0AJ, United Kingdom }

\author{Xiao Liang Hu}
\affiliation{ Thomas Young Centre and London Centre for Nanotechnology, 17--19 Gordon Street, London WC1H 0AH, United Kingdom }
\alsoaffiliation{ Department of Physics and Astronomy, University College London, Gower Street, London WC1E 6BT, United Kingdom }

\author{Sandro Sorella}
\affiliation{ SISSA--International School for Advanced Studies, Via Bonomea 26, 34136 Trieste, Italy }
\alsoaffiliation{ INFM Democritos National Simulation Center, 34151 Trieste, Italy }

\author{Dario Alf\`e}
\affiliation{ Thomas Young Centre and London Centre for Nanotechnology, 17--19 Gordon Street, London WC1H 0AH, United Kingdom }
\alsoaffiliation{ Department of Earth Sciences, University College London, Gower Street, London WC1E 6BT, United Kingdom }

\author{Angelos Michaelides}
\email{angelos.michaelides@ucl.ac.uk}
\affiliation{ Thomas Young Centre and London Centre for Nanotechnology, 17--19 Gordon Street, London WC1H 0AH, United Kingdom }
\alsoaffiliation{ Department of Physics and Astronomy, University College London, Gower Street, London WC1E 6BT, United Kingdom }

\title{
Toward Accurate Adsorption Energetics on Clay Surfaces 
}

\begin{document}

\begin{abstract}
Clay minerals are ubiquitous in nature and the manner in which they interact with their surroundings has
important industrial and environmental implications. Consequently, a molecular level understanding of
the adsorption of molecules on clay surfaces is crucial. In this regard computer simulations play an
important role, yet the accuracy of widely used empirical force field (FF) and density functional
theory (DFT) exchange-correlation functionals is often unclear in adsorption systems dominated
by weak interactions. Herein 
we present results from quantum Monte-Carlo (QMC) for water and methanol adsorption on the
prototypical clay kaolinite. To the best 
of our knowledge, this is the first time QMC has been used
to investigate adsorption at a complex, natural surface such as a clay. As well as being valuable in their own right,
the QMC benchmarks obtained provide reference data against which the performance of cheaper DFT methods can 
be tested. Indeed using various DFT exchange-correlation functionals yields a very broad range of adsorption
energies, and it is unclear {\em a priori} which evaluation is better. QMC reveals that in the systems considered 
here it is essential to account for van der Waals (vdW) dispersion forces since this alters both the absolute and
relative adsorption energies of water and methanol. We show, via FF simulations, that incorrect relative
energies can lead to significant changes in the interfacial densities of water and methanol solutions at the kaolinite interface.
Despite the clear improvements offered by the vdW-corrected and the vdW-inclusive functionals, absolute adsorption energies are often overestimated, suggesting that the treatment of vdW forces in DFT is not yet a solved problem.
\end{abstract}

\maketitle

\begin{center}
{\bf Graphical TOC Entry}\\
\fbox{\includegraphics[width=3.25in]{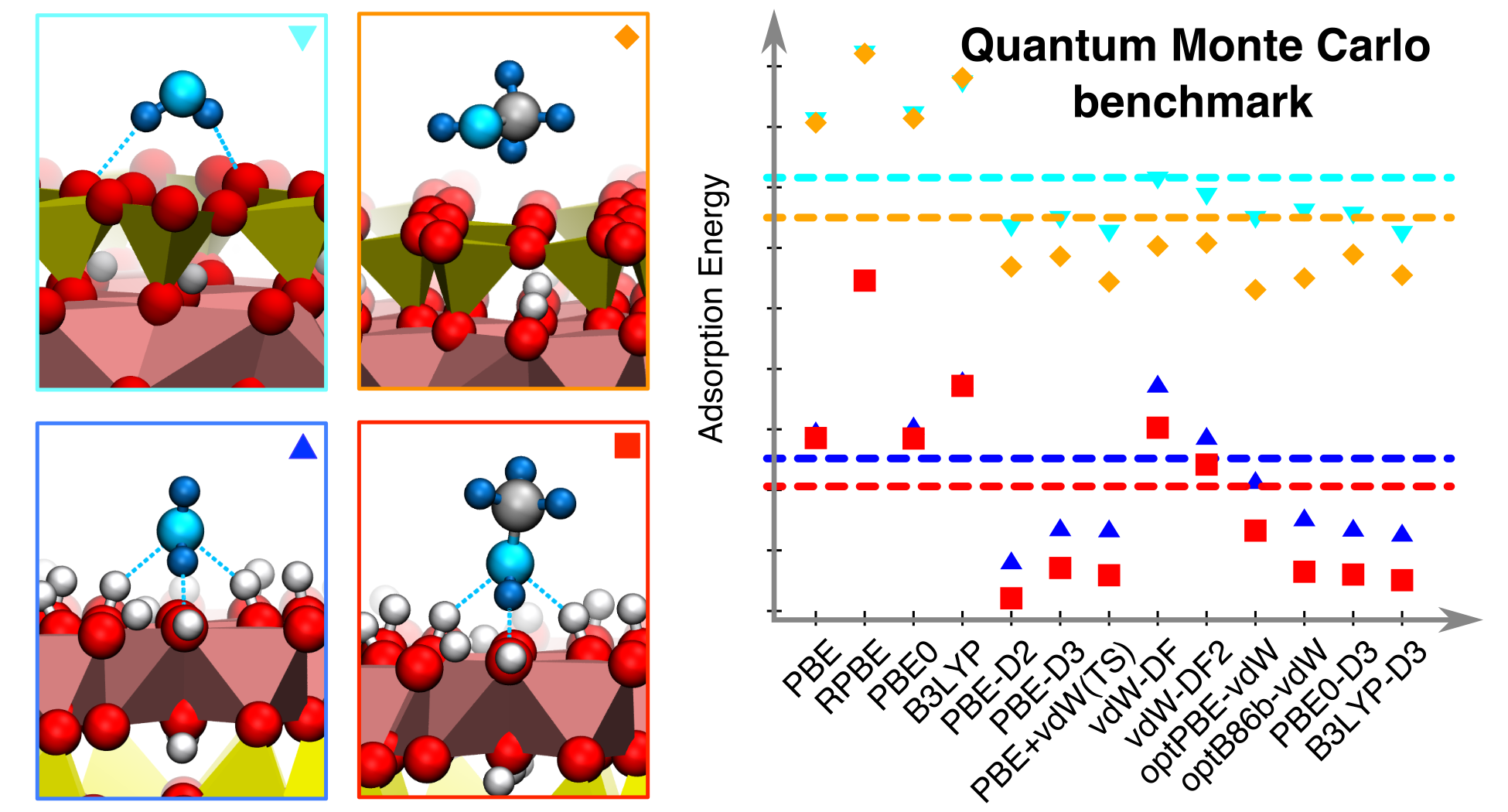}}
\end{center}

\section{Introduction}
\label{sec:introduction}


The accurate treatment of the adsorption of molecules on surfaces is a major challenge of materials modelling, with important applications in nanotechnologies and science: heterogeneous catalysis, sensors, corrosion, lubrication, friction and coatings, to name but a few. 
An important case to study is that of clays.
Clay minerals are natural aluminosilicates that find uses in a wide variety of fields such as medicine,
adhesives, paints and oil drilling.\cite{murray:clay-uses,
ipcs:clay-health, usgs:bentonite, kaolinite-medicine1,
kaolinite-medicine2, kaolinite-viruses1, clay-viruses2, clay-viruses3,
clays-serum} They also act as catalysts to ice nucleation in the atmosphere\cite{murray:review}
and help cleanse soils and groundwater
through adsorption of pollutants. A clear understanding of how molecules interact
with the surfaces of clays is of the utmost importance to understand,
improve and control such processes.

Reliable reference data from theory and simulation is of intrinsic value and often important as a complement to
experiments.\cite{campbell_experimental_2009, campbell_enthalpies_2013}
Computer simulations of water-surface interactions, at the molecular level, are often based on force fields (FF) and density functional theory (DFT) approaches.\cite{Carrasco:2012iu, bjorneholm_water_2016, striolo_carbon-water_2016} 
Although these techniques are incredibly powerful and useful, there are cases where their accuracy is not satisfactory.
FF potentials have parameters that have to be fit in order to reproduce experimental results or higher lever theoretical benchmarks, and this is not always straightforward.
DFT is traditionally more accurate than FFs but at a larger computational cost.
Unfortunately, DFT results are highly sensitive to the choice of the exchange-correlation (XC) functional used and nowadays
there are countless XC functionals to choose from.\cite{challengesDFT:2012,burke_perspective_2012}
In the field of materials science, the description of weak bonding interactions, and in particular London dispersion forces, is one of the most important challenges.
Immense progress has been made in this area recently,\cite{jiri:review,Grimme_Gerit_chemrev2016}
however, there is no rigorous way to systematically improve XC functionals and as a result validation with alternative methods is needed.
Of the various high level reference methods available,\cite{foulkes01, reviewQC, CCSDT:Rev2007, DMRG:2002, FCIQMC:JCP2009, FCIQMC:Nat2013, AFQMC:Zhang2003, LRDMC:prl2005, casula10} quantum Monte-Carlo (QMC) is a powerful
approach for obtaining benchmark values for solids, surfaces and large molecular systems.
QMC, within the fixed node diffusion Monte-Carlo (DMC) approach has already been used to tackle interesting materials science problems that have been beyond the reach of DFT (see e.g. Refs.~\citenum{Ice:prl2011, Morales:bulkwat:2014, Cox:2014, Benali:2014,AlHamdani:hBN:2015, gillan15,AFQMC:CoGraf, Morales:perspective2014, Mazzola:nat2014, Mazzola:prl2015, Zen-liquidwat, Chen:jcp2014, Wagner:2013, Wagner:prb2014}).
This has provided reference data which has exposed shortcomings in existing FF models and DFT XC functionals, which in 
turn aids the development of such approaches.


\begin{figure*}[t!]
   \centering
   \includegraphics[width=4.5in]{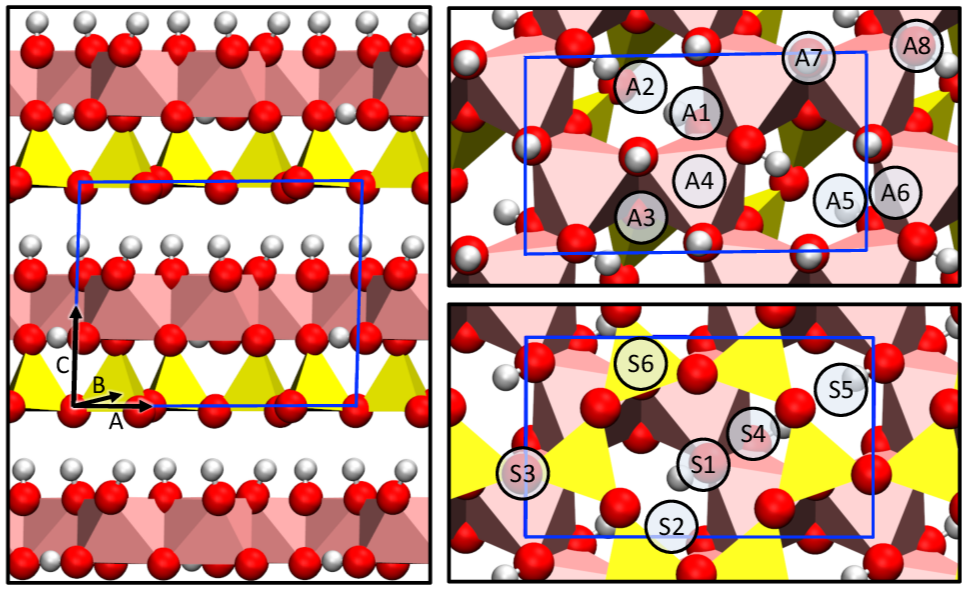}
   \caption{  
   Representation of the kaolinite structure. The hydrogens 
      are sketched in white, the oxygens in red, the silicons by
      yellow tetrahedra, and the aluminiums by pink octahedra. The
      conventional unit cell is indicated by the blue line. The figure
      on the left illustrates the layered bulk. The figures on the right
      are the hydroxyl-terminated face (top), and the
      silicate-terminated face (bottom). Various adsorption sites on
      the hydroxyl- and silicate-terminated face are labeled.
      }
   \label{fig:kaolinite_bulk}
\end{figure*}

In this paper we will use two QMC approaches to investigate molecular adsorption on a clay surface, namely DMC and lattice regularized diffusion Monte Carlo\cite{LRDMC:prl2005,casula10} (LRDMC).
%
The particular clay we will examine is
kaolinite ($\ce{Al4Si4O10(OH)8}$), as shown in Fig.~\ref{fig:kaolinite_bulk}. Since the first outline of the kaolinite crystal structure by
Pauling in 1930,\cite{pnas:kaolinite-pauling} numerous structural
studies using X-ray and neutron powder diffraction,\cite{Exp_X-ray,
Exp_neutron1, Exp_neutron2, Exp_neutron3} X-ray single crystal,
\cite{Exp_SingleCrystal} and electron diffraction methods,
\cite{Exp_ElectronDiffraction} as well as theoretical studies
\cite{cygan:AmMin, tunega:jcp-137_2012, schafer:jcp-B, jmc:civalleri}
have been carried out on kaolinite. 
Consequently it is one of the most suitable
aluminosilicate clays to assess the performance of various
theoretical methods. In addition, when looking at adsorption processes
on kaolinte, cleavage along the (001) basal plane leads to exposure of either aluminate
and silicate faces (Fig.~\ref{fig:kaolinite_bulk}). The aluminate ($\ce{AlOH}$) face
is terminated in hydroxyl groups and as a result is
regarded as hydrophilic, whereas the silicate ($\ce{SiO2}$) face which exposes saturated $\ce{Si-O}$ groups,
is considered to be hydrophobic. The distinct chemical nature of these two surfaces means
that adsorbates will interact differently with them, making kaolinite an interesting case study
for understanding the role of vdW forces on the adsorption
at clay mineral surfaces.

In what follows, we will provide benchmark values for the adsorption of water and methanol molecules at the pristine hydroxyl- and silicate-terminated (001) faces of kaolinite, by using DMC and LRDMC.
Then, we probe DFT XC functionals by considering a range of generalized gradient approximation (GGA) functionals,  hybrid functionals and
dispersion corrected density functionals which account for vdW forces.
In the case of molecules adsorbed on kaolinite we find that the bare GGAs and hybrids are quite unreliable: as expected adsorption energies are underestimated, but more importantly, the relative adsorption energies of water versus methanol do not even agree with QMC. 
Moreover, on the silicate-terminated face the molecules hardly bind at all and move quite far from the surface during geometry optimizations.
Accounting for vdW forces improves adsorption energies significantly and stabilizes the structures. However most of the vdW-corrected and vdW-inclusive functionals predict adsorption energies which are slightly too large compared to QMC.
This indicated that there remains room for improvement in terms of how vdW forces are handled in DFT.

The remainder of this paper is organized as follows. In Section~\ref{sec:setup} we outline the key computational details of our simulations. Since a range of techniques has been used, for brevity the more detailed descriptions of the computational setups are provided in the Supporting Information.  
QMC and DFT evaluations of the adsorption of water and methanol  at both (001) faces of kaolinite are reported and discussed in Section~\ref{sec:results}. 
Finally, we summarize our results and draw conclusions in Section~\ref{sec:conclusion}.

\section{Methods and Computational Setup}
\label{sec:setup}

\subsection{Adsorption Energy Evaluation}

Adsorption was examined on a single layer of kaolinite, a system with 2D periodicity along the A and B axes as indicated in Fig.~\ref{fig:kaolinite_bulk}.
The simulated supercell was $1\times2$ the conventional unit cell of bulk kaolinite (\textit{ca.} 10.38~\AA~$\times$~9.01~\AA) and comprises of 8 aluminiums, 8 silicons, 36 oxygens and 16 hydrogens for the kaolinite slab, plus the atoms of the adsorbed molecule. 
Note that with \textit{ca.} 300 electrons the simulations are large for QMC calculations.

The adsorption energy, $E^{M@X}_{\textrm{ads}}$, of the molecule $M$ at the face $X$ of the kaolinite layer, 
where $M$ can  either be the water ($\ce{H2O}$) or the  methanol ($\ce{MeOH}$), 
and $X$ can be the hydroxyl-terminated ($\ce{AlOH}$) face or the silicate-terminated ($\ce{SiO2}$) face, 
can be evaluated in two ways:
the first method, hereafter called {\em complex-minus-fragments}, is computed as
\begin{equation}
  \label{eqn:Eads1}
  E^{{M@X}}_{\textrm{ads}}=E_{\ce{slab}+{M@X}}- E_{{M}} - E_{\ce{slab}}
\end{equation}
where $E_{\ce{slab}+{M@X}}$ is the total energy for the system with $M$ at the $X$-face of the kaolinite slab, and $E_{\ce{slab}}$ and $E_{{M}}$ are the total energies of the isolated slab and the isolated molecule ${M}$, respectively;
the second method, hereafter called {\em complex-minus-far}, is computed as
\begin{equation}
  \label{eqn:Eads2}
  E^{{M@X}}_{\textrm{ads}}=E_{\ce{slab}+{M@X}}- E_{\ce{slab}-M} 
\end{equation}
where $E_{\ce{slab}-M}$ is the total energy of a system where the kaolinite slab and the molecule $M$ are far enough apart that their interaction is negligible.
The two methods are equivalent if and only if: 
(i)  the size-effects due to the periodicity of the system are negligible; and 
(ii)  the electronic structure calculations are performed with methods that are exactly size-consistent.
If these conditions are not satisfied in general we have that 
$E_{\ce{slab-M}} \ne E_{\ce{M}} + E_{\ce{slab}}$, 
meaning that Eqs.~\ref{eqn:Eads1} and \ref{eqn:Eads2} provide different evaluations of the adsorption energies.
In particular, whenever size-effects are detected, the \textit{complex-minus-far} method usually benefits from a larger error cancellation.
On the other hand, in cases where size-effects are negligible and electronic structure methods are size-consistent, there are no residual interactions between the molecule and the periodic partners, then the \textit{complex-minus-fragments} method is usually to be preferred.
The reason for that is the computational cost: for a system with $N$ electrons the computation is proportional to $N^\gamma$, with $\gamma > 1$ (\textit{e.g.}, in DFT $\gamma$ is typically between 2 and 3, and in QMC between 3 and 4), so the cost for calculations of $E_{\ce{slab}}$ and $E_{\ce{M}}$ are cheaper than $E_{\ce{slab-M}}$.
Moreover, when several adsorption energies need to be evaluated, $E_{\ce{slab}}$ is calculated only once, whereas a different calculation of $E_{\ce{slab-M}}$ has to be performed for each molecule $\ce{M}$.

\subsection{QMC calculations}

The two QMC approaches used are DMC and LRDMC. 
They are both projection Monte Carlo methods: they can access  the electronic ground state energy of the system by iteratively projecting an initial trial wave function $\psi_T$ into the ground state, with the constraint that the projected wave function $\Phi$ has the same nodal surface of an appropriately chosen guiding function $\psi_G$ (fixed node approximation).\cite{Reynolds:1982en,foulkes01}
Both the trial and the guiding wave functions are parametrized functions, and they have to be the best approximation of the ground state that we can provide (given the constraint of their ansatz). Thus, usually they are taken such that  $\psi_T=\psi_G=\psi_\text{VMC}$, where $\psi_\text{VMC}$ is the best function obtained within a variational Monte Carlo approach, with the variational parameters optimized in order to minimize either the variational energy or the variance.
Whenever $\psi_G$ has the exact nodal surface, the approach is exact, otherwise it gives the best approximation of the ground state given the fixed node constraint.

In projection Monte Carlo approaches there is a second approximation in how the projection is performed, and it is different in DMC and LRDMC.
The projection in DMC comes from the imaginary time Scr\"{o}dinger equation; it is implemented as an imaginary time evolution, where a time-step  $\tau$ has to be chosen.
The chosen $\tau$ is a trade-off between accuracy and computational cost:
exact results are obtained for $\tau \to 0$, 
but the computational cost is $\propto 1/\tau$.
The finite time-step error can be controlled by performing several calculations with different values of  $\tau$ and finally extrapolating to the continuum limit $\tau\to 0$. However, in big systems like those considered here, the extrapolation is impractical and sometimes unfeasible or unreliable,\cite{sizeconsDMC} but it is sufficient to consider the results for a $\tau$ small enough that the expected finite-time error is smaller than the required accuracy.
Here, we have chosen $\tau$  in order to have an expected time-step error smaller than the stochastic error of the evaluations, see 
Section~SI in the Supporting Information. 
On the other hand, 
LRDMC is based on the spatial discretization of the molecular Hamiltonian on a lattice of mesh size $a$, and it resorts to the projection scheme used also in the Green function Monte Carlo algorithm.\cite{Sorella:2000p17651,Buonaura:1998p25304}
The error induced by the finite mesh size $a$ is analogous to the time-step error appearing in standard DMC calculations. 
%
%
LRDMC preserves the variational principle even when used in combination with nonlocal pseudopotentials\cite{LRDMC:prl2005} (PPs), and it is size-consistent for any value of the mesh $a$,
maintaining its efficiency even  for systems with a large number of electrons.\cite{casula10}

Both DMC and LRDMC  provide excellent benchmark values for weakly interacting systems, as established in a number of studies.~\cite{Ice:prl2011,santra_on_2013,Quigley:Ice_0_i_Ih:jcp2014,Morales:bulkwat:2014,Cox:2014,Benali:2014,AlHamdani:hBN:2015,gillan15,AFQMC:CoGraf,Morales:perspective2014,Zen-liquidwat}
We used here a standard setup, described in detail in the Supporting Information. 
The stochastic error associated with the QMC evaluations of the adsorption energy is {\em ca.}~20~meV.
The systems under consideration are too large 
for a QMC-based structural optimization, even at the variational level, so the reference structures were those obtained from the PBE-D3 functional, as described in Section~\ref{sec:DFTopt}. 
As we will see in Section~\ref{sec:results}, PBE-D3 configurations are in good agreement with those obtained by all the other vdW-inclusive functionals, thus the bias given by the use of PBE-D3 configurations is expected to be small compared to the stochastic error of the DMC evaluation.

There is one aspect of QMC simulations that deserves special care in this specific system, namely the finite-size errors (FSEs).\cite{Lin:qmctwistavg:pre2001, Chiesa:size_effects:prl2006, KZK:prl2008, drummond_finite-size_2008, kspecial.Sandro}
QMC is a many-body method, and in contrast to (effective) one-particle methods such as DFT, QMC cannot simply exploit Bloch's theorem in calculations for extended periodic systems.
FSEs can be taken into account by performing simulations in larger periodic supercells, through the twist-average method,\cite{Lin:qmctwistavg:pre2001} through corrections to the Ewald energy,\cite{Chiesa:size_effects:prl2006} or the Kwee, Zhang, Krakauer (KZK) method.\cite{KZK:prl2008} 
In this work we have used the KZK method
(see Section~S3 in the Supporting Information).

\subsection{DFT calculations}\label{sec:DFTsetup}

There is by now an almost limitless variety of DFT XC functionals that we could examine.\cite{burke:jcp-perspective} 
Here we restrict ourselves to:
LDA functional; \cite{LDA_PerdewZunger}
two GGA functionals, PBE,\cite{PBE, PBE_Erratum} 
RPBE;\cite{RPBE} 
two hybrid functionals, PBE0,\cite{PBE0} B3LYP;\cite{Vosko:1980ui,LEE:1988ub,Becke:1993vx,STEPHENS:1994vd}
three  vdW-corrected PBE functionals:
PBE-D2\cite{DFT-D2}, PBE-D3\cite{DFT-D3} (both from Grimme, D3 correction with ``zero-dampling''\cite{Grimme:BJdamping}),
and PBE+vdW(TS) from Tkatchenko and Scheffler;\cite{DFT-TS}
two vdW-corrected hybrid functionals, PBE0-D3, B3LYP-D3\cite{DFT-D3} (both from Grimme\cite{DFT-D3});
and four self-consistent non-local functionals 
(often called vdW-inclusive functionals),
the original vdW-DF from Dion (also named revPBE-vdW),\cite{revPBE, vdW-DF}
the second generation vdW-DF2,\cite{vdW-DF2}
as well as   
optPBE-vdW\cite{klimes-vdW-DF} 
and 
optB86b-vdW\cite{optB86b-vdW} 
from Klime\v{s} {\em et al.} 
We stress that the latter four vdW-inclusive functionals are actually based on GGAs and basically differ from the vdW-corrected GGA functionals (e.g., PBE-D2, PBE-D3 and PBE+vdW(TS)) only in the way the dispersion energy is approximated.\cite{jiri:review, Grimme_Gerit_chemrev2016}
Other functionals and vdW-corrections have been tested, and results obtained using a comprehensive set of approaches is reported in Table~S1 of the Supporting Information.
Adsorption energies were evaluated using the {\em complex-minus-fragments} method, see Eq.~\ref{eqn:Eads1},
but the results are the same as those obtained with the \textit{complex-minus-far method}, as expected.
Further details about the setup of the DFT calculations are reported in Section~S4 of the Supporting Information. 

\subsection{Molecular Dynamics Simulations}

We also performed a series of molecular dynamics simulations using
classical force fields for aqueous water-methanol solutions on kaolinite. The kaolinite slab was modeled as a single sheet
of kaolinite (approx. $31\times 36$ \AA) using the CLAYFF
force field,\cite{cygan:clayff} and the OH bond lengths were
constrained using the P-LINCS algorithm.\cite{hess:lincs} Above this
slab 538 TIP4P/2005 \cite{vega:tip4p-2005} water and 230 OPLS/UA
methanol\cite{jorgensen1986optimized} molecules were randomly placed in order
to create a liquid film on the kaolinite surface.
The standard Lorentz-Berthelot mixing rules
were used to compute cross-interactions, except to adjust the
adsorption energies as detailed below. Using the GROMACS 4.5
simulation package,\cite{gromacs4} constant volume and temperature
dynamics were propagated using a leap-frog integrator and a
Nos\`{e}-Hoover chain thermostat, along with replica-exchange amongst
eight replicas with temperatures ranging from 275--310\,K in 5\,K
intervals. Real-space interactions were truncated at 9\,\AA{} with
corrections to the energy applied and particle-mesh Ewald was used to
account for long-ranged electrostatics \cite{pme1,pme2} with the
corrections for the slab geometry of the system.\cite{yeh:2dEwald}
A time step of 2\,fs was used and molecular dynamics simulations performed for at least
11\,ns, with the first nanosecond being disregarded as equilibration.

Adsorption energies were computed after geometry optimization using
the \emph{complex-minus-fragments} method. This was done in three
ways: First, the adsorption energy was computed by applying the standard
mixing rules. This yielded adsorption energies of $642$\,meV and
$640$\,meV for water and methanol, respectively, and $\Delta E_\textrm{ads} =
-2$\,meV; Second, the strength of the Lennard-Jones interaction between
the \ce{CH3} group of methanol and the oxygen atoms of the kaolinite OH
groups was adjusted such that $\Delta E_\textrm{ads}$ matched that of PBE; Finally,
the same was done to match $\Delta E_\textrm{ads}$ obtained by DMC.

\section{Results}
\label{sec:results}

\subsection{ Reference structures for water and methanol adsorbed on kaolinite }\label{sec:DFTopt}

Water adsorption on the hydroxyl-terminated face of kaolinite has been studied experimentally\cite{Costanzo, Lipsicas, Khalfi} and theoretically,\cite{Kaolinite_Hu0, Kaolinite_Hu1, Kaolinite_Hu2, Tunega_1, Tunega_2, Tunega_3, Patey_FFsim_2009} whereas adsorption on the silicate-terminated face is less well studied. Very little is known about methanol adsorption on either face.
In the following, the most stable adsorption structures identified for water and methanol on the two kaolinite surfaces are presented. 

\begin{figure*}[tb]
\centering 
\includegraphics[width=6.0in]{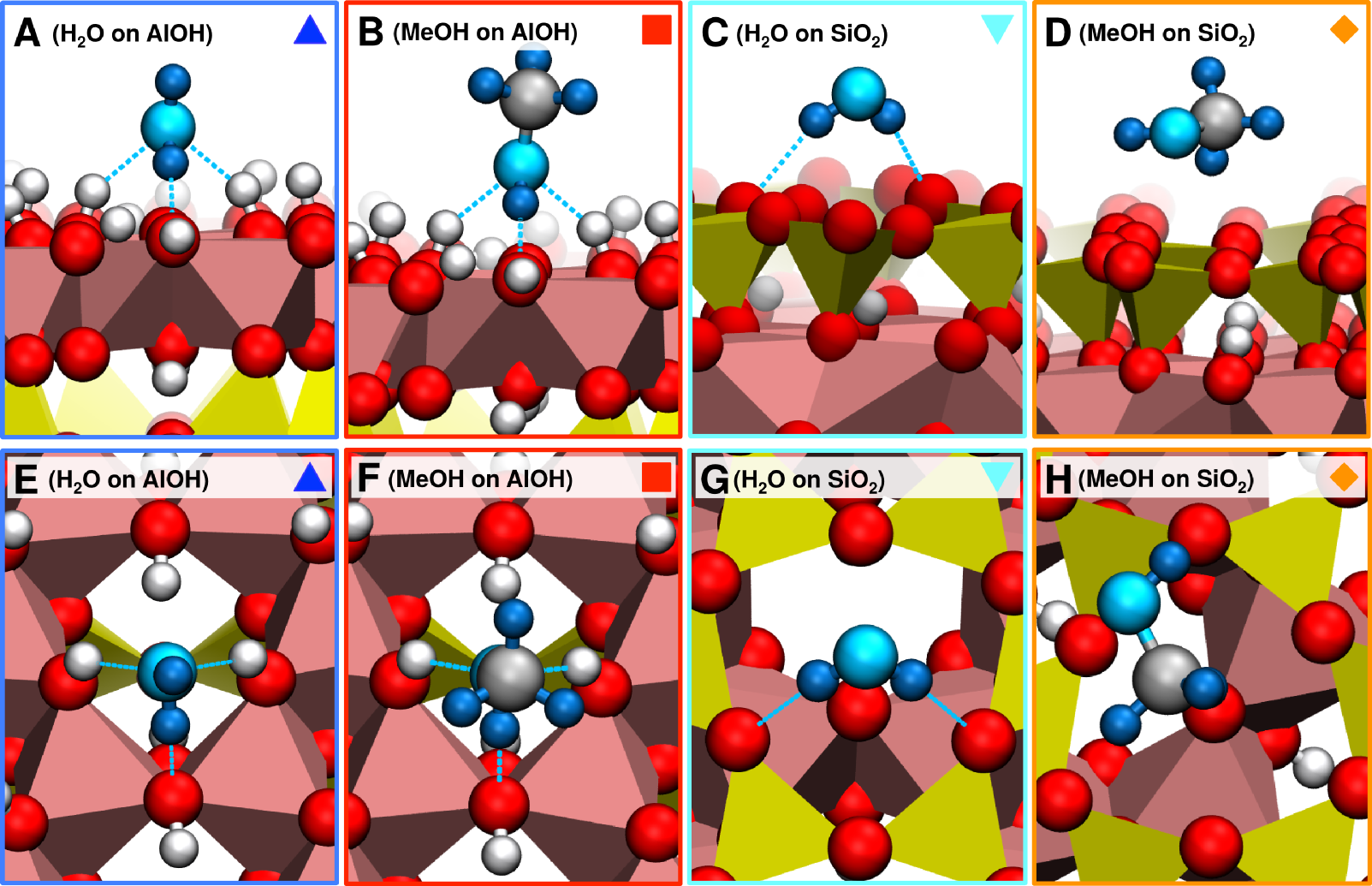}
\caption{Adsorption of water and methanol on the hydroxyl-terminated and the silicate-terminated faces of kaolinite (side view in first row, top view in second row). Geometries are relaxed using the PBE-D3 functional and have been taken as reference for the other calculations. The adsorbed molecule on kaolinite is depicted in cyan and gray and the H-bonds are represented by the blue dashed lines.}
\label{fig:ads_refgeo}
\end{figure*}

\begin{figure}[htbp]
\centering 
\includegraphics[width=3.in]{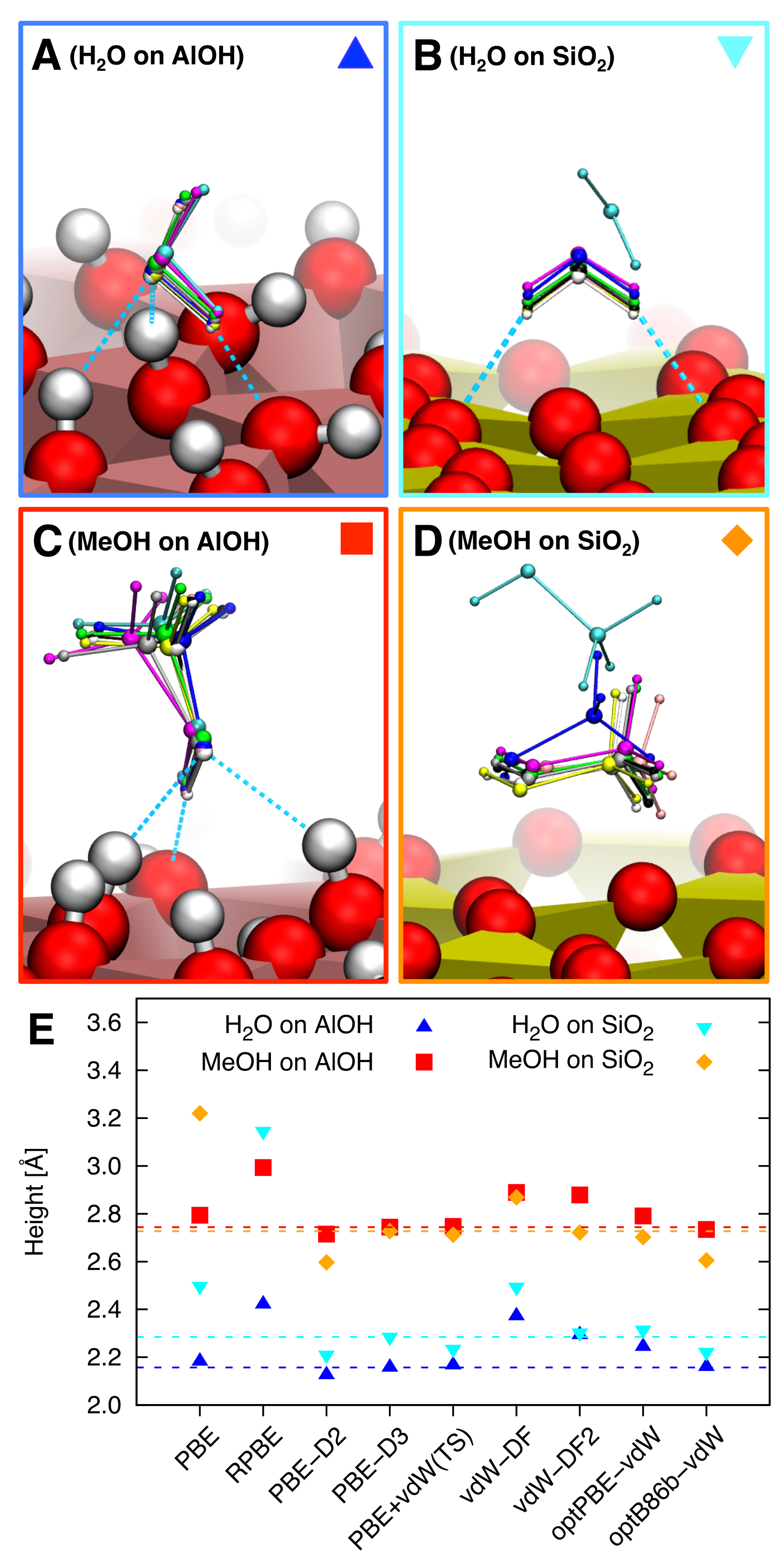}
\caption{
Panels A, B, C and D show the most stable DFT structures for the adsorption of water and methanol on the hydroxyl- and silicate-terminated faces of kaolinite, provided by the different XC functionals considered.
The color scheme for the various functionals is:
blue for PBE, cyan for RPBE, white for PBE-D2, black for PBE-D3 (that is also the reference for QMC calculations), pink for  PBE+vdW(TS), violet for vdW-DF, green for vdW-DF2, gray for optPBE-vdW, and yellow for optB86b-vdW.  
Panel E shows the height of the center-of-mass of the adsorbed molecules from the average surface plane defined by the surface oxygens 
for the different XC functionals.
The four dashed horizontal lines correspond to the values for the reference PBE-D3 structures. }
\label{fig:ads_conf}
\end{figure}

On each surface a range of adsorption sites was considered, as indicated in Fig. \ref{fig:kaolinite_bulk}.
According to the number of H-bonds formed between the adsorbate and the surface, the adsorption sites can be classified into three categories: threefold, twofold and onefold sites.
The most stable configurations obtained, using DFT with the PBE-D3 functional, are shown in Figs.~\ref{fig:ads_refgeo}A--\ref{fig:ads_refgeo}H.
These structures have been taken as the reference for DMC, LRDMC and the other DFT calculations.
In addition, 
starting from the reference PBE-D3 structures, we have relaxed the geometries for each of the different functionals considered, as shown in Figs.~\ref{fig:ads_conf}A--\ref{fig:ads_conf}D.
Fig.~\ref{fig:ads_conf}E compares the distance of the molecules from the slab as obtained with different functionals.

Concerning water at the hydroxyl-terminated face (\ce{H2O} on \ce{AlOH}), all structures initially put in twofold and onefold sites (A5-A8) moved to the threefold site A1. This preference for the A1 site agrees with previous DFT studies with local\cite{Tunega_2} and semi-local\cite{Kaolinite_Hu1} XC functionals.  
In the most stable configuration, shown in Figs.~\ref{fig:ads_refgeo}A and \ref{fig:ads_refgeo}E, the $C_2$ axis of the water
molecule lies almost parallel to the plane of the surface. 
The water molecule donates one H-bond (OH-distance of 1.69~\AA) to and accepts two H-bonds
(2.01 and 2.04~\AA) from the surface (PBE-D3 values). 

Let us now consider the adsorption of methanol at the hydroxyl-terminated face (\ce{MeOH} on \ce{AlOH}). 
One way of viewing methanol is as a water molecule with one of its hydrogen atoms replaced by a methyl group. 
This leads to two possible types of interaction with the surface: (i) hydrogen bond formation with the hydroxyl functional group; and (ii) dispersion interactions arising from the $-\ce{CH3}$ group. 
All calculations in which the methanol began parallel to the surface ended with the methanol perpendicular to the surface, maximizing the distance between the $-\ce{CH3}$ group and the kaolinite. The $-\ce{CH3}$ group can therefore be considered a `spectator group' that does not participate directly in the adsorption on the surface. 
The adsorption of methanol is therefore very similar to that of water and indeed, we find that A1 is the most favorable site, with the methanol donating one H-bond to and accepting two from the surface. 
As was the case for the water structure, the H-bond donated by the methanol is much stronger than the two H-bonds it accepts:
1.68~vs.~1.97 and~2.03~\AA, respectively, with the PBE-D3 functional. 
The most stable configuration is shown in Figs.~\ref{fig:ads_refgeo}B and \ref{fig:ads_refgeo}F.

As noted above, adsorption of water at the silicate-terminated face (\ce{H2O} on \ce{SiO2}) is less well studied than adsorption at the hydroxyl-terminated face.\cite{Patey_FFsim_2009, Tunega_1, Tunega_2, Tunega_3} 
Of the six adsorption sites (S1-S6) considered here, the onefold S5 site turned out to be the most stable at the GGA level, and the twofold S1 generally is the most stable for the  vdW-corrected and vdW-inclusive functionals. 
The PBE-D3 structure is depicted in Figs.~\ref{fig:ads_refgeo}C and \ref{fig:ads_refgeo}G. 

The most stable structure found for methanol at the silicate-terminated face (\ce{MeOH} on \ce{SiO2})
using the PBE-D3 functional is shown in Figs.~\ref{fig:ads_refgeo}D and \ref{fig:ads_refgeo}H. 
The leading interaction here is dispersion; there is no H-bond-like interaction because the OH group of the methanol is parallel to the surface of the slab.

\subsection{Benchmark results from DMC and LRDMC}\label{sec:DMC}

\begin{table*}[htbp]													
\caption{ 
DMC and LRDMC evaluations (in meV) of the 
adsorption energy of water and methanol molecules on the hydroxyl- and silicate-terminated faces of kaolinite, and the water minus methanol difference,  
$\Delta E_\textrm{ads}=E^{{\ce{H2O}}}_{\textrm{ads}} - E^{{\ce{MeOH}}}_{\textrm{ads}}$, 
for each face of kaolinite ($\Delta E_\textrm{ads}$  is positive when methanol is more strongly adsorbed, negative otherwise).
As discussed in the text, bare DMC and LRDMC results are affected by finite-size errors (see Section~S3 and Table~S2 in the Supporting Information) 
that we have estimated and corrected the adsorption energies for accordingly.
In addition, bare LRDMC evaluations are affected by an unphysical dipole-dipole interaction between the periodic slabs (because in this case 2D periodicity was not available and we had to use 3D periodicity), 
thus we have included a dipole interaction correction.  
LRDMC simulations have not been performed for adsorption on the $\ce{SiO2}$ face.
\\ }													
\label{tab:QMCadsene}													
\begin{tabular}{ l  c c c  c   c c c }													
\hline
\hline													
	&	\multicolumn{3}{ c }{hydroxyl-terminated face}				&~~&	\multicolumn{3}{ c }{silicate-terminated face}					\\
	\cline{2-4}\cline{6-8}
	&	$\ce{H2O}$	&	$\ce{MeOH}$	&	$\Delta E_\textrm{ads}$ &&	$\ce{H2O}$	&	$\ce{MeOH}$	&	$\Delta E_\textrm{ads}$	\\
\hline													
\hline
bare DMC (Eq.~\ref{eqn:Eads2})	&	-632$\pm$18	&	-677$\pm$18	&	45$\pm$25	&&	-172$\pm$23	&	-236$\pm$18	&	64$\pm$29	\\
FSE correction		&	-16	&	-17	&	+1	&&	-12	&	-14	&	+2	\\
{\bf corrected DMC} 	&	{\bf -648$\pm$18}	&	{\bf -694$\pm$18}	&	{\bf 46$\pm$25}	&&	{\bf -184$\pm$23}	&	{\bf -250$\pm$18}	&	{\bf 66$\pm$29}	\\
\hline													
bare LRDMC  (Eq.~\ref{eqn:Eads1})	&	-674$\pm$14	&	-736$\pm$13	&	64$\pm$13	&&		&		&		\\
FSE correction		&	+35	&	+73	&	-38	&&		&		&		\\
dipole correction 	&	-36	&	-38	&	 +2	&&		&		&		\\
{\bf corrected LRDMC}	&	{\bf -675$\pm$14}	&	 {\bf -701$\pm$13}	&	{\bf 26$\pm$13}	&&		&		&		\\
\hline
\hline													
\end{tabular}													
\end{table*}													

The DMC and LRDMC results for water and methanol adsorption on the two faces of kaolinite are reported in Table~\ref{tab:QMCadsene}.
As mentioned in the previous section, ``bare'' DMC and LRDMC evaluations have to be corrected for finite-size effects and in our LRDMC simulations there is also an unphysical dipole interaction between slabs due to the 3D periodicity employed. 
The DMC calculations have been performed with 2D periodicity and so do not suffer from the latter problem.
The bare and corrected results are reported in Table~\ref{tab:QMCadsene}.
From this it can be seen that our best estimates of the adsorption energy of water on the hydroxyl-terminated face are 
$-648 \pm 18$~meV with DMC and 
$-675\pm 14$~meV with LRDMC.
For methanol our best estimates of the adsorption energy are 
$-694 \pm 18$~meV with DMC and 
$-701\pm 13$~meV with LRDMC.
We notice that we are in the chemisorption regime both for water and for methanol, although the adsorption energy of methanol is slightly larger.
Note that for both molecules the DMC and LRDMC evaluations are in good agreement, with the differences falling within the stochastic error of the evaluations.
This shows that fixed-node projection QMC schemes are robust approaches: they are only slightly affected by the actual computational setup and implementation. Nonetheless, two slightly different adsorption energies for each case are obtained, and we should choose only one of them to use as our benchmark. We feel that in the specific case considered here the DMC values are likely to be more reliable since they have been obtained in 2D; as opposed to the LRDMC results which have been corrected for the dipole in the 3D cell. 
Moreover, 
the reported LRDMC evaluations use Eq.~\ref{eqn:Eads1}, which has larger FSE than the reported DMC evaluations, which use Eq.~\ref{eqn:Eads2}.

Having compared the results of the two QMC approaches on the hydroxyl-terminated face, we have only performed a DMC evaluation on the silicate-terminated face.
The DMC adsorption energy at the silicate-terminated face is $-184\pm 23$~meV for water and $-250\pm 18$~meV for methanol.
The methanol adsorbs more strongly than water, as for the hydroxyl-terminated face, but in this case the adsorption is weaker, and we are in the physisorption regime.

\subsection{Evaluation of DFT XC functionals: Adsorption energies and structures}
\label{sec:AlOHface}
We now examine how the various DFT XC functionals considered in this study perform for water and methanol adsorption on the two faces of kaolinite.

\begin{table*}[htbp]													
\caption{ Adsorption energy of water, $E^\ce{H2O}_\textrm{ads}$, and of methanol, $E^\ce{MeOH}_\textrm{ads}$, on the hydroxyl-terminated face of kaolinite, and adsorption energy difference, 
$\Delta E_\textrm{ads}=E^{\ce{H2O}}_{\textrm{ads}} - E^{\ce{MeOH}}_{\textrm{ads}}$,
between water and methanol, obtained with DMC and several DFT XC functionals. The best performing functional is indicated in bold. All energy values are in meV. Energies have been obtained on PBE-D3 optimized structures but in parenthesis we also report the adsorption energies when geometries are fully relaxed consistently for each  GGA and GGA+vdW functional. \\ }													
\label{tab:adseneAlOH}													
\begin{tabular}{  l   c c c   }													
\hline
\hline													
	&	\multicolumn{3}{ c }{hydroxyl-terminated face}					\\
	\cline{2-4}
Method	&	$E^\ce{H2O}_\textrm{ads}$	&	$E^\ce{MeOH}_\textrm{ads}$	&	$\Delta E_\textrm{ads}$	\\
\hline													
\hline
DMC	&	-648$\pm$18	&	-694$\pm$18	&	46$\pm$25	\\
\hline													
LDA	& -1102 &	-1138 &	36 \\
\multicolumn{4}{l}{\em GGA functionals} \\
PBE              & -607(-608)   &  -614(-616)   &       7(8)    \\
RPBE             & -360(-381)   &  -354(-380)   &      -6(-1)   \\
\multicolumn{4}{l}{\em hybrid functionals} \\
PBE0 & -599&	-615&	16 \\
B3LYP & -524&	-528&	4\\
\multicolumn{4}{l}{\em GGA+vdW functionals} \\
PBE-D2           & -822(-826)   &  -879(-882)   &      57(56)   \\
PBE-D3           & -767   &  -829   &      62   \\
PBE+vdW(TS)           & -769(-764)   &  -841(-833)   &      72(69)   \\
vdW-DF           & -530(-566)   &  -597(-635)   &      66(69)   \\
{\bf vdW-DF2}          & {\bf -616}(-641)   &  {\bf -658}(-686)   &      {\bf 42}(44)   \\
optPBE-vdW       & -689(-699)   &  -767(-779)   &      78(80)   \\
optB86b-vdW      & -751(-752)   &  -835(-836)   &      84(85)   \\
\multicolumn{4}{l}{\em hybrid+vdW functionals} \\
PBE0-D3 & -768&	-840&	72 \\
B3LYP-D3 & -776&	-849&	72 \\
\hline			
\hline										
\end{tabular}													
\end{table*}

\begin{figure}[htbp]
\centering 
\includegraphics[width=2.8in]{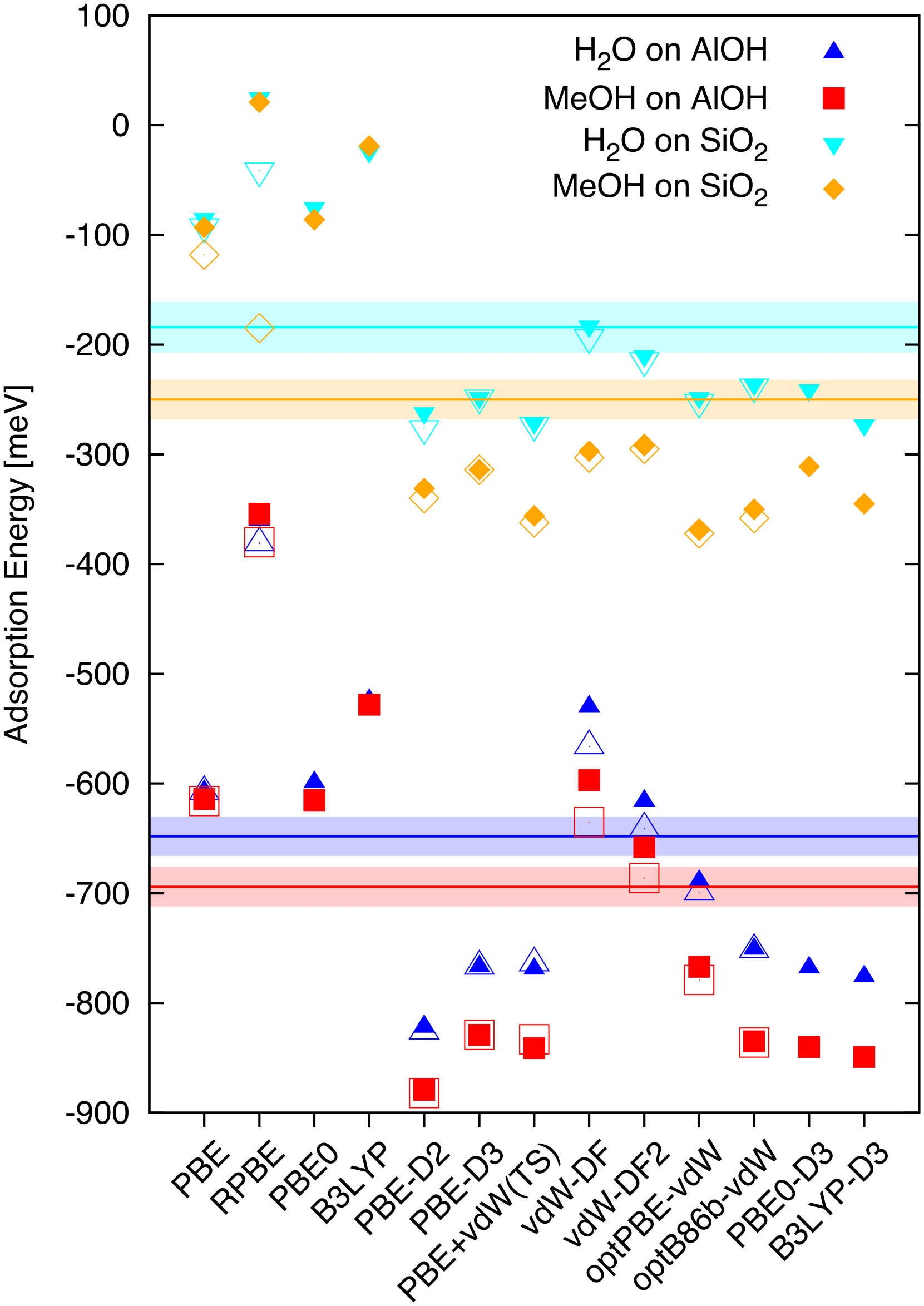}
\caption{
Adsorption energies on kaolinite obtained by various XC functionals and DMC, for 
water on the hydroxyl-face ($\ce{H2O}$ on $\ce{AlOH}$, blue upper triangles),   
methanol on the hydroxyl-face ($\ce{MeOH}$ on $\ce{AlOH}$, red squares), 
water on the silicate-face ($\ce{H2O}$ on $\ce{SiO2}$, cyan lower triangles),  and
methanol on the silicate-face ($\ce{MeOH}$ on $\ce{SiO2}$, orange diamonds).
Filled points represent the values with the reference structures (obtained using PBE-D3) and empty points  (reported only for GGA and GGA+vdW functionals) correspond to relaxed structures for the specific functional. The solid lines are the reference DMC adsorption energies and the shaded areas show 
the stochastic error.}
\label{fig:ads}
\end{figure}

{\bf I. Water adsorption on the hydroxyl-terminated face of kaolinite}:
In Table~\ref{tab:adseneAlOH} and Fig.~\ref{fig:ads} we summarize the adsorption energies obtained with the different density functionals. 
At the GGA level PBE and RPBE give significantly different adsorption energies, with RPBE providing a value that is roughly 50\% that of PBE.
In line with the smaller adsorption energy we also see that the bonds the molecule makes with the surface with the RPBE functional are considerably longer than what is obtained with PBE. 
Specifically, with RPBE the two H-bonds accepted from the surface are 2.30~\AA~ and  2.36~\AA, and the one donated is 1.81~\AA, versus 2.03~\AA, 2.06~\AA~ and 1.70~\AA~ with PBE.
Including dispersion interactions does not drastically change the geometry of the adsorbed water monomer at the hydroxyl-terminated face:
the bond lengths at the PBE-D2, PBE-D3,  PBE+vdW(TS) and opt-B86b-vdW level are slightly shortened, but they remain within 0.05~\AA~ of the PBE structure.
PBE-D2 predicts the shortest distance from the surface, and the shortest H-bonds.
The other functionals 
give H-bond lengths between the values provided by PBE and RPBE.
From the shortest to the longest interaction distance, the functionals are ranked in the following order:
PBE-D2 $<$ PBE-D3 $\sim$ optb86b-vdW $\sim$  PBE+vdW(TS) $<$ PBE $<$ optPBE-vdW $<$ vdW-DF2 $<$ vdW-DF $<$ RPBE.
This trend roughly follows the sequence of adsorption energy predicted by the functionals and is also consistent with previous studies of DFT XC functionals for hydrogen bonded systems.\cite{javi:wetting, JCP:Javier, JCC:Gross,santra_on_2013}
Relaxation from the PBE-D3 geometry  (performed for all the GGA and GGA+vdW functionals) results in rather small increases in adsorption energies. The maximum difference is observed for vdW-DF, with an increase of 36~meV upon relaxation. 

A comparison with the QMC adsorption energies shows that vdW-DF2 and optPBE-vdW yield the best agreement, with the former providing a slightly underestimated adsorption energy (by $-32 \pm 18$~meV) and the latter a slightly overestimated one (by $41 \pm 18$~meV).
It also appears that the two GGA  functionals (PBE and RPBE), the two hybrid functionals (PBE0 and B3LYP), and vdW-DF underestimate the interaction energy, whereas all the other functionals (PBE-D2, PBE-D3,  PBE+vdW(TS), optB86b-vdW,  PBE0-D3 and B3LYP-D3) overestimate the interaction.
 In particular, evaluations of the adsorption using vdW-corrected hybrid functionals do not seem to improve significantly compared to the GGA+vdW approaches.

\begin{figure*}[htbp]
\centering 
\includegraphics[width=5in]{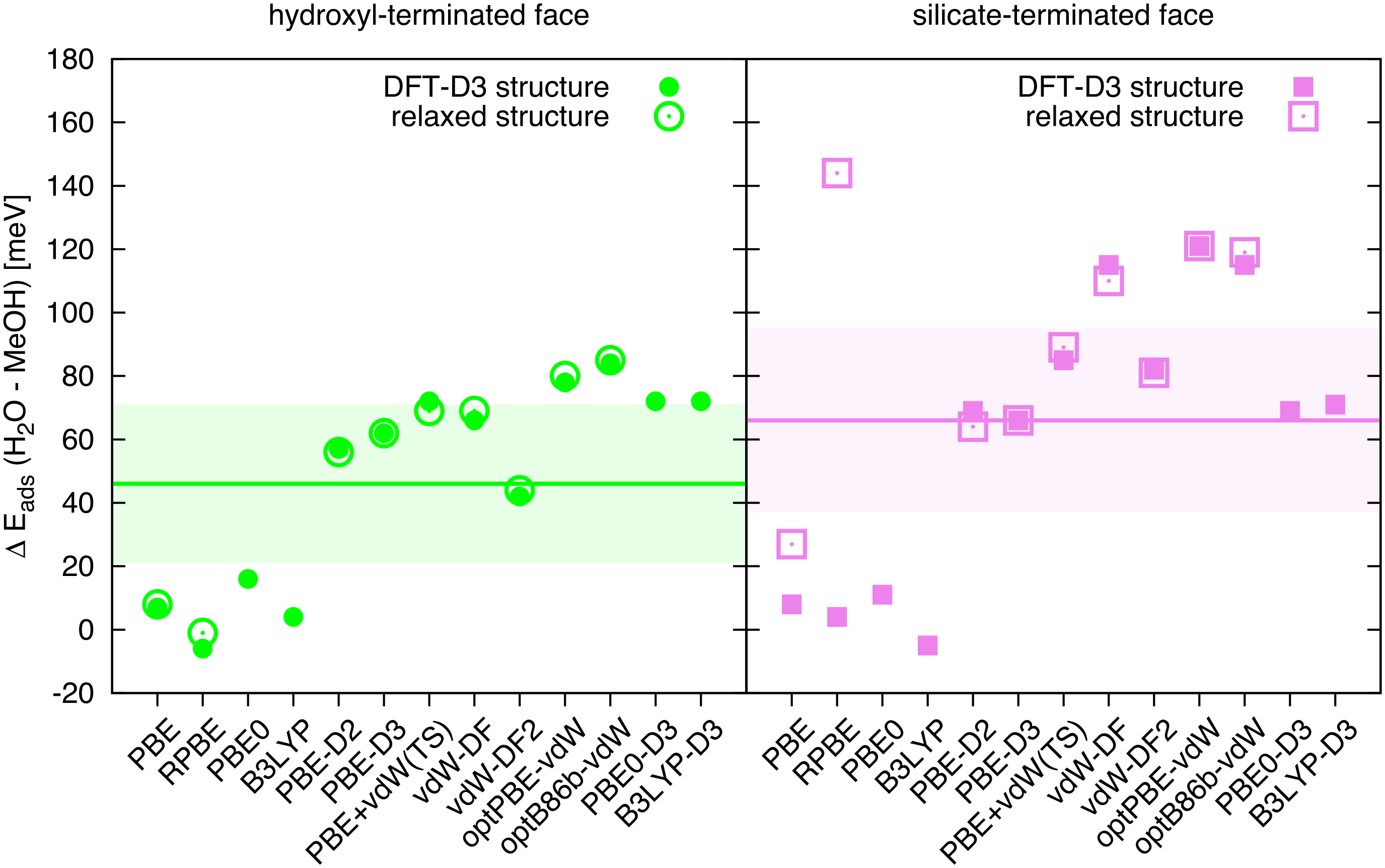}
\caption{
Difference in adsorption energy, between water and methanol on the hydroxyl-terminated (left panel) and silicate-terminated (right panel) faces of kaolinite, as obtained from various XC functionals and DMC. 
Positive values mean that methanol binds more strongly than water.
Filled points represent the values for the configurations optimized using PBE-D3 and empty points  (reported only for GGA and GGA+vdW functionals) correspond to relaxed structures for the specific functional. The solid lines are the reference DMC adsorption energies and the shaded areas show 
the stochastic error. }
\label{fig:dads}
\end{figure*}

{\bf II. Methanol adsorption on the hydroxyl-terminated face of kaolinite}:
A  careful investigation shows that the ordering of the functionals according to the H-bond lengths and to the adsorption energy is almost the same as that observed for water.
The only exception is vdW-DF, which for methanol gives a larger adsorption energy than that obtained with PBE.
The comparison with DMC confirms, as for water, that the best performing functionals are vdW-DF2 and optPBE-vdW.
Again the GGA functionals and vdW-DF underestimate the adsorption energy, while all the other functionals (PBE-D2, PBE-D3,  PBE+vdW(TS), optB86b-vdW, PBE0-D3 and B3LYP-D3) overestimate the interaction.
 As for the previous system, vdW-corrected hybrid functionals do not seem to improve significantly with respect the GGA+vdW approaches.

Before discussing adsorption on the silicate face of kaolinite, we briefly compare water and methanol adsorption.
An important finding from the results presented in Table~\ref{tab:adseneAlOH} is that, with the GGA functionals the adsorption energies of water and methanol are similar  
(e.g. $E^{\ce{H2O}}_{\mathrm{ads,PBE}} = -608$~meV and
$E^{\ce{MeOH}}_{\mathrm{ads,PBE}} = -616$~meV), 
but upon inclusion of dispersion interactions methanol is stabilized to a greater extent 
(e.g. $E^{\ce{H2O}}_{\mathrm{ads,optPBE-vdW}} = -699$~meV and
$E^{\ce{MeOH}}_{\mathrm{ads,optPBE-vdW}} = -779$~meV). 
This is apparent in Fig.~\ref{fig:dads}, where the difference in adsorption between water and methanol is plotted. 
Therefore, even though the methyl group is considered a spectator, its vdW interaction with the surface is non-negligible and it is clearly desirable to properly account for dispersion interactions in these systems.
DMC confirms the reliability of the vdW-inclusive functionals on this issue as DMC also finds that methanol binds more strongly than water.

{\bf III. Water adsorption on the silicate-terminated face of kaolinite}:
In Table~\ref{tab:adseneSiO2} we present the results from all the functionals for adsorption on the silicate-terminated face. 
Irrespective of which functional is used the adsorption energies obtained are in the physisorption regime.
Consequently, the inclusion of vdW forces is expected to have a more obvious impact than on the hydroxyl-terminated face.

\begin{table*}[htbp]	
\caption{ Adsorption energy of water, $E^\ce{H2O}_\textrm{ads}$, and of methanol, $E^\ce{MeOH}_\textrm{ads}$, on the silicate-terminated face of kaolinite, and adsorption energy difference, 
$\Delta E_\textrm{ads}=E^{\ce{H2O}}_{\textrm{ads}} - E^{\ce{MeOH}}_{\textrm{ads}}$,
between water and methanol, obtained with DMC and several DFT XC functionals. The best performing functionals are indicated in bold. All energy values are in meV. Energies have been obtained on PBE-D3 optimized structures but in parenthesis we also report the adsorption energies when geometries are fully relaxed consistently for each   GGA and GGA+vdW functional. \\ }													
\label{tab:adseneSiO2}													
\begin{tabular}{  l   c c c   }													
\hline
\hline													
	&	\multicolumn{3}{ c }{silicate-terminated face }					\\
	\cline{2-4}
Method	&	$E^\ce{H2O}_\textrm{ads}$	&	$E^\ce{MeOH}_\textrm{ads}$	&	$\Delta E_\textrm{ads}$	\\
\hline													
\hline
DMC	 &	-184$\pm$23	&	-250$\pm$18	&	66$\pm$29	\\
\hline		
LDA	& -295&	-315&	20 \\
\multicolumn{4}{l}{\em GGA functionals} \\											
PBE             & -85( -92)&      -93(-118)&        8( 27)\\
RPBE            &  25( -41)&       21(-185)&        4(144)\\
\multicolumn{4}{l}{\em hybrid functionals} \\
PBE0 & -75&	-86&	11 \\
B3LYP & -24&	-19&	-5\\
\multicolumn{4}{l}{\em GGA+vdW functionals} \\
PBE-D2          &-262(-276)&     -331(-340)&       69( 64)\\
{\bf PBE-D3}          &-248&     -314&       {\bf 66}\\
PBE+vdW(TS)          &-271(-273)&     -356(-362)&       85( 89)\\
{\bf vdW-DF}          &{\bf-183}(-193)&     -297(-303)&      115(110)\\
{\bf vdW-DF2}         &-210(-214)&     {\bf -292}(-295)&       82( 81)\\
optPBE-vdW      &-248(-252)&     -369(-372)&      121(121)\\
optB86b-vdW     &-236(-238)&     -350(-358)&      115(119)\\
\multicolumn{4}{l}{\em hybrid+vdW functionals} \\
PBE0-D3 	& -241&	-311&	69 \\
B3LYP-D3 	& -273&	-345&	71 \\
\hline			
\hline										
\end{tabular}													
\end{table*}

As on the hydroxyl-terminated face, RPBE gives an adsorption energy that is noticeably less exothermic than PBE. 
In the case of the vdW-DFs we see an across-the-board stabilization relative to the GGA functionals. Like at the hydroxyl-terminated face, vdW-DF and vdW-DF2 give the weakest adsorption energy of the vdW-DFs. 
The PBE-D2 and  PBE+vdW(TS) functionals give the strongest overall adsorption energy, with 276~meV and 273~meV respectively, followed by optPBE-vdW, PBE-D3 and optB86b-vdW with values close to 250~meV. 
Overall, the spread of the vdW-based evaluations is much smaller than for the hydroxyl-terminated face.
Whereas at the hydroxyl-terminated face the adsorption structure was altered only moderately upon inclusion of vdW, at the silicate-terminated face more significant changes are observed. Specifically, the GGA functionals predict the molecule to be much further away from the surface than of the vdW inclusive functionals do. 
This difference  is also reflected in Fig.~\ref{fig:ads}, if we consider the difference between the adsorption energies of the GGA functionals at the PBE-D3 geometry and when the structures are relaxed. 
On the other hand, the geometries provided by the vdW-inclusive approaches are in very good agreement with PBE-D3, so $E_\textrm{ads}$ evaluated on either the PBE-D3 geometry or on the relaxed structures are similar.

Comparison with DMC supports the general reliability of the  vdW-corrected and vdW-inclusive approaches over the  bare GGA and hybrid functionals.
However, it also shows that almost all the  GGA+vdW and the two hybrid+vdW functionals overestimate the adsorption energy. 
Similar overestimates have been seen recently for physisorbed water on hexagonal boron-nitride.\cite{AlHamdani:hBN:2015}
On this surface the best performance is seen for the vdW-DF and the vdW-DF2 functionals, 
both of them being in agreement with DMC, given the DMC stochastic error.
It is also interesting to note that even though water still binds preferentially to the hydroxyl-terminated face, the relative adsorption strengths are significantly altered: 
the ratio 
$E_{\mathrm{ads}}^{\ce{H2O}@\ce{AlOH}}/E_{\mathrm{ads}}^{\ce{H2O}@\ce{SiO2}} $
is $6.6$ for PBE, 
$9.3$ for RPBE,
\textit{ca.} 3 for the vdW-inclusive functionals,
and $3.5\pm 0.5$ at the DMC level.

{\bf IV. Methanol adsorption on the silicate-terminated face of kaolinite}:
Similar to water, we again expect dispersion interactions to play more of a role at the silicate-terminated than at the hydroxyl-terminated face. 
Indeed, we again observe big differences between functionals including or not the vdW interaction.
In the most stable structure found 
using the PBE-D3 functional there is no hydrogen bond-like interaction and methanol is parallel to the surface of the slab (see Fig.~\ref{fig:ads_refgeo}D).
The geometry at the GGA level has the methanol molecule found at a much larger distance from the surface, as depicted in Fig.~\ref{fig:ads_conf}D.

As on the hydroxyl-terminated face, the degree of stabilization due to dispersion interactions is greater for methanol than it is for water. 
At the GGA level, water and methanol bind with similar interaction strengths (methanol binds more strongly by 27~meV at the PBE level), but  when vdW is accounted in the functional we observe that methanol binds more
strongly by 64~meV (for PBE-D2) to 121~meV (for optPBE-vdW), as shown in Fig.~\ref{fig:dads}.

The comparison with DMC shows that in this case the GGA functionals underestimate the adsorption energy, and that all the  GGA+vdW and hybrid+vdW functionals overestimate $E_\textrm{ads}$.
The best agreement is again obtained for the vdW-DF and vdW-DF2 functionals,
the former overestimating the interaction by $47 \pm 18$~meV and the latter by $42 \pm 18$~meV.
The ratio
$E_{\mathrm{ads}}^{\ce{MeOH}@\ce{AlOH}}/E_{\mathrm{ads}}^{\ce{MeOH}@\ce{SiO2}} $
is 5.2 for PBE, 
2.1 for RPBE,
between 2.1 and 2.6 for the  vdW-corrected and vdW-inclusive functionals,
and $2.8\pm 0.2$ at the DMC level.

\begin{figure}[!htp]
  \centering
\includegraphics[width=2.8in]{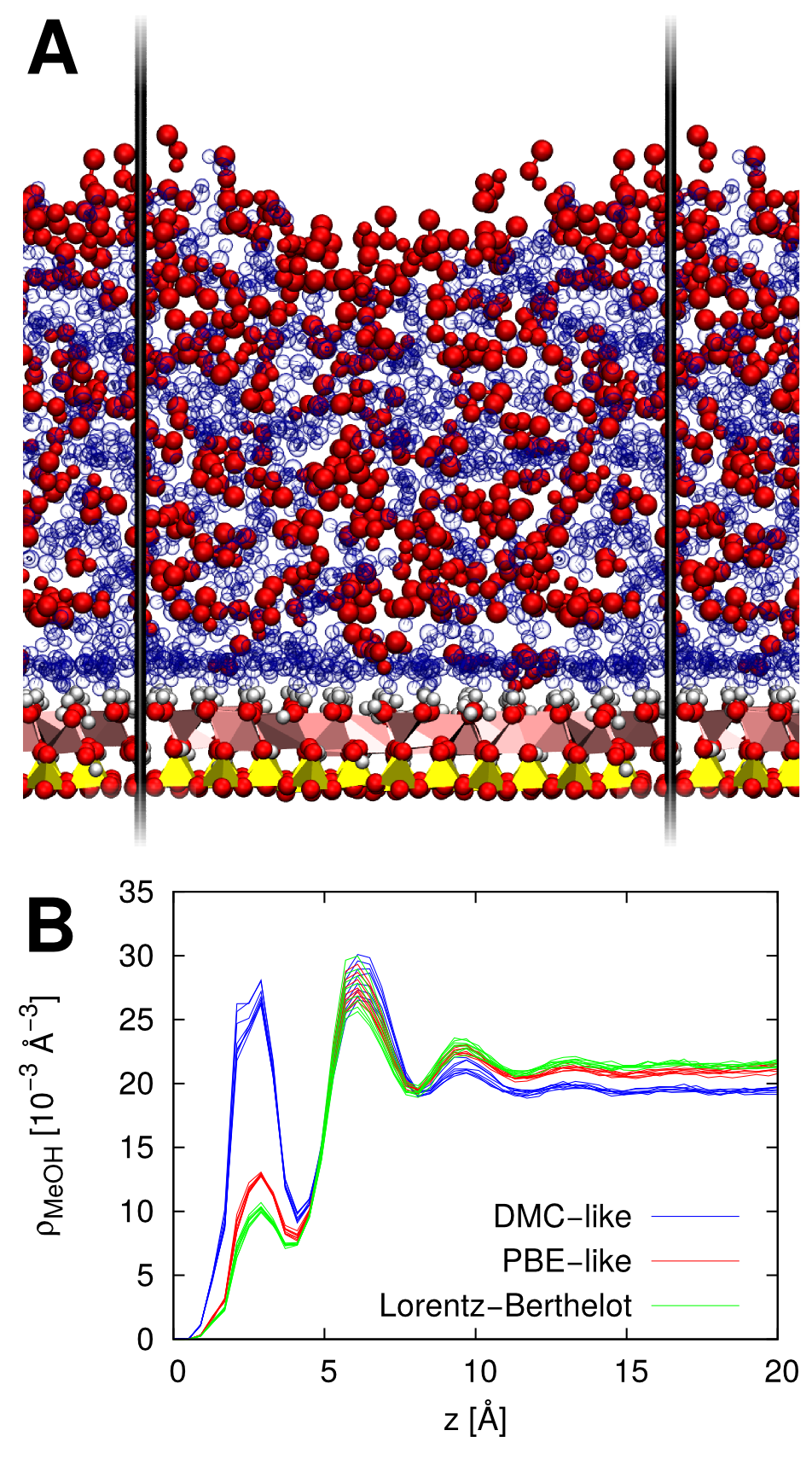}
\caption{
Molecular dynamics simulations of a water methanol mixture on the hydroxyl-terminated face of kaolinite. 
Panel A: 
Snapshot of the simulation system, where the methanol molecules are shown in red and the water molecules as glassy blue circles, and the kaolinite slab is the same as in Fig.~\ref{fig:kaolinite_bulk}.
The black lines show part of the periodic simulation cell boundaries.
Panel B: 
Density profiles of methanol above the kaolinite surface, for the eight replicas with temperatures ranging from 275--310\,K, obtained for values of $\Delta E_\textrm{ads}$ (namely, the difference between adsorption energy of water and methanol, see text) corresponding to Lorentz-Bethelot, PBE and DMC. ``z'' is the distance from the average position of the top layer of oxygen atoms in the kaolinite surface. It is clear that an appropriate choice of $\Delta E_\textrm{ads}$ can significantly affect the density of the water-methanol solution at the interface. 
}
  \label{fig:mdfig}
\end{figure}

\section{Discussion and Conclusions}
\label{sec:conclusion}

In this paper we have used QMC to examine the adsorption of water and methanol on the hydroxyl- and silicate-terminated (001) faces of kaolinite.
The QMC results on the hydroxyl-terminated face have been obtained independently with two different fixed-node projection QMC methods: DMC and LRDMC. The two methods differ in terms of algorithms (DMC is based on a time-discretization approximation, LRDMC on a space-discretization approximation), implementation (DMC calculations have been performed using the CASINO code; LRDMC using the TurboRVB code) and setup (for instance, different PPs, basis sets and Jastrow terms). 
Nonetheless both approaches produce results in good agreement, with the small differences between the approaches coming within the stochastic error of the evaluations.

QMC results indicate that both water and methanol adsorb on the hydroxyl-terminated face,  forming  three H-bonds, with an interaction energy larger than 0.6~eV.
The adsorption on the silicate-terminated face is much weaker, smaller than 0.3~eV.
In both cases the methanol binds slightly more strongly than water.

As discussed, 
the QMC results provide a benchmark that can help to further understanding of how other computationally cheaper methods perform for adsorption.
Specifically, we have compared them with the results provided by a selection of commonly used XC functionals in DFT (covering GGA, hybrid, vdW-corrected GGA, vdW-corrected hybrid, and vdW-inclusive functionals).
This shows that the  vdW-corrected and vdW-inclusive functionals predict adsorption energies that are considerably larger than those calculated using the  bare GGA or hybrid functionals, but the degree of stabilization is system dependent. 
As discussed,  in the systems under consideration in this work the QMC references indicate that  bare GGA and hybrid based predictions are often underestimated, whereas  approaches that account for the vdW interaction yield results in qualitative agreement with QMC, although the absolute value of the adsorption energy can be overestimated, particularly on the silicate-terminated face.
Overall, the best results are provided by vdW-DF2,  and among the vdW-corrected approaches we notice good performance from PBE-D3. 
Inclusion of exact exchange does not appear to lead to any improvement for the systems considered here,
for instance results from PBE-D3 and PBE0-D3 are almost identical.
The  GGA+vdW functionals,  although based on GGA, perform better than the bare GGAs also in terms of geometries. 
Indeed, on the silicate-terminated face (where the interaction is weaker) structure relaxation performed with   the vdW-corrected and vdW-inclusive functionals leads to very similar configurations, whereas with the GGAs the adsorbates sometimes strayed away from the surface.
Looking forward there certainly still seems to  be scope for further improvements in the treatment of these systems with DFT.
Of the functional considered vdW-DF2 offers the best performance but it does not convincingly deliver chemical accuracy for all four adsorption scenarios considered.  
Approaches such as Hamada's revised vdW-DF2 functional\cite{PhysRevB.89.121103} or Tkatchenko's many body dispersion\cite{PhysRevLett.108.236402} would be interesting to explore.

The comparison between the adsorption of water and methanol is also interesting. At the GGA level there is very little difference in adsorption energies, whereas methanol becomes more strongly bound when vdW interactions are accounted for. 
As clay minerals can cleanse ground water through the uptake of pollutants, the relative adsorption energies with  respect to water is a highly important quantity. 
Even for methanol, which is one of the simplest organic molecules able
to form a hydrogen bond, we see that including vdW
interactions can significantly alter the adsorption energy relative to
water; on the hydroxyl-terminated face, water and methanol bind with
similar energies, but inclusion of dispersion forces tips the balance
in favor of methanol. 

Before closing we note that we have examined the consequences of 
altering the relative interaction strength of water and methanol with kaolinite 
through a series of classical molecular dynamics simulations of liquid water-methanol solutions on kaolinite. 
The results of these simulations are shown in Figs.~\ref{fig:mdfig}A--\ref{fig:mdfig}B. Specifically in Fig.~\ref{fig:mdfig}B we show results obtained with water and methanol interaction parameters that use standard Lorentz-Berthelot mixing rules, values matching PBE or values matching DMC.
As can be seen Fig.~\ref{fig:mdfig}B, with the
DMC value of $\Delta E_\textrm{ads}$ the adsorption of methanol yields a density profile with a much more
pronounced first peak compared to the $\Delta E_\textrm{ads}$ corresponding to PBE or standard Lorentz-Berthelot mixing rules. 
Thus we see that the standard approach for exploring aqueous solutions at a clay surface with force fields leads to a rather poor description of the interface.
This effect is likely to
become even more significant as the size of the organic tail of the
adsorbate increases and demonstrates the importance of an accurate
modeling of dispersion interactions 
when exploring wet interfaces of environmental relevance. 
The ability to accurately incorporate non-local dispersion interactions is therefore extremely important if one aims to model environmentally relevant adsorption processes on kaolinite and other clays.

\begin{suppinfo}
In Supporting Information we report a detailed descriptions of the computational setups for the calculations performed in this work.
Moreover, in Table~S1 of the Supporting Information we report the adsorption energy obtained using a larger set of DFT functionals (including LDA, GGA, meta-GGA and hybrid type) and of vdW corrections.
Finally, we report  the structures, in .cif file format, used for the DMC and LRDMC calculations. 
\end{suppinfo}

\begin{acknowledgement}
AZ and AM's work has been sponsored by the Air Force Office of Scientific Research, Air Force Material Command, USAF, under grant number FA8655-12-1-2099 and by the European Research Council under the European Union's Seventh Framework Programme (FP/2007-2013)/ERC Grant Agreement No. 616121 (HeteroIce project). AM is also supported by the Royal Society through a Wolfson Research merit Award. 
Calculations were performed on the U.K. national service ARCHER, the UK's national high-performance computing service, which is funded by the Office of Science and Technology through EPSRC's High End Computing Programme, grant number EP/K038249/1. This research also used resources of the Oak Ridge Leadership Computing Facility located in the Oak Ridge National Laboratory, which is supported by the Office of Science of the Department of Energy under Contract No. DE-AC05-00OR22725. 
\end{acknowledgement}

\newpage

\setcounter{section}{0}
\renewcommand{\thesection}{S\arabic{section}}%
\setcounter{table}{0}
\renewcommand{\thetable}{S\arabic{table}}%
\setcounter{figure}{0}
\renewcommand{\thefigure}{S\arabic{figure}}%

\section*{Supporting Information}

We report here a detailed descriptions of the computational setups for the calculations performed in this work.
In particular, Section~\ref{app:DMC} is about DMC, Section~\ref{app:LRDMC} about LRDMC, Section~\ref{app:FSE} about the finite-size effects in QMC, and Section~\ref{app:DFT} about DFT.

Moreover, in table \ref{tab:adsene} we report the adsorption energy obtained, for the reference PBE-D3 optimized structures (used for QMC),  using several DFT functionals, including LDA, GGA, meta-GGA and hybrid type.
GGA functionals used include the Perdew-Burke-Ernzerhof (PBE) \cite{PBE, PBE_Erratum}, RPBE \cite{RPBE} and Becke-Lee-Yang-Parr (BLYP) \cite{PhysRevA.38.3098, PhysRevB.37.785}.
SCAN is a meta-GGA functional \cite{sun_accurate_2016}.
Hybrid functionals considered are
the PBE0 \cite{PBE0}, HSE06 \cite{HSE06} and B3LYP \cite{Vosko:1980ui,LEE:1988ub,Becke:1993vx,STEPHENS:1994vd} functionals.
Van der Waals corrections were accounted in several different schemes.
The first is the D2 and D3 corrections of Grimme \cite{grimme_D2, grimme_D3},
and in parenthesis 
BJ indicates a Becke and Johnson damping (instead of the zero-dampling) \cite{Grimme:BJdamping}
and ATM indicates the use of a nonadditive three-body dispersion term of Axilrod-Teller-Muto type.
The second is the corrections of Tkatchenko and Scheffler (+vdW(TS)) \cite{PRL09:Scheffler}, also with the self consitent screening (+vdW(TS-scs)) \cite{PhysRevLett.108.236402}. %
Results for the van der Waals functionals vdW-DF, vdW-DF2, optPBE-vdW and optB86b-vdW are also reported \cite{revPBE, vdW-DF, klimes-vdW-DF, optB86b-vdW}.

Finally, we provide the structures, in .cif file format, used for the DMC and LRDMC calculations. As described in the main paper, these structures have been obtained through a DFT-base relaxation, using the PBE-D3 functional.

\begin{table*}[!p]
\caption{ 
Adsorption Energy [meV] of water (W) and methanol (M) molecules on the hydroxyl-terminated ($\ce{AlOH}$) and silicate-terminated ($\ce{SiO2}$) faces of kaolinite. 
All evaluations have been performed on the reference structures obtained from PBE-D3 based relaxation, and used also for the DMC calculations. Functionals are defined in the text.
}
\label{tab:adsene}
{\footnotesize
\begin{tabular}{| l  c c c c |}
\hline
	&	W@AlOH	&	M@AlOH	&	W@SiO$_2$	&	M@SiO$_2$	\\
\hline
LDA	&	-1102	&	-1138	&	-295	&	-315	\\
\hline									
PBE	&	-607	&	-614	&	-85	&	-93	\\
PBE-D2	&	-822	&	-879	&	-262	&	-331	\\
PBE-D3	&	-767	&	-828	&	-248	&	-314	\\
PBE-D3(ATM)	&	-757	&	-814	&	-241	&	-300	\\
PBE-D3(BJ)	&	-766	&	-827	&	-245	&	-313	\\
PBE-D3(BJ,ATM)	&	-757	&	-812	&	-237	&	-299	\\
PBE+vdW(TS)	&	-769	&	-841	&	-271	&	-356	\\
\hline						               			
RPBE	&	-360	&	-354	&	25	&	21	\\
RPBE-D3	&	-697	&	-749	&	-261	&	-328	\\
RPBE-D3(ATM)	&	-688	&	-735	&	-253	&	-314	\\
RPBE-D3(BJ)	&	-721	&	-783	&	-266	&	-343	\\
RPBE-D3(BJ,ATM)	&	-712	&	-768	&	-258	&	-330	\\
\hline									
BLYP	&	-470	&	-463	&	10	&	24	\\
BLYP-D2	&	-815	&	-888	&	-273	&	-357	\\
BLYP-D3	&	-764	&	-830	&	-293	&	-368	\\
BLYP-D3(ATM)	&	-755	&	-816	&	-285	&	-354	\\
BLYP-D3(BJ)	&	-756	&	-832	&	-276	&	-359	\\
BLYP-D3(BJ,ATM)	&	-747	&	-817	&	-268	&	-345	\\
\hline									
vdW-DF	&	-530	&	-597	&	-183	&	-297	\\
vdW-DF2	&	-616	&	-658	&	-210	&	-292	\\
optPBE-vdW	&	-689	&	-767	&	-248	&	-369	\\
optB86b-vdW	&	-751	&	-835	&	-236	&	-350	\\
\hline									
SCAN	&	-707	&	-728	&	-160	&	-181	\\
SCAN-D3	&	-777	&	-838	&	-226	&	-284	\\
SCAN-D3(ATM)	&	-768	&	-823	&	-218	&	-270	\\
SCAN-D3(BJ)	&	-805	&	-866	&	-258	&	-322	\\
SCAN-D3(BJ,ATM)	&	-796	&	-852	&	-250	&	-308	\\
\hline									
PBE0	&	-599	&	-615	&	-75	&	-86	\\
PBE0-D2	&	-771	&	-827	&	-217	&	-277	\\
PBE0-D3	&	-768	&	-840	&	-241	&	-311	\\
PBE0-D3(ATM)	&	-759	&	-825	&	-233	&	-297	\\
PBE0-D3(BJ)	&	-750	&	-823	&	-233	&	-307	\\
PBE0-D3(BJ,ATM)	&	-741	&	-808	&	-225	&	-293	\\
PBE0+vdW(TS-scs)	&	-784	&	-869	&	-255	&	-333	\\
\hline									
HSE06	&	-596	&	-608	&	-70	&	-73	\\
HSE06-D3	&	-716	&	-776	&	-203	&	-255	\\
HSE06-D3(ATM)	&	-707	&	-762	&	-195	&	-242	\\
HSE06-D3(BJ)	&	-736	&	-809	&	-221	&	-294	\\
HSE06-D3(BJ,ATM)	&	-727	&	-794	&	-213	&	-280	\\
HSE06+vdW(TS)	&	-758	&	-835	&	-255	&	-335	\\
\hline									
B3LYP	&	-524	&	-528	&	-24	&	-19	\\
B3LYP-D2	&	-826	&	-900	&	-272	&	-352	\\
B3LYP-D3	&	-776	&	-849	&	-273	&	-345	\\
B3LYP-D3(ATM)	&	-767	&	-834	&	-265	&	-331	\\
B3LYP-D3(BJ)	&	-764	&	-840	&	-265	&	-343	\\
B3LYP-D3(BJ,ATM)	&	-755	&	-826	&	-257	&	-330	\\
B3LYP+vdW(TS-scs)	&	-799	&	-893	&	-278	&	-343	\\
\hline									
\end{tabular}
}
\end{table*}

\section{DMC setup}\label{app:DMC}

DMC calculations were performed using the {\sc CASINO} code.\cite{casino,CASINO2.3}
A Slater-Jastrow wavefunction ansatz is used as a guiding function for the importance sampling DMC calculations.
Trail and Needs PPs\cite{PPurl,trail05_NCHF,trail05_SRHF} are used for all atoms in the system. 
The Slater determinant is obtained from DFT-LDA plane-wave calculations using 
{\sc PWSCF}\cite{pwscf}
with a 600~Ry energy cutoff. 
The resulting molecular orbitals were expanded in terms of B-splines.\cite{splinesQMC} 
The Jastrow factor contains electron-electron, electron-nucleus, and electron-electron-nucleus terms, which have been optimized, within a variational Monte Carlo (VMC) scheme, 
via minimization of the VMC variance.
The DMC calculations were performed within the fixed-node approximation 
and non-local terms in the PPs were handled by using the locality approximation.
Umrigar {\em et al.}'s scheme\cite{UMRIGAR:1993} for the branching-drift-diffusion process is adopted, with a time-step $\tau$ of 0.005~a.u.
The adsorption energies were evaluated using the \textit{complex-minus-far} method, see Eq.~2 in main paper. 
For the calculation of 
$E_{\ce{slab}+{M@X}}$ and $E_{\ce{slab}-M}$
we used 2D periodicity (Ewald summation in two dimensions), as implemented in {\sc CASINO},
and the $E_{\ce{slab}-M}$ has been evaluated with the molecule $M >16$~\AA{} from the slab of kaolinite.
The reliability of this setup for evaluations of weak interactions in similar systems has  already been assessed in previous studies.~\cite{Ice:prl2011,santra_on_2013,Quigley:Ice_0_i_Ih:jcp2014,Morales:bulkwat:2014,Cox:2014,Benali:2014,AlHamdani:hBN:2015,gillan15,AFQMC:CoGraf,Morales:perspective2014,Zen-liquidwat}


This choice of the \textit{complex-minus-far} method is explained  by two considerations: 
(i)
DMC calculations (as well as any many-body electronic structure method) are affected by size-effects due to the periodicity of the system that are much larger that those of DFT, and they are  sizeable also in the big supercells considered here;\cite{Lin:qmctwistavg:pre2001,Chiesa:size_effects:prl2006,KZK:prl2008,drummond_finite-size_2008}
and
(ii)
DMC simulations with finite time step $\tau$, with the commonly adopted Umrigar {\em et al.}'s modifications\cite{UMRIGAR:1993} of the  Green function in the proximity of the nodal surface, will result in a violation of the size-consistency, that is recovered only in the limit of $\tau\to 0$.\footnote{
It is worth  mentioning that in a recent work\cite{sizeconsDMC} the size-consistency issue of DMC at finite $\tau$ has been considerably improved by introducing an improved branching-drift-diffusion algorithm, so that now much larger $\tau$ can be used. 
However, the calculations  presented here were performed before the aforementioned developments.
}
A complete $\tau\to 0$ extrapolation is out of the reach for the big systems considered here, but we verified in a model subsystem that $\tau=0.005$~a.u. gives a finite time-step error on the estimated $E^{{M-X}}_{\textrm{ads}}$ that  can be considered as negligible. In particular, the finite time-step error on  $E^{{M-X}}_{\textrm{ads}}$ is negligible in comparison with the \textit{ca.} 20 meV stochastic error of the evaluation, provided that we use the \textit{complex-minus-far} method, which benefits from an almost perfect error cancellation. The \textit{complex-minus-fragments} method is instead much more affected by the finite time-step bias, so it would require a $\tau$ that is at least one order of magnitude smaller.

For each configuration considered, we used a target population of 204,800 walkers across 20,480 physical nodes (1280 computational nodes on Titan supercomputer) and  we simulated more that 15,000 DMC time steps: the first 1,000 for equilibration; the remaining averaged using the ``blocking'' method\cite{blockingmethod} with a bin of 512.
This setup guarantees evaluations of the absolute energy of each configuration with an associated stochastic error of $\sim 13$~meV, giving an error of $\sim 20$~meV on the adsorption energies.

\section{LRDMC setup}\label{app:LRDMC}

The LRDMC results reported have been obtained with the  {\sc TurboRVB} package developed by Sorella and coworkers.\cite{TurboRVB} 
The setup for the LRDMC calculations  slightly differs from that of the DMC calculations as a consequence of the different implementations of the algorithms in {\sc TurboRVB} and {\sc CASINO}.
The main differences are in the variational wave function:
orbitals in  {\sc TurboRVB} are expressed in terms of localized basis functions (of Gaussian, Slater and other types), and the Jastrow factor is parametrized differently.
The wave function ansatz implemented in  {\sc TurboRVB} is the Jastrow Antisymmetrized Geminal Power, which includes the Slater-Jastrow ansatz as a special case.\cite{Zen:2013is}
Moreover, the parameters of the variational wave function (to be used as the guiding function for the LRDMC calculation) are optimized in order to minimize the variational energy. 
For a more detailed description of the wave function implemented in {\sc TurboRVB} see Ref.  \citenum{Zen:2013is}.

In this work we have used a Slater-Jastrow wave function.
Core electrons of C, O, Al and Si atoms have been described via
scalar-relativistic energy-consistent Hartree-Fock PPs of Burkatzki {\em et al.}\cite{Burkatzki:2007p25447}
Consistent with the choice of the PP, the basis set has been obtained starting from the Burkatzki {\em et al.} VTZ basis set.\cite{filippi-basis} 
For use in the {\sc TurboRVB} package, we have uncontracted the basis set and removed the almost redundant exponents and those too small or too large in value (keeping these would imply a much slower and inefficient optimization of the wave-function parameters, with an almost negligible energy gain), obtaining (7s,2p) for the H atoms, (11s,11p,2d) for C, (10s,11p,2d) for O, 
(11s,11p,2d,1f) for Al and Si.
The coefficients of the molecular orbitals have then been optimized by performing a DFT-LDA calculation, using the DFT code included in the  {\sc TurboRVB} package.\cite{TurboPREP}
The Jastrow factor used here consists of both homogeneous and non-homogeneous terms that account for the electron-electron, electron-nucleus and electron-electron-nucleus interactions. The non-homogeneous terms are expressed in terms of atomic orbitals, which are expanded in terms of (2s,2p) basis for H atoms, (3s,2p,1d) for C and O atoms, (3s,2p,2d) for Al and Si atoms.
The exponents of the Jastrow atomic orbitals have been fixed to the values obtained from the optimization in a smaller model system. 
All the other parameters of the Jastrow factor have been optimized for each specific configuration.
%
In the LRDMC we used a mesh $a$ of 0.4~a.u.

Since we are evaluating the adsorption energy in weakly interacting systems, we expect that the above setup leads to unbiased results, in particular with respect to the choice of the basis set and the LRDMC mesh $a$. 
See \textit{e.g.} Ref.~\citenum{Zen-liquidwat} for results on the water dimer.
However, here we have also checked directly whether the above setup leads to unbiased results for the kaolinite plus water/methanol system, by performing tests on model systems (i.e., a molecule bound to a cluster representing kaolinite).
We observed that the adsorption energies obtained with  $a=0.4$~a.u. are the same, within error bars, as those obtained with $a=0.1$ and 0.2~a.u.
Moreover, by performing additional calculations with different basis sets, we ascertained that with the chosen setup we have no basis-set bias at the LRDMC level. 

{\sc TurboRVB} allows one to perform calculations with  open conditions or with 3D periodic boundary conditions.
Unfortunately, 2D periodic boundary conditions are not yet implemented in the code.
This makes the use of the \textit{complex-minus-far} evaluation of the adsorption energy problematic. 
On the other hand, LRDMC is size consistent for any mesh $a$. Thus, we have chosen the \textit{complex-minus-fragments} approach here, see Eq.~1 in the main paper. 
As described in Section~\ref{app:FSE}, we have used the KZK method to evaluate the finite-size effects (Table~\ref{tab:kzk}) and LRDMC evaluations have been corrected accordingly.
The effect of the dipole interactions perpendicular to the slab have been corrected on the basis of DFT-based estimates of the effect. Namely, in the reported LRDMC calculations there is a vacuum between the slabs of $\sim 15$~\AA, for which DFT predicts that the adsorption energy without dipole correction would result in a  36~meV underestimation for water on the hydroxyl-terminated face, and a 38~meV underestimation for methanol.


\section{Finite-size effects in QMC}\label{app:FSE}

FSEs in our QMC evaluations have been evaluated using the KZK method.\cite{KZK:prl2008}
This method is pretty simple and computationally cheap: we have to perform two DFT calculations for each complex considered, the first with the LDA functional and the second with the KZK functional (which mimics the FSE in many-body electronic structure calculations). 
The difference between the two evaluations provides an estimate of the FSE in QMC.
KZK corrections have been calculated using the implementation in PWSCF\cite{pwscf} due to E. Sola, as reported in Ref.~\citenum{Binnie:2009ba},
with the setup previously described to generate the Slater determinant of the guiding function.
%
In this way we have seen that FSEs are relatively large in the absolute energy of the complexes (KZK predicts an underestimation of $\sim 1.3$~eV on the absolute energy), but there is a huge error cancellation in the evaluation of the adsorption energy:
according to KZK the QMC evaluation with the \textit{complex-minus-far} method leads to an underestimate $\left| E^{{M@X}}_{\textrm{ads}} \right|$ of 12 - 17~meV, depending on the surface and on the molecule considered (Table~\ref{tab:kzk}).
On the other hand, if the \textit{complex-minus-fragments} method is used, error cancellation is slightly worse and the correction has the opposite sign: 
$\left| E^{\ce{H_2O}@X}_{\textrm{ads}} \right|$
is overestimated by $\sim 35$~meV ($X$ being either the hydroxyl-terminated face, $\ce{AlOH}$, or the silicate-terminated face, $\ce{SiO2}$), whereas 
$\left| E^{\ce{MeOH}@X}_{\textrm{ads}} \right|$
is overestimated by $\sim 75$~meV.
Thus, the \textit{complex-minus-far} method is to be preferred in terms of FSEs. 

\begin{table}[htb]                                                    
\caption{ 
The following table summarizes the
KZK\cite{KZK:prl2008} evaluations of the FSE on the adsorption energy of the different systems considered, either by using the \textit{complex-minus-fragments} evaluation, 
Eq.~1 in the main paper, 
or the \textit{complex-minus-far}, 
Eq.~2 in the main paper. 
Here, a positive value implies that the absolute value of $E^{M@X}_{\textrm{ads}}$ is overestimated if the bare QMC value is taken; a negative value indicates an underestimation.
}                                                                                                     
\label{tab:kzk}                                                      
\begin{tabular}{l  c c }
\hline \hline
	& \multicolumn{2}{c}{\em complex-minus-} \\
	&	{\em fragments}	&	{\em far}	\\
	&	Eq.~1	&	Eq.~2	\\
\hline					
$\ce{H_2O}@\ce{AlOH}$	&	+35	&	-16	\\
$\ce{MeOH}@\ce{AlOH}$	&	+73	&	-17	\\
$\ce{H_2O}@\ce{SiO2}$	&	+38	&	-12	\\
$\ce{MeOH}@\ce{SiO2}$	&	+76	&	-14	\\
\hline \hline
\end{tabular}
\end{table}

\section{DFT setup}\label{app:DFT}

DFT calculations were performed using the plane-wave code
{\footnotesize VASP 5.4}.\cite{VASP1, VASP2, VASP3, VASP4} 
Calculations using the van der Waals density functionals were carried out self-consistently using the approach of Rom\'{a}n-P\'{e}rez and Soler\cite{perez&soler:implementation} as implemented in {\footnotesize VASP} by Klime\v{s} \emph{et al.}.\cite{optB86b-vdW}  
Electron-core interactions were described using the projector-augmented wave\cite{PAW1, PAW2} (PAW) potentials supplied with {\footnotesize VASP}. 
PBE PAW potentials for all functionals were used, with the exception of LDA where LDA PAW was used. 
It has been shown on a range of systems for the vdW functionals that this approximation with the PAW potentials does not introduce significant errors in the energies and structures.\cite{optB86b-vdW, JPCM:Graziano} 

Adsorption energies were evaluated using the \textit{complex-minus-fragments} method, see Eq.~1 in the main paper, 
because we have close-shell systems, where DFT is exactly size-consistent, and the system is large enough to have negligible size-effects. 
%
$E_{\ce{M}}$ was calculated at the \(\Gamma\)-point by isolating a single molecule in a \(20\times20\times20\) \AA~box. 
For the calculation of 
$E_{\ce{slab}}$ and  $E_{\ce{slab}+M@X}$
we used three dimensional periodicity with a vacuum region between slabs of \textit{ca.} 15~\AA, and  the dipole interaction across the slab was corrected with the scheme of Neugebauer and Scheffler,\cite{Dipole1, Dipole2} in order to mimic a 2D system, and
$\Gamma$-point sampling of reciprocal space.
For all adsorption calculations, a plane-wave energy cut-off of 500 eV was used. 
During structure optimizations all atoms were fully relaxed until the forces were reduced below $\mathrm{10^{-3}}$ eV/\AA.

\newpage

\begin{multicols}{2}

{\scriptsize

\begin{verbatim}
# water molecule

data_water
_symmetry_Int_Tables_number        1
_cell_length_a                     20.000000
_cell_length_b                     20.000000
_cell_length_c                     20.000000
_cell_angle_alpha                  90.000000
_cell_angle_beta                   90.000000
_cell_angle_gamma                  90.000000

loop_
_symmetry_equiv_pos_site_id
_symmetry_equiv_pos_as_xyz
   1   x,y,z

loop_
_atom_site_label
_atom_site_type_symbol
_atom_site_fract_x
_atom_site_fract_y
_atom_site_fract_z
_atom_site_occupancy
O1 O 0.16139 0.22133 0.33630 1.0000
H1 H 0.14296 0.20583 0.37851 1.0000
H2 H 0.14227 0.19157 0.30298 1.0000
\end{verbatim}
\newpage
\begin{verbatim}
# methanol molecule

data_methanol
_symmetry_Int_Tables_number        1
_cell_length_a                     20.000000
_cell_length_b                     20.000000
_cell_length_c                     20.000000
_cell_angle_alpha                  90.000000
_cell_angle_beta                   90.000000
_cell_angle_gamma                  90.000000

loop_
_symmetry_equiv_pos_site_id
_symmetry_equiv_pos_as_xyz
   1   x,y,z

loop_
_atom_site_label
_atom_site_type_symbol
_atom_site_fract_x
_atom_site_fract_y
_atom_site_fract_z
_atom_site_occupancy
O1 O 0.15721 0.22210 0.33789 1.0000
H1 H 0.14581 0.19214 0.30141 1.0000
H2 H 0.06137 0.21459 0.38181 1.0000
H3 H 0.12872 0.24160 0.43372 1.0000
H4 H 0.12187 0.15537 0.41185 1.0000
C1 C 0.11476 0.20696 0.39355 1.0000
\end{verbatim}
\newpage
\begin{verbatim}
# Kaolinite slab 

data_KAOslab
_symmetry_Int_Tables_number        1
_cell_length_a                     10.384527
_cell_length_b                     9.011475
_cell_length_c                     22.250826
_cell_angle_alpha                  90.000000
_cell_angle_beta                   90.000000
_cell_angle_gamma                  90.000000

loop_
_symmetry_equiv_pos_site_id
_symmetry_equiv_pos_as_xyz
   1   x,y,z

loop_
_atom_site_label
_atom_site_type_symbol
_atom_site_fract_x
_atom_site_fract_y
_atom_site_fract_z
_atom_site_occupancy
Al1 Al 0.14470 0.49573 0.15530 1.0000
Al2 Al 0.14194 0.83066 0.15527 1.0000
Al3 Al 0.39470 0.99573 0.15530 1.0000
Al4 Al 0.39194 0.33066 0.15527 1.0000
Al5 Al 0.64470 0.49573 0.15530 1.0000
Al6 Al 0.64194 0.83066 0.15527 1.0000
Al7 Al 0.89470 0.99573 0.15530 1.0000
Al8 Al 0.89194 0.33066 0.15527 1.0000
Si1 Si 0.06153 0.34406 0.03339 1.0000
Si2 Si 0.06981 0.67178 0.03336 1.0000
Si3 Si 0.31153 0.84406 0.03339 1.0000
Si4 Si 0.31981 0.17178 0.03336 1.0000
Si5 Si 0.56153 0.34406 0.03339 1.0000
Si6 Si 0.56981 0.67178 0.03336 1.0000
Si7 Si 0.81153 0.84406 0.03339 1.0000
Si8 Si 0.81981 0.17178 0.03336 1.0000
O1 O 0.04590 0.35865 0.10686 1.0000
O2 O 0.08879 0.66353 0.10683 1.0000
O3 O 0.09137 0.50668 0.00490 1.0000
O4 O 0.17404 0.22476 0.01699 1.0000
O5 O 0.17675 0.78321 0.00480 1.0000
O6 O 0.29590 0.85865 0.10686 1.0000
O7 O 0.33879 0.16353 0.10683 1.0000
O8 O 0.34137 0.00668 0.00490 1.0000
O9 O 0.42404 0.72476 0.01699 1.0000
O10 O 0.42675 0.28321 0.00480 1.0000
O11 O 0.04652 0.96964 0.10954 1.0000
O12 O 0.45538 0.16288 0.19990 1.0000
O13 O 0.98712 0.47498 0.19984 1.0000
O14 O 0.98506 0.84663 0.20102 1.0000
O15 O 0.29652 0.46964 0.10954 1.0000
O16 O 0.20538 0.66288 0.19990 1.0000
O17 O 0.23712 0.97498 0.19984 1.0000
O18 O 0.23506 0.34663 0.20102 1.0000
O19 O 0.54590 0.35865 0.10686 1.0000
O20 O 0.58879 0.66353 0.10683 1.0000
O21 O 0.59137 0.50668 0.00490 1.0000
O22 O 0.67404 0.22476 0.01699 1.0000
O23 O 0.67675 0.78321 0.00480 1.0000
O24 O 0.79590 0.85865 0.10686 1.0000
O25 O 0.83879 0.16353 0.10683 1.0000
O26 O 0.84137 0.00668 0.00490 1.0000
O27 O 0.92404 0.72476 0.01699 1.0000
O28 O 0.92675 0.28321 0.00480 1.0000
O29 O 0.54652 0.96964 0.10954 1.0000
O30 O 0.95538 0.16288 0.19990 1.0000
O31 O 0.48712 0.47498 0.19984 1.0000
O32 O 0.48506 0.84663 0.20102 1.0000
O33 O 0.79652 0.46964 0.10954 1.0000
O34 O 0.70538 0.66288 0.19990 1.0000
O35 O 0.73712 0.97498 0.19984 1.0000
O36 O 0.73506 0.34663 0.20102 1.0000
H1 H 0.08797 0.05188 0.08919 1.0000
H2 H 0.45862 0.16200 0.24326 1.0000
H3 H 0.98379 0.47943 0.24316 1.0000
H4 H 0.43751 0.75333 0.20231 1.0000
H5 H 0.33797 0.55188 0.08919 1.0000
H6 H 0.20861 0.66200 0.24326 1.0000
H7 H 0.23379 0.97944 0.24315 1.0000
H8 H 0.18751 0.25333 0.20231 1.0000
H9 H 0.58797 0.05188 0.08919 1.0000
H10 H 0.95860 0.16200 0.24326 1.0000
H11 H 0.48379 0.47943 0.24316 1.0000
H12 H 0.93751 0.75333 0.20231 1.0000
H13 H 0.83797 0.55188 0.08919 1.0000
H14 H 0.70861 0.66200 0.24326 1.0000
H15 H 0.73380 0.97942 0.24316 1.0000
H16 H 0.68751 0.25333 0.20231 1.0000
\end{verbatim}
\newpage
\begin{verbatim}
# Kaolinite slab + 1 water at AlOH

data_AlOH+W
_symmetry_Int_Tables_number        1
_cell_length_a                     10.384527
_cell_length_b                     9.011475
_cell_length_c                     22.250826
_cell_angle_alpha                  90.000000
_cell_angle_beta                   90.000000
_cell_angle_gamma                  90.000000

loop_
_symmetry_equiv_pos_site_id
_symmetry_equiv_pos_as_xyz
   1   x,y,z

loop_
_atom_site_label
_atom_site_type_symbol
_atom_site_fract_x
_atom_site_fract_y
_atom_site_fract_z
_atom_site_occupancy
Al1 Al 0.14572 0.49631 0.15560 1.0000
Al2 Al 0.13591 0.83014 0.15686 1.0000
Al3 Al 0.38792 0.99569 0.15503 1.0000
Al4 Al 0.39838 0.32793 0.15529 1.0000
Al5 Al 0.64615 0.49650 0.15677 1.0000
Al6 Al 0.63612 0.83040 0.15517 1.0000
Al7 Al 0.88697 0.99664 0.15555 1.0000
Al8 Al 0.89416 0.32876 0.15696 1.0000
Si1 Si 0.06027 0.34270 0.03414 1.0000
Si2 Si 0.06643 0.67123 0.03420 1.0000
Si3 Si 0.30754 0.84349 0.03416 1.0000
Si4 Si 0.31668 0.17140 0.03359 1.0000
Si5 Si 0.56059 0.34305 0.03344 1.0000
Si6 Si 0.56681 0.67089 0.03381 1.0000
Si7 Si 0.80777 0.84363 0.03385 1.0000
Si8 Si 0.81709 0.17168 0.03405 1.0000
O1 O 0.04871 0.35378 0.10797 1.0000
O2 O 0.08554 0.66233 0.10762 1.0000
O3 O 0.08838 0.50611 0.00604 1.0000
O4 O 0.17215 0.22484 0.01549 1.0000
O5 O 0.17297 0.78273 0.00531 1.0000
O6 O 0.28947 0.85954 0.10744 1.0000
O7 O 0.33241 0.16435 0.10721 1.0000
O8 O 0.33885 0.00574 0.00560 1.0000
O9 O 0.42033 0.72372 0.01851 1.0000
O10 O 0.42509 0.28198 0.00579 1.0000
O11 O 0.04181 0.97011 0.11110 1.0000
O12 O 0.45249 0.15700 0.20115 1.0000
O13 O 0.99614 0.46321 0.20134 1.0000
O14 O 0.97613 0.84603 0.20082 1.0000
O15 O 0.29793 0.46861 0.11091 1.0000
O16 O 0.20064 0.66425 0.20058 1.0000
O17 O 0.22945 0.98055 0.20185 1.0000
O18 O 0.24370 0.34713 0.20337 1.0000
O19 O 0.31793 0.50780 0.29778 1.0000
O20 O 0.54903 0.35564 0.10711 1.0000
O21 O 0.58739 0.66159 0.10711 1.0000
O22 O 0.58838 0.50623 0.00503 1.0000
O23 O 0.67242 0.22441 0.01561 1.0000
O24 O 0.67300 0.78324 0.00500 1.0000
O25 O 0.79036 0.85898 0.10722 1.0000
O26 O 0.83312 0.16477 0.10758 1.0000
O27 O 0.83922 0.00624 0.00580 1.0000
O28 O 0.92044 0.72405 0.01810 1.0000
O29 O 0.92484 0.28278 0.00613 1.0000
O30 O 0.54084 0.96827 0.10913 1.0000
O31 O 0.94273 0.15777 0.20305 1.0000
O32 O 0.49242 0.46912 0.20034 1.0000
O33 O 0.47755 0.84342 0.19956 1.0000
O34 O 0.79856 0.47105 0.11157 1.0000
O35 O 0.70418 0.66798 0.20153 1.0000
O36 O 0.72581 0.98081 0.20041 1.0000
O37 O 0.74310 0.35774 0.20324 1.0000
H1 H 0.08347 0.05041 0.08963 1.0000
H2 H 0.54442 0.14179 0.20692 1.0000
H3 H 0.97003 0.50916 0.23850 1.0000
H4 H 0.43035 0.74998 0.20012 1.0000
H5 H 0.33870 0.55206 0.09103 1.0000
H6 H 0.22698 0.65502 0.24237 1.0000
H7 H 0.24703 0.96039 0.24386 1.0000
H8 H 0.20421 0.24999 0.20838 1.0000
H9 H 0.30872 0.46548 0.33763 1.0000
H10 H 0.28393 0.43190 0.26859 1.0000
H11 H 0.58085 0.04622 0.08564 1.0000
H12 H 0.02510 0.14579 0.22270 1.0000
H13 H 0.46928 0.49269 0.24163 1.0000
H14 H 0.92584 0.75488 0.20258 1.0000
H15 H 0.84016 0.55412 0.09188 1.0000
H16 H 0.68596 0.66634 0.24419 1.0000
H17 H 0.73951 0.96439 0.24297 1.0000
H18 H 0.70795 0.27353 0.22449 1.0000
\end{verbatim}
\newpage
\begin{verbatim}
# Kaolinite slab + water at SiO2

data_SiO2+W
_symmetry_Int_Tables_number        1
_cell_length_a                     10.384527
_cell_length_b                     9.011475
_cell_length_c                     22.250826
_cell_angle_alpha                  90.000000
_cell_angle_beta                   90.000000
_cell_angle_gamma                  90.000000

loop_
_symmetry_equiv_pos_site_id
_symmetry_equiv_pos_as_xyz
   1   x,y,z

loop_
_atom_site_label
_atom_site_type_symbol
_atom_site_fract_x
_atom_site_fract_y
_atom_site_fract_z
_atom_site_occupancy
Al1 Al 0.34494 0.47202 0.59779 1.0000
Al2 Al 0.34550 0.80745 0.59805 1.0000
Al3 Al 0.59946 0.97205 0.59703 1.0000
Al4 Al 0.59381 0.30859 0.59651 1.0000
Al5 Al 0.84692 0.47407 0.59645 1.0000
Al6 Al 0.84800 0.80630 0.59697 1.0000
Al7 Al 0.09764 0.97474 0.59829 1.0000
Al8 Al 0.09436 0.30672 0.59685 1.0000
Si1 Si 0.26537 0.31908 0.47519 1.0000
Si2 Si 0.27234 0.64832 0.47582 1.0000
Si3 Si 0.51447 0.81984 0.47540 1.0000
Si4 Si 0.52354 0.14750 0.47476 1.0000
Si5 Si 0.76535 0.32042 0.47501 1.0000
Si6 Si 0.77299 0.64823 0.47498 1.0000
Si7 Si 0.01538 0.81978 0.47575 1.0000
Si8 Si 0.02332 0.14733 0.47520 1.0000
O1 O 0.25025 0.33147 0.54851 1.0000
O2 O 0.29007 0.64073 0.54925 1.0000
O3 O 0.29484 0.48295 0.44711 1.0000
O4 O 0.37890 0.20231 0.45758 1.0000
O5 O 0.37943 0.75908 0.44717 1.0000
O6 O 0.49910 0.83520 0.54896 1.0000
O7 O 0.54280 0.14029 0.54824 1.0000
O8 O 0.54531 0.98223 0.44686 1.0000
O9 O 0.62708 0.70068 0.45972 1.0000
O10 O 0.63063 0.25932 0.44584 1.0000
O11 O 0.25004 0.94710 0.55250 1.0000
O12 O 0.15629 0.14351 0.64189 1.0000
O13 O 0.18757 0.44910 0.64315 1.0000
O14 O 0.19238 0.83436 0.64502 1.0000
O15 O 0.49790 0.44779 0.55153 1.0000
O16 O 0.40536 0.63911 0.64270 1.0000
O17 O 0.44354 0.95207 0.64190 1.0000
O18 O 0.43704 0.32191 0.64224 1.0000
O19 O 0.51495 0.47869 0.34562 1.0000
O20 O 0.74774 0.33569 0.54828 1.0000
O21 O 0.79290 0.64013 0.54852 1.0000
O22 O 0.79543 0.48300 0.44667 1.0000
O23 O 0.87738 0.20049 0.45888 1.0000
O24 O 0.87996 0.76057 0.44725 1.0000
O25 O 0.00181 0.83324 0.54934 1.0000
O26 O 0.04070 0.13964 0.54872 1.0000
O27 O 0.04530 0.98220 0.44716 1.0000
O28 O 0.12658 0.69931 0.45869 1.0000
O29 O 0.13021 0.25905 0.44674 1.0000
O30 O 0.75217 0.94813 0.55227 1.0000
O31 O 0.66236 0.14095 0.64267 1.0000
O32 O 0.68890 0.45446 0.64117 1.0000
O33 O 0.69055 0.82226 0.64212 1.0000
O34 O 0.99969 0.44678 0.55111 1.0000
O35 O 0.91385 0.63793 0.64223 1.0000
O36 O 0.94427 0.94201 0.64450 1.0000
O37 O 0.93427 0.32447 0.64144 1.0000
H1 H 0.43587 0.49030 0.36860 1.0000
H2 H 0.56271 0.40337 0.36764 1.0000
H3 H 0.29371 0.03197 0.53517 1.0000
H4 H 0.64106 0.14224 0.68505 1.0000
H5 H 0.13288 0.53615 0.64680 1.0000
H6 H 0.64211 0.72970 0.64444 1.0000
H7 H 0.53957 0.53111 0.53199 1.0000
H8 H 0.40122 0.63497 0.68605 1.0000
H9 H 0.43696 0.95803 0.68512 1.0000
H10 H 0.38879 0.22929 0.64480 1.0000
H11 H 0.79145 0.02621 0.52857 1.0000
H12 H 0.15692 0.14701 0.68527 1.0000
H13 H 0.68856 0.45203 0.68458 1.0000
H14 H 0.14078 0.75647 0.66274 1.0000
H15 H 0.03868 0.52300 0.52633 1.0000
H16 H 0.89666 0.64275 0.68496 1.0000
H17 H 0.89743 0.01709 0.66646 1.0000
H18 H 0.87914 0.23718 0.64143 1.0000
\end{verbatim}
\newpage
\begin{verbatim}
# Kaolinite slab + 1 methanol at AlOH

data_AlOH+M
_symmetry_Int_Tables_number        1
_cell_length_a                     10.384527
_cell_length_b                     9.011475
_cell_length_c                     22.250826
_cell_angle_alpha                  90.000000
_cell_angle_beta                   90.000000
_cell_angle_gamma                  90.000000

loop_
_symmetry_equiv_pos_site_id
_symmetry_equiv_pos_as_xyz
   1   x,y,z

loop_
_atom_site_label
_atom_site_type_symbol
_atom_site_fract_x
_atom_site_fract_y
_atom_site_fract_z
_atom_site_occupancy
Al1 Al 0.14576 0.49695 0.15682 1.0000
Al2 Al 0.13620 0.83081 0.15827 1.0000
Al3 Al 0.38826 0.99635 0.15634 1.0000
Al4 Al 0.39861 0.32858 0.15643 1.0000
Al5 Al 0.64618 0.49724 0.15812 1.0000
Al6 Al 0.63646 0.83107 0.15645 1.0000
Al7 Al 0.88729 0.99738 0.15689 1.0000
Al8 Al 0.89426 0.32946 0.15834 1.0000
Si1 Si 0.06062 0.34334 0.03547 1.0000
Si2 Si 0.06664 0.67194 0.03554 1.0000
Si3 Si 0.30782 0.84413 0.03548 1.0000
Si4 Si 0.31704 0.17200 0.03483 1.0000
Si5 Si 0.56096 0.34371 0.03471 1.0000
Si6 Si 0.56711 0.67158 0.03511 1.0000
Si7 Si 0.80806 0.84432 0.03514 1.0000
Si8 Si 0.81747 0.17235 0.03538 1.0000
O1 O 0.04890 0.35433 0.10929 1.0000
O2 O 0.08555 0.66311 0.10898 1.0000
O3 O 0.08872 0.50681 0.00742 1.0000
O4 O 0.17255 0.22559 0.01678 1.0000
O5 O 0.17323 0.78338 0.00665 1.0000
O6 O 0.28976 0.86018 0.10877 1.0000
O7 O 0.33273 0.16493 0.10845 1.0000
O8 O 0.33918 0.00632 0.00687 1.0000
O9 O 0.42057 0.72427 0.01984 1.0000
O10 O 0.42551 0.28255 0.00704 1.0000
O11 O 0.04222 0.97089 0.11252 1.0000
O12 O 0.94257 0.15853 0.20438 1.0000
O13 O 0.99667 0.46313 0.20290 1.0000
O14 O 0.97623 0.84676 0.20219 1.0000
O15 O 0.29802 0.46932 0.11210 1.0000
O16 O 0.20059 0.66472 0.20193 1.0000
O17 O 0.23000 0.98091 0.20323 1.0000
O18 O 0.24377 0.34761 0.20436 1.0000
O19 O 0.32008 0.50831 0.29792 1.0000
O20 O 0.54933 0.35624 0.10837 1.0000
O21 O 0.58776 0.66230 0.10841 1.0000
O22 O 0.58878 0.50692 0.00633 1.0000
O23 O 0.67284 0.22510 0.01692 1.0000
O24 O 0.67326 0.78397 0.00629 1.0000
O25 O 0.79074 0.85963 0.10852 1.0000
O26 O 0.83342 0.16546 0.10892 1.0000
O27 O 0.83959 0.00693 0.00712 1.0000
O28 O 0.92066 0.72467 0.01936 1.0000
O29 O 0.92521 0.28342 0.00744 1.0000
O30 O 0.54113 0.96887 0.11040 1.0000
O31 O 0.45282 0.15774 0.20242 1.0000
O32 O 0.49249 0.46964 0.20156 1.0000
O33 O 0.47790 0.84405 0.20088 1.0000
O34 O 0.79878 0.47179 0.11299 1.0000
O35 O 0.70450 0.66866 0.20291 1.0000
O36 O 0.72596 0.98140 0.20166 1.0000
O37 O 0.74314 0.35863 0.20455 1.0000
H1 H 0.08411 0.05156 0.09137 1.0000
H2 H 0.02361 0.14725 0.22523 1.0000
H3 H 0.96819 0.51351 0.23872 1.0000
H4 H 0.43056 0.75071 0.20132 1.0000
H5 H 0.33868 0.55257 0.09204 1.0000
H6 H 0.22846 0.65515 0.24354 1.0000
H7 H 0.24753 0.96106 0.24527 1.0000
H8 H 0.20417 0.25066 0.20974 1.0000
H9 H 0.28518 0.43487 0.26800 1.0000
H10 H 0.21075 0.42564 0.37009 1.0000
H11 H 0.35398 0.52546 0.38850 1.0000
H12 H 0.36530 0.34050 0.35995 1.0000
H13 H 0.58113 0.04688 0.08693 1.0000
H14 H 0.54484 0.14278 0.20802 1.0000
H15 H 0.46691 0.49571 0.24238 1.0000
H16 H 0.92549 0.75595 0.20387 1.0000
H17 H 0.84027 0.55464 0.09310 1.0000
H18 H 0.68550 0.66707 0.24550 1.0000
H19 H 0.73942 0.96524 0.24424 1.0000
H20 H 0.70816 0.27500 0.22623 1.0000
C1 C 0.31163 0.44589 0.35679 1.0000
\end{verbatim}
\newpage
\begin{verbatim}
# Kaolinite slab + methanol at SiO2

data_SiO2+M
_symmetry_Int_Tables_number        1
_cell_length_a                     10.384527
_cell_length_b                     9.011475
_cell_length_c                     22.250826
_cell_angle_alpha                  90.000000
_cell_angle_beta                   90.000000
_cell_angle_gamma                  90.000000

loop_
_symmetry_equiv_pos_site_id
_symmetry_equiv_pos_as_xyz
   1   x,y,z

loop_
_atom_site_label
_atom_site_type_symbol
_atom_site_fract_x
_atom_site_fract_y
_atom_site_fract_z
_atom_site_occupancy
Al1 Al 0.34955 0.47089 0.55292 1.0000
Al2 Al 0.33739 0.80651 0.55226 1.0000
Al3 Al 0.58833 0.97216 0.55153 1.0000
Al4 Al 0.59787 0.30477 0.55258 1.0000
Al5 Al 0.84959 0.47072 0.55322 1.0000
Al6 Al 0.83727 0.80591 0.55265 1.0000
Al7 Al 0.08856 0.97223 0.55199 1.0000
Al8 Al 0.09794 0.30455 0.55286 1.0000
Si1 Si 0.26199 0.31881 0.43025 1.0000
Si2 Si 0.26742 0.64674 0.43099 1.0000
Si3 Si 0.50895 0.81930 0.43005 1.0000
Si4 Si 0.51872 0.14741 0.43012 1.0000
Si5 Si 0.76207 0.31888 0.43029 1.0000
Si6 Si 0.76823 0.64626 0.43126 1.0000
Si7 Si 0.00841 0.81955 0.43081 1.0000
Si8 Si 0.01850 0.14735 0.43042 1.0000
O1 O 0.25044 0.33115 0.50393 1.0000
O2 O 0.28811 0.63702 0.50416 1.0000
O3 O 0.28960 0.48238 0.40186 1.0000
O4 O 0.37418 0.20115 0.41176 1.0000
O5 O 0.37304 0.75923 0.40177 1.0000
O6 O 0.48983 0.83684 0.50348 1.0000
O7 O 0.53460 0.14089 0.50365 1.0000
O8 O 0.54067 0.98186 0.40207 1.0000
O9 O 0.62086 0.69680 0.41740 1.0000
O10 O 0.62669 0.25856 0.40221 1.0000
O11 O 0.24211 0.94516 0.50666 1.0000
O12 O 0.15131 0.13346 0.59892 1.0000
O13 O 0.19536 0.44591 0.59762 1.0000
O14 O 0.17864 0.82003 0.59666 1.0000
O15 O 0.50091 0.44657 0.50747 1.0000
O16 O 0.40646 0.64289 0.59711 1.0000
O17 O 0.42875 0.95814 0.59760 1.0000
O18 O 0.44601 0.32930 0.59951 1.0000
O19 O 0.53310 0.74667 0.28620 1.0000
O20 O 0.75040 0.33127 0.50399 1.0000
O21 O 0.79015 0.63654 0.50446 1.0000
O22 O 0.79041 0.48215 0.40200 1.0000
O23 O 0.87392 0.20046 0.41214 1.0000
O24 O 0.87348 0.75926 0.40212 1.0000
O25 O 0.99098 0.83578 0.50410 1.0000
O26 O 0.03460 0.14058 0.50393 1.0000
O27 O 0.04050 0.98162 0.40230 1.0000
O28 O 0.12105 0.69936 0.41525 1.0000
O29 O 0.12673 0.25782 0.40233 1.0000
O30 O 0.74257 0.94466 0.50671 1.0000
O31 O 0.65107 0.13335 0.59860 1.0000
O32 O 0.69512 0.44573 0.59758 1.0000
O33 O 0.67798 0.81970 0.59641 1.0000
O34 O 0.00125 0.44609 0.50782 1.0000
O35 O 0.90633 0.64258 0.59769 1.0000
O36 O 0.92880 0.95775 0.59786 1.0000
O37 O 0.94605 0.32902 0.59978 1.0000
H1 H 0.44174 0.76998 0.28705 1.0000
H2 H 0.49894 0.52906 0.31796 1.0000
H3 H 0.64986 0.56473 0.28257 1.0000
H4 H 0.50820 0.54735 0.23768 1.0000
H5 H 0.28284 0.02341 0.48362 1.0000
H6 H 0.74209 0.11675 0.60619 1.0000
H7 H 0.19087 0.44386 0.64096 1.0000
H8 H 0.63078 0.72613 0.59762 1.0000
H9 H 0.54115 0.52922 0.48681 1.0000
H10 H 0.40334 0.64038 0.64051 1.0000
H11 H 0.44478 0.94044 0.63992 1.0000
H12 H 0.40963 0.23682 0.61370 1.0000
H13 H 0.78268 0.02213 0.48300 1.0000
H14 H 0.24224 0.11661 0.60668 1.0000
H15 H 0.68946 0.44500 0.64089 1.0000
H16 H 0.13153 0.72642 0.59796 1.0000
H17 H 0.04187 0.52978 0.48809 1.0000
H18 H 0.90463 0.63977 0.64110 1.0000
H19 H 0.94440 -0.05969 0.64023 1.0000
H20 H 0.90959 0.23694 0.61434 1.0000
C1 C 0.54639 0.58874 0.28079 1.0000
\end{verbatim}
}

\end{multicols}


\begin{mcitethebibliography}{139}
\providecommand*\natexlab[1]{#1}
\providecommand*\mciteSetBstSublistMode[1]{}
\providecommand*\mciteSetBstMaxWidthForm[2]{}
\providecommand*\mciteBstWouldAddEndPuncttrue
  {\def\EndOfBibitem{\unskip.}}
\providecommand*\mciteBstWouldAddEndPunctfalse
  {\let\EndOfBibitem\relax}
\providecommand*\mciteSetBstMidEndSepPunct[3]{}
\providecommand*\mciteSetBstSublistLabelBeginEnd[3]{}
\providecommand*\EndOfBibitem{}
\mciteSetBstSublistMode{f}
\mciteSetBstMaxWidthForm{subitem}{(\alph{mcitesubitemcount})}
\mciteSetBstSublistLabelBeginEnd
  {\mcitemaxwidthsubitemform\space}
  {\relax}
  {\relax}

\bibitem[Murray(2000)]{murray:clay-uses}
Murray,~H.~H. {Traditional and new applications for kaolin, smectite, and
  palygorskite: a general overview}. \emph{Appl. Clay. Sci.} \textbf{2000},
  \emph{17}, 207--221\relax
\mciteBstWouldAddEndPuncttrue
\mciteSetBstMidEndSepPunct{\mcitedefaultmidpunct}
{\mcitedefaultendpunct}{\mcitedefaultseppunct}\relax
\EndOfBibitem
\bibitem[{International Programme on Chemical Safety}(2005)]{ipcs:clay-health}
{International Programme on Chemical Safety}, \emph{Environmental Health
  Criteria 231: Bentonite, Kaolin and Selected Clay Minerals}; World Health
  Organization: Geneva, 2005\relax
\mciteBstWouldAddEndPuncttrue
\mciteSetBstMidEndSepPunct{\mcitedefaultmidpunct}
{\mcitedefaultendpunct}{\mcitedefaultseppunct}\relax
\EndOfBibitem
\bibitem[Hosterman and Patterson(1992)Hosterman, and Patterson]{usgs:bentonite}
Hosterman,~J.~W.; Patterson,~S.~H. {Bentonite and fuller's earth resources of
  the United States}. \emph{U.S. Geological Survey Professional Papers}
  \textbf{1992}, \emph{1522}, 1\relax
\mciteBstWouldAddEndPuncttrue
\mciteSetBstMidEndSepPunct{\mcitedefaultmidpunct}
{\mcitedefaultendpunct}{\mcitedefaultseppunct}\relax
\EndOfBibitem
\bibitem[Voinot \latin{et~al.}(2012)Voinot, Fischer, B{\oe}uf, Schmidt,
  Delval-Dubois, Reichardt, Liewig, Chaumande, Ehret-Sabatier, Lignot, and
  Angel]{kaolinite-medicine1}
Voinot,~F.; Fischer,~C.; B{\oe}uf,~A.; Schmidt,~C.; Delval-Dubois,~V.;
  Reichardt,~F.; Liewig,~N.; Chaumande,~B.; Ehret-Sabatier,~L.; Lignot,~J.-H.
  \latin{et~al.}  {Effects of controlled ingestion of kaolinite (5\%) on food
  intake, gut morphology and in vitro motility in rats}. \emph{Fundam. Clin.
  Pharmacology} \textbf{2012}, \emph{26}, 565--576\relax
\mciteBstWouldAddEndPuncttrue
\mciteSetBstMidEndSepPunct{\mcitedefaultmidpunct}
{\mcitedefaultendpunct}{\mcitedefaultseppunct}\relax
\EndOfBibitem
\bibitem[Wallace \latin{et~al.}(1975)Wallace, Headley, and
  Weber]{kaolinite-medicine2}
Wallace,~W.~E.; Headley,~L.~C.; Weber,~K.~C. {Dipalmitoyl lecithin surfactant
  adsorption by kaolin dust in vitro}. \emph{J. Colloid Interface Sci.}
  \textbf{1975}, \emph{51}, 535--537\relax
\mciteBstWouldAddEndPuncttrue
\mciteSetBstMidEndSepPunct{\mcitedefaultmidpunct}
{\mcitedefaultendpunct}{\mcitedefaultseppunct}\relax
\EndOfBibitem
\bibitem[Chrysikopoulos and Syngouna(2012)Chrysikopoulos, and
  Syngouna]{kaolinite-viruses1}
Chrysikopoulos,~C.~V.; Syngouna,~V.~I. {Attachment of bacteriophages MS2 and
  $\Phi$X174 onto kaolinite and montmorillonite: Extended-DLVO interactions}.
  \emph{Colloids Surf., B} \textbf{2012}, \emph{92}, 74--83\relax
\mciteBstWouldAddEndPuncttrue
\mciteSetBstMidEndSepPunct{\mcitedefaultmidpunct}
{\mcitedefaultendpunct}{\mcitedefaultseppunct}\relax
\EndOfBibitem
\bibitem[Lipson and Stotzky(1983)Lipson, and Stotzky]{clay-viruses2}
Lipson,~S.~M.; Stotzky,~G. {Adsorption of reovirus to clay minerals: effects of
  cation-exchange capacity, cation saturation, and surface area}. \emph{Appl.
  Environ. Microbiol.} \textbf{1983}, \emph{46}, 673--682\relax
\mciteBstWouldAddEndPuncttrue
\mciteSetBstMidEndSepPunct{\mcitedefaultmidpunct}
{\mcitedefaultendpunct}{\mcitedefaultseppunct}\relax
\EndOfBibitem
\bibitem[Schiffenbauer and Stotzky(1982)Schiffenbauer, and
  Stotzky]{clay-viruses3}
Schiffenbauer,~M.; Stotzky,~G. {Adsorption of coliphages T1 and T7 to clay
  minerals}. \emph{Appl. Environ. Microbiol.} \textbf{1982}, \emph{43},
  590--596\relax
\mciteBstWouldAddEndPuncttrue
\mciteSetBstMidEndSepPunct{\mcitedefaultmidpunct}
{\mcitedefaultendpunct}{\mcitedefaultseppunct}\relax
\EndOfBibitem
\bibitem[Inouye and Kono(1972)Inouye, and Kono]{clays-serum}
Inouye,~S.; Kono,~R. {Effect of a Modified Kaolin Treatment on Serum
  Immunoglobulins}. \emph{Appl. Microbiol.} \textbf{1972}, \emph{23}, 203\relax
\mciteBstWouldAddEndPuncttrue
\mciteSetBstMidEndSepPunct{\mcitedefaultmidpunct}
{\mcitedefaultendpunct}{\mcitedefaultseppunct}\relax
\EndOfBibitem
\bibitem[Murray \latin{et~al.}(2012)Murray, O{'}Sullivan, Atkinson, and
  Webb]{murray:review}
Murray,~B.~J.; O{'}Sullivan,~D.; Atkinson,~J.~D.; Webb,~M.~E. {Ice nucleation
  by particles immersed in supercooled cloud droplets}. \emph{Chem. Soc. Rev.}
  \textbf{2012}, \emph{41}, 6519--6554\relax
\mciteBstWouldAddEndPuncttrue
\mciteSetBstMidEndSepPunct{\mcitedefaultmidpunct}
{\mcitedefaultendpunct}{\mcitedefaultseppunct}\relax
\EndOfBibitem
\bibitem[Campbell and Lytken(2009)Campbell, and
  Lytken]{campbell_experimental_2009}
Campbell,~C.~T.; Lytken,~O. Experimental measurements of the energetics of
  surface reactions. \emph{Surf. Sci.} \textbf{2009}, \emph{603},
  1365--1372\relax
\mciteBstWouldAddEndPuncttrue
\mciteSetBstMidEndSepPunct{\mcitedefaultmidpunct}
{\mcitedefaultendpunct}{\mcitedefaultseppunct}\relax
\EndOfBibitem
\bibitem[Campbell and Sellers(2013)Campbell, and
  Sellers]{campbell_enthalpies_2013}
Campbell,~C.~T.; Sellers,~J. R.~V. Enthalpies and {Entropies} of {Adsorption}
  on {Well}-{Defined} {Oxide} {Surfaces}: {Experimental} {Measurements}.
  \emph{Chem. Rev.} \textbf{2013}, \emph{113}, 4106--4135\relax
\mciteBstWouldAddEndPuncttrue
\mciteSetBstMidEndSepPunct{\mcitedefaultmidpunct}
{\mcitedefaultendpunct}{\mcitedefaultseppunct}\relax
\EndOfBibitem
\bibitem[Carrasco \latin{et~al.}(2012)Carrasco, Hodgson, and
  Michaelides]{Carrasco:2012iu}
Carrasco,~J.; Hodgson,~A.; Michaelides,~A. {A molecular perspective of water at
  metal interfaces}. \emph{Nat. Mater.} \textbf{2012}, \emph{11},
  667--674\relax
\mciteBstWouldAddEndPuncttrue
\mciteSetBstMidEndSepPunct{\mcitedefaultmidpunct}
{\mcitedefaultendpunct}{\mcitedefaultseppunct}\relax
\EndOfBibitem
\bibitem[Bj\"{o}rneholm \latin{et~al.}(2016)Bj\"{o}rneholm, Hansen, Hodgson,
  Liu, Limmer, Michaelides, Pedevilla, Rossmeisl, Shen, Tocci, Tyrode, Walz,
  Werner, and Bluhm]{bjorneholm_water_2016}
Bj\"{o}rneholm,~O.; Hansen,~M.~H.; Hodgson,~A.; Liu,~L.-M.; Limmer,~D.~T.;
  Michaelides,~A.; Pedevilla,~P.; Rossmeisl,~J.; Shen,~H.; Tocci,~G.
  \latin{et~al.}  Water at {Interfaces}. \emph{Chem. Rev.} \textbf{2016},
  \emph{116}, 7698--7726\relax
\mciteBstWouldAddEndPuncttrue
\mciteSetBstMidEndSepPunct{\mcitedefaultmidpunct}
{\mcitedefaultendpunct}{\mcitedefaultseppunct}\relax
\EndOfBibitem
\bibitem[Striolo \latin{et~al.}(2016)Striolo, Michaelides, and
  Joly]{striolo_carbon-water_2016}
Striolo,~A.; Michaelides,~A.; Joly,~L. The {Carbon}-{Water} {Interface}:
  {Modeling} {Challenges} and {Opportunities} for the {Water}-{Energy} {Nexus}.
  \emph{Annu. Rev. Chem. Biomol. Eng.} \textbf{2016}, \emph{7}, 533--556\relax
\mciteBstWouldAddEndPuncttrue
\mciteSetBstMidEndSepPunct{\mcitedefaultmidpunct}
{\mcitedefaultendpunct}{\mcitedefaultseppunct}\relax
\EndOfBibitem
\bibitem[Cohen \latin{et~al.}(2012)Cohen, Mori-Sanchez, and
  Yang]{challengesDFT:2012}
Cohen,~A.~J.; Mori-Sanchez,~P.; Yang,~W. {Challenges for Density Functional
  Theory}. \emph{Chem. Rev.} \textbf{2012}, \emph{112}, 289--320\relax
\mciteBstWouldAddEndPuncttrue
\mciteSetBstMidEndSepPunct{\mcitedefaultmidpunct}
{\mcitedefaultendpunct}{\mcitedefaultseppunct}\relax
\EndOfBibitem
\bibitem[Burke(2012)]{burke_perspective_2012}
Burke,~K. Perspective on density functional theory. \emph{J. Chem. Phys.}
  \textbf{2012}, \emph{136}, 150901\relax
\mciteBstWouldAddEndPuncttrue
\mciteSetBstMidEndSepPunct{\mcitedefaultmidpunct}
{\mcitedefaultendpunct}{\mcitedefaultseppunct}\relax
\EndOfBibitem
\bibitem[Klime\v{s} and Michaelides(2012)Klime\v{s}, and
  Michaelides]{jiri:review}
Klime\v{s},~J.; Michaelides,~A. Desorption of water during the drying of clay
  minerals. Enthalpy and entropy variation. \emph{J. Chem. Phys.}
  \textbf{2012}, \emph{137}, 120901\relax
\mciteBstWouldAddEndPuncttrue
\mciteSetBstMidEndSepPunct{\mcitedefaultmidpunct}
{\mcitedefaultendpunct}{\mcitedefaultseppunct}\relax
\EndOfBibitem
\bibitem[Grimme \latin{et~al.}(2016)Grimme, Hansen, Brandenburg, and
  Bannwarth]{Grimme_Gerit_chemrev2016}
Grimme,~S.; Hansen,~A.; Brandenburg,~J.~G.; Bannwarth,~C. Dispersion-Corrected
  Mean-Field Electronic Structure Methods. \emph{Chem. Rev.} \textbf{2016},
  \emph{116}, 5105--5154\relax
\mciteBstWouldAddEndPuncttrue
\mciteSetBstMidEndSepPunct{\mcitedefaultmidpunct}
{\mcitedefaultendpunct}{\mcitedefaultseppunct}\relax
\EndOfBibitem
\bibitem[Foulkes \latin{et~al.}(2001)Foulkes, Mitas, Needs, and
  Rajagopal]{foulkes01}
Foulkes,~W. M.~C.; Mitas,~L.; Needs,~R.~J.; Rajagopal,~G. {Quantum Monte Carlo
  simulations of solids}. \emph{Rev. Mod. Phys.} \textbf{2001}, \emph{73},
  33--83\relax
\mciteBstWouldAddEndPuncttrue
\mciteSetBstMidEndSepPunct{\mcitedefaultmidpunct}
{\mcitedefaultendpunct}{\mcitedefaultseppunct}\relax
\EndOfBibitem
\bibitem[Ochsenfeld \latin{et~al.}(2007)Ochsenfeld, Kussmann, and
  Lambrecht]{reviewQC}
Ochsenfeld,~C.; Kussmann,~J.; Lambrecht,~D.~S. \emph{Rev. Comput. Chem.}; John
  Wiley {\&} Sons, Inc., 2007; pp 1--82\relax
\mciteBstWouldAddEndPuncttrue
\mciteSetBstMidEndSepPunct{\mcitedefaultmidpunct}
{\mcitedefaultendpunct}{\mcitedefaultseppunct}\relax
\EndOfBibitem
\bibitem[Bartlett and Musia{\l}(2007)Bartlett, and Musia{\l}]{CCSDT:Rev2007}
Bartlett,~R.; Musia{\l},~M. {Coupled-cluster theory in quantum chemistry}.
  \emph{Rev. Mod. Phys.} \textbf{2007}, \emph{79}, 291--352\relax
\mciteBstWouldAddEndPuncttrue
\mciteSetBstMidEndSepPunct{\mcitedefaultmidpunct}
{\mcitedefaultendpunct}{\mcitedefaultseppunct}\relax
\EndOfBibitem
\bibitem[Chan and Head-Gordon(2002)Chan, and Head-Gordon]{DMRG:2002}
Chan,~G. K.-L.; Head-Gordon,~M. {Highly correlated calculations with a
  polynomial cost algorithm: A study of the density matrix renormalization
  group}. \emph{J. Chem. Phys.} \textbf{2002}, \emph{116}, 4462\relax
\mciteBstWouldAddEndPuncttrue
\mciteSetBstMidEndSepPunct{\mcitedefaultmidpunct}
{\mcitedefaultendpunct}{\mcitedefaultseppunct}\relax
\EndOfBibitem
\bibitem[Booth \latin{et~al.}(2009)Booth, Thom, and Alavi]{FCIQMC:JCP2009}
Booth,~G.~H.; Thom,~A. J.~W.; Alavi,~A. {Fermion Monte Carlo without fixed
  nodes: A game of life, death, and annihilation in Slater determinant space}.
  \emph{J. Chem. Phys.} \textbf{2009}, \emph{131}, 054106\relax
\mciteBstWouldAddEndPuncttrue
\mciteSetBstMidEndSepPunct{\mcitedefaultmidpunct}
{\mcitedefaultendpunct}{\mcitedefaultseppunct}\relax
\EndOfBibitem
\bibitem[Booth \latin{et~al.}(2013)Booth, Gr{\"u}neis, Kresse, and
  Alavi]{FCIQMC:Nat2013}
Booth,~G.~H.; Gr{\"u}neis,~A.; Kresse,~G.; Alavi,~A. {Towards an exact
  description of electronic wavefunctions in real solids}. \emph{Nature}
  \textbf{2013}, \emph{493}, 365--370\relax
\mciteBstWouldAddEndPuncttrue
\mciteSetBstMidEndSepPunct{\mcitedefaultmidpunct}
{\mcitedefaultendpunct}{\mcitedefaultseppunct}\relax
\EndOfBibitem
\bibitem[Zhang and Krakauer(2003)Zhang, and Krakauer]{AFQMC:Zhang2003}
Zhang,~S.; Krakauer,~H. {Quantum Monte Carlo Method using Phase-Free Random
  Walks with Slater Determinants}. \emph{Phys. Rev. Lett.} \textbf{2003},
  \emph{90}, 136401\relax
\mciteBstWouldAddEndPuncttrue
\mciteSetBstMidEndSepPunct{\mcitedefaultmidpunct}
{\mcitedefaultendpunct}{\mcitedefaultseppunct}\relax
\EndOfBibitem
\bibitem[Casula \latin{et~al.}(2005)Casula, Filippi, and
  Sorella]{LRDMC:prl2005}
Casula,~M.; Filippi,~C.; Sorella,~S. {Diffusion Monte Carlo method with lattice
  regularization}. \emph{Phys. Rev. Lett.} \textbf{2005}, \emph{95},
  100201\relax
\mciteBstWouldAddEndPuncttrue
\mciteSetBstMidEndSepPunct{\mcitedefaultmidpunct}
{\mcitedefaultendpunct}{\mcitedefaultseppunct}\relax
\EndOfBibitem
\bibitem[Casula \latin{et~al.}(2010)Casula, Moroni, Sorella, and
  Filippi]{casula10}
Casula,~M.; Moroni,~S.; Sorella,~S.; Filippi,~C. {Size-consistent variational
  approaches to nonlocal pseudopotentials: Standard and lattice regularized
  diffusion Monte Carlo methods revisited}. \emph{J. Chem. Phys.}
  \textbf{2010}, \emph{132}, 154113\relax
\mciteBstWouldAddEndPuncttrue
\mciteSetBstMidEndSepPunct{\mcitedefaultmidpunct}
{\mcitedefaultendpunct}{\mcitedefaultseppunct}\relax
\EndOfBibitem
\bibitem[Santra \latin{et~al.}(2011)Santra, Klime{\v s}, Alf{\`e}, Tkatchenko,
  Slater, Michaelides, Car, and Scheffler]{Ice:prl2011}
Santra,~B.; Klime{\v s},~J.; Alf{\`e},~D.; Tkatchenko,~A.; Slater,~B.;
  Michaelides,~A.; Car,~R.; Scheffler,~M. {Hydrogen Bonds and van der Waals
  Forces in Ice at Ambient and High Pressures}. \emph{Phys. Rev. Lett.}
  \textbf{2011}, \emph{107}, 185701\relax
\mciteBstWouldAddEndPuncttrue
\mciteSetBstMidEndSepPunct{\mcitedefaultmidpunct}
{\mcitedefaultendpunct}{\mcitedefaultseppunct}\relax
\EndOfBibitem
\bibitem[Morales \latin{et~al.}(2014)Morales, Gergely, McMinis, McMahon, Kim,
  and Ceperley]{Morales:bulkwat:2014}
Morales,~M.~A.; Gergely,~J.~R.; McMinis,~J.; McMahon,~J.~M.; Kim,~J.;
  Ceperley,~D.~M. {Quantum Monte Carlo Benchmark of Exchange-Correlation
  Functionals for Bulk Water}. \emph{J. Chem. Theory Comput.} \textbf{2014},
  \emph{10}, 2355--2362\relax
\mciteBstWouldAddEndPuncttrue
\mciteSetBstMidEndSepPunct{\mcitedefaultmidpunct}
{\mcitedefaultendpunct}{\mcitedefaultseppunct}\relax
\EndOfBibitem
\bibitem[Cox \latin{et~al.}(2014)Cox, Towler, Alf{\`e}, and
  Michaelides]{Cox:2014}
Cox,~S.~J.; Towler,~M.~D.; Alf{\`e},~D.; Michaelides,~A. {Benchmarking the
  performance of density functional theory and point charge force fields in
  their description of sI methane hydrate against diffusion Monte Carlo}.
  \emph{J. Chem. Phys.} \textbf{2014}, \emph{140}, 174703\relax
\mciteBstWouldAddEndPuncttrue
\mciteSetBstMidEndSepPunct{\mcitedefaultmidpunct}
{\mcitedefaultendpunct}{\mcitedefaultseppunct}\relax
\EndOfBibitem
\bibitem[Benali \latin{et~al.}(2014)Benali, Shulenburger, Romero, Kim, and von
  Lilienfeld]{Benali:2014}
Benali,~A.; Shulenburger,~L.; Romero,~N.~A.; Kim,~J.; von Lilienfeld,~O.~A.
  {Application of Diffusion Monte Carlo to Materials Dominated by van der Waals
  Interactions}. \emph{J. Chem. Theory Comput.} \textbf{2014}, \emph{10},
  3417--3422\relax
\mciteBstWouldAddEndPuncttrue
\mciteSetBstMidEndSepPunct{\mcitedefaultmidpunct}
{\mcitedefaultendpunct}{\mcitedefaultseppunct}\relax
\EndOfBibitem
\bibitem[Al-Hamdani \latin{et~al.}(2015)Al-Hamdani, Ma, Alf{\`e}, von
  Lilienfeld, and Michaelides]{AlHamdani:hBN:2015}
Al-Hamdani,~Y.~S.; Ma,~M.; Alf{\`e},~D.; von Lilienfeld,~O.~A.; Michaelides,~A.
  {Communication: Water on hexagonal boron nitride from diffusion Monte Carlo}.
  \emph{J. Chem. Phys.} \textbf{2015}, \emph{142}, 181101\relax
\mciteBstWouldAddEndPuncttrue
\mciteSetBstMidEndSepPunct{\mcitedefaultmidpunct}
{\mcitedefaultendpunct}{\mcitedefaultseppunct}\relax
\EndOfBibitem
\bibitem[Gillan \latin{et~al.}(2015)Gillan, Alf\`{e}, and Manby]{gillan15}
Gillan,~M.~J.; Alf\`{e},~D.; Manby,~F.~R. {Energy benchmarks for methane-water
  systems from quantum Monte Carlo and second-order M{\o}ller-Plesset
  calculations}. \emph{J. Chem. Phys.} \textbf{2015}, \emph{143}, 102812\relax
\mciteBstWouldAddEndPuncttrue
\mciteSetBstMidEndSepPunct{\mcitedefaultmidpunct}
{\mcitedefaultendpunct}{\mcitedefaultseppunct}\relax
\EndOfBibitem
\bibitem[Virgus \latin{et~al.}(2012)Virgus, Purwanto, Krakauer, and
  Zhang]{AFQMC:CoGraf}
Virgus,~Y.; Purwanto,~W.; Krakauer,~H.; Zhang,~S. {Ab initiomany-body study of
  cobalt adatoms adsorbed on graphene}. \emph{Phys. Rev. B} \textbf{2012},
  \emph{86}, 241406\relax
\mciteBstWouldAddEndPuncttrue
\mciteSetBstMidEndSepPunct{\mcitedefaultmidpunct}
{\mcitedefaultendpunct}{\mcitedefaultseppunct}\relax
\EndOfBibitem
\bibitem[Morales \latin{et~al.}(2014)Morales, Clay, Pierleoni, and
  Ceperley]{Morales:perspective2014}
Morales,~M.; Clay,~R.; Pierleoni,~C.; Ceperley,~D. {First Principles Methods: A
  Perspective from Quantum Monte Carlo}. \emph{Entropy} \textbf{2014},
  \emph{16}, 287--321\relax
\mciteBstWouldAddEndPuncttrue
\mciteSetBstMidEndSepPunct{\mcitedefaultmidpunct}
{\mcitedefaultendpunct}{\mcitedefaultseppunct}\relax
\EndOfBibitem
\bibitem[Mazzola \latin{et~al.}(2014)Mazzola, Yunoki, and
  Sorella]{Mazzola:nat2014}
Mazzola,~G.; Yunoki,~S.; Sorella,~S. {Unexpectedly high pressure for molecular
  dissociation in liquid hydrogen by electronic simulation}. \emph{Nat.
  Commun.} \textbf{2014}, \emph{5}, 3487\relax
\mciteBstWouldAddEndPuncttrue
\mciteSetBstMidEndSepPunct{\mcitedefaultmidpunct}
{\mcitedefaultendpunct}{\mcitedefaultseppunct}\relax
\EndOfBibitem
\bibitem[Mazzola and Sorella(2015)Mazzola, and Sorella]{Mazzola:prl2015}
Mazzola,~G.; Sorella,~S. {Distinct Metallization and Atomization Transitions in
  Dense Liquid Hydrogen}. \emph{Phys. Rev. Lett.} \textbf{2015}, \emph{114},
  105701\relax
\mciteBstWouldAddEndPuncttrue
\mciteSetBstMidEndSepPunct{\mcitedefaultmidpunct}
{\mcitedefaultendpunct}{\mcitedefaultseppunct}\relax
\EndOfBibitem
\bibitem[Zen \latin{et~al.}(2015)Zen, Luo, Mazzola, Guidoni, and
  Sorella]{Zen-liquidwat}
Zen,~A.; Luo,~Y.; Mazzola,~G.; Guidoni,~L.; Sorella,~S. {Ab initio molecular
  dynamics simulation of liquid water by quantum Monte Carlo}. \emph{J. Chem.
  Phys.} \textbf{2015}, \emph{142}, 144111\relax
\mciteBstWouldAddEndPuncttrue
\mciteSetBstMidEndSepPunct{\mcitedefaultmidpunct}
{\mcitedefaultendpunct}{\mcitedefaultseppunct}\relax
\EndOfBibitem
\bibitem[Chen \latin{et~al.}(2014)Chen, Ren, Li, Alf{\`e}, and
  Wang]{Chen:jcp2014}
Chen,~J.; Ren,~X.; Li,~X.-Z.; Alf{\`e},~D.; Wang,~E. {On the room-temperature
  phase diagram of high pressure hydrogen: An ab initio molecular dynamics
  perspective and a diffusion Monte Carlo study}. \emph{J. Chem. Phys.}
  \textbf{2014}, \emph{141}, 024501\relax
\mciteBstWouldAddEndPuncttrue
\mciteSetBstMidEndSepPunct{\mcitedefaultmidpunct}
{\mcitedefaultendpunct}{\mcitedefaultseppunct}\relax
\EndOfBibitem
\bibitem[Wagner(2013)]{Wagner:2013}
Wagner,~L.~K. {Quantum Monte Carlo for Ab Initiocalculations of energy-relevant
  materials}. \emph{Int. J. Quantum Chem.} \textbf{2013}, \emph{114},
  94--101\relax
\mciteBstWouldAddEndPuncttrue
\mciteSetBstMidEndSepPunct{\mcitedefaultmidpunct}
{\mcitedefaultendpunct}{\mcitedefaultseppunct}\relax
\EndOfBibitem
\bibitem[Wagner and Abbamonte(2014)Wagner, and Abbamonte]{Wagner:prb2014}
Wagner,~L.~K.; Abbamonte,~P. Effect of electron correlation on the electronic
  structure and spin-lattice coupling of high-${T}_{c}$ cuprates: Quantum Monte
  Carlo calculations. \emph{Phys. Rev. B} \textbf{2014}, \emph{90},
  125129\relax
\mciteBstWouldAddEndPuncttrue
\mciteSetBstMidEndSepPunct{\mcitedefaultmidpunct}
{\mcitedefaultendpunct}{\mcitedefaultseppunct}\relax
\EndOfBibitem
\bibitem[Pauling(1930)]{pnas:kaolinite-pauling}
Pauling,~L. \emph{Proc. Natl. Acad. Sci. USA} \textbf{1930}, \emph{16},
  578--582\relax
\mciteBstWouldAddEndPuncttrue
\mciteSetBstMidEndSepPunct{\mcitedefaultmidpunct}
{\mcitedefaultendpunct}{\mcitedefaultseppunct}\relax
\EndOfBibitem
\bibitem[Bish and {von Dreele}(1989)Bish, and {von Dreele}]{Exp_X-ray}
Bish,~D.~L.; {von Dreele},~R.~B. Rietveld refinement of non-hydrogen atomic
  positions in kaolinite. \emph{Clays Clay Miner.} \textbf{1989}, \emph{37},
  289--296\relax
\mciteBstWouldAddEndPuncttrue
\mciteSetBstMidEndSepPunct{\mcitedefaultmidpunct}
{\mcitedefaultendpunct}{\mcitedefaultseppunct}\relax
\EndOfBibitem
\bibitem[Adams(1983)]{Exp_neutron1}
Adams,~J.~M. Hydrogen atom positions in kaolinite by neutron profile
  refinement. \emph{Clays Clay Miner.} \textbf{1983}, \emph{31}, 352--356\relax
\mciteBstWouldAddEndPuncttrue
\mciteSetBstMidEndSepPunct{\mcitedefaultmidpunct}
{\mcitedefaultendpunct}{\mcitedefaultseppunct}\relax
\EndOfBibitem
\bibitem[Young and Hewat(1988)Young, and Hewat]{Exp_neutron2}
Young,~R.~A.; Hewat,~A.~W. Verification of the triclinic crystal structure of
  kaolinite. \emph{Clays Clay Miner.} \textbf{1988}, \emph{36}, 225--232\relax
\mciteBstWouldAddEndPuncttrue
\mciteSetBstMidEndSepPunct{\mcitedefaultmidpunct}
{\mcitedefaultendpunct}{\mcitedefaultseppunct}\relax
\EndOfBibitem
\bibitem[Bish(1993)]{Exp_neutron3}
Bish,~D.~L. Rietveld refinement of the kaolinite structure at 1.5 K.
  \emph{Clays Clay Miner.} \textbf{1993}, \emph{41}, 738--744\relax
\mciteBstWouldAddEndPuncttrue
\mciteSetBstMidEndSepPunct{\mcitedefaultmidpunct}
{\mcitedefaultendpunct}{\mcitedefaultseppunct}\relax
\EndOfBibitem
\bibitem[Neder \latin{et~al.}(1999)Neder, M.~Burghammer, Schulz, Bram, and
  Fiedler]{Exp_SingleCrystal}
Neder,~R.~B.; M.~Burghammer,~T.~G.; Schulz,~H.; Bram,~A.; Fiedler,~S.
  Refinement of the kaolinite structure from single-crystal synchrotron data.
  \emph{Clays Clay Miner.} \textbf{1999}, \emph{47}, 487--494\relax
\mciteBstWouldAddEndPuncttrue
\mciteSetBstMidEndSepPunct{\mcitedefaultmidpunct}
{\mcitedefaultendpunct}{\mcitedefaultseppunct}\relax
\EndOfBibitem
\bibitem[Zvyagim(1960)]{Exp_ElectronDiffraction}
Zvyagim,~B.~B. Electron-diffraction determination of the structure of
  kaolinite. \emph{Soviet Phys. Crystallogr.} \textbf{1960}, \emph{5},
  32--42\relax
\mciteBstWouldAddEndPuncttrue
\mciteSetBstMidEndSepPunct{\mcitedefaultmidpunct}
{\mcitedefaultendpunct}{\mcitedefaultseppunct}\relax
\EndOfBibitem
\bibitem[Hobbs \latin{et~al.}(1997)Hobbs, Cygan, Nagy, Schultz, and
  Sears]{cygan:AmMin}
Hobbs,~J.~D.; Cygan,~T.~R.; Nagy,~K.~L.; Schultz,~P.~A.; Sears,~M.~P. All-atom
  ab initio energy minimization of the kaolinite crystal structure. \emph{Am.
  Mineral.} \textbf{1997}, \emph{82}, 657--662\relax
\mciteBstWouldAddEndPuncttrue
\mciteSetBstMidEndSepPunct{\mcitedefaultmidpunct}
{\mcitedefaultendpunct}{\mcitedefaultseppunct}\relax
\EndOfBibitem
\bibitem[Tunega \latin{et~al.}(2012)Tunega, Bu\v{c}ko, and
  Zaoui]{tunega:jcp-137_2012}
Tunega,~D.; Bu\v{c}ko,~T.; Zaoui,~A. Assessment of ten DFT methods in
  predicting structures of sheet silicates: Importance of dispersion
  corrections. \emph{J. Chem. Phys.} \textbf{2012}, \emph{137}, 114105\relax
\mciteBstWouldAddEndPuncttrue
\mciteSetBstMidEndSepPunct{\mcitedefaultmidpunct}
{\mcitedefaultendpunct}{\mcitedefaultseppunct}\relax
\EndOfBibitem
\bibitem[Teppen \latin{et~al.}(1997)Teppen, Rasmussen, Bertsch, Miller, and
  Sch\"{a}fer]{schafer:jcp-B}
Teppen,~B.~J.; Rasmussen,~K.; Bertsch,~P.~M.; Miller,~D.~M.; Sch\"{a}fer,~L.
  Molecular Dynamics Modeling of Clay Minerals. 1. Gibbsite, Kaolinite,
  Pyrophyllite, and Beidellite. \emph{J. Phys. Chem. B} \textbf{1997},
  \emph{101}, 1579--1587\relax
\mciteBstWouldAddEndPuncttrue
\mciteSetBstMidEndSepPunct{\mcitedefaultmidpunct}
{\mcitedefaultendpunct}{\mcitedefaultseppunct}\relax
\EndOfBibitem
\bibitem[Ugliengo \latin{et~al.}(2009)Ugliengo, Zicovich-Wilson, Tosoni, and
  Civalleri]{jmc:civalleri}
Ugliengo,~P.; Zicovich-Wilson,~C.~M.; Tosoni,~S.; Civalleri,~B. Role of
  dispersive interactions in layered materials: a periodic B3LYP and B3LYP-D*
  study of Mg(OH)2, Ca(OH)2 and kaolinite. \emph{J. Mater. Chem.}
  \textbf{2009}, \emph{19}, 2564--2572\relax
\mciteBstWouldAddEndPuncttrue
\mciteSetBstMidEndSepPunct{\mcitedefaultmidpunct}
{\mcitedefaultendpunct}{\mcitedefaultseppunct}\relax
\EndOfBibitem
\bibitem[Reynolds \latin{et~al.}(1982)Reynolds, Ceperley, Alder, and
  Lester]{Reynolds:1982en}
Reynolds,~P.~J.; Ceperley,~D.~M.; Alder,~B.~J.; Lester,~W.~A. {Fixed-node
  quantum Monte Carlo for molecules}. \emph{J. Chem. Phys.} \textbf{1982},
  \emph{77}, 5593--5603\relax
\mciteBstWouldAddEndPuncttrue
\mciteSetBstMidEndSepPunct{\mcitedefaultmidpunct}
{\mcitedefaultendpunct}{\mcitedefaultseppunct}\relax
\EndOfBibitem
\bibitem[Zen \latin{et~al.}(2016)Zen, Sorella, Gillan, Michaelides, and
  Alf\`e]{sizeconsDMC}
Zen,~A.; Sorella,~S.; Gillan,~M.~J.; Michaelides,~A.; Alf\`e,~D. { Boosting the
  accuracy and speed of quantum Monte Carlo: size-consistency and time-step. }.
  \emph{Phys. Rev. B} \textbf{2016}, \emph{93}, 241118(R)\relax
\mciteBstWouldAddEndPuncttrue
\mciteSetBstMidEndSepPunct{\mcitedefaultmidpunct}
{\mcitedefaultendpunct}{\mcitedefaultseppunct}\relax
\EndOfBibitem
\bibitem[Sorella and Capriotti(2000)Sorella, and Capriotti]{Sorella:2000p17651}
Sorella,~S.; Capriotti,~L. {Green function Monte Carlo with stochastic
  reconfiguration: An effective remedy for the sign problem}. \emph{Phys. Rev.
  B} \textbf{2000}, \emph{61}, 2599--2612\relax
\mciteBstWouldAddEndPuncttrue
\mciteSetBstMidEndSepPunct{\mcitedefaultmidpunct}
{\mcitedefaultendpunct}{\mcitedefaultseppunct}\relax
\EndOfBibitem
\bibitem[Buonaura and Sorella(1998)Buonaura, and Sorella]{Buonaura:1998p25304}
Buonaura,~M.; Sorella,~S. {Numerical study of the two-dimensional Heisenberg
  model using a Green function Monte Carlo technique with a fixed number of
  walkers}. \emph{Phys. Rev. B} \textbf{1998}, \emph{57}, 11446--11456\relax
\mciteBstWouldAddEndPuncttrue
\mciteSetBstMidEndSepPunct{\mcitedefaultmidpunct}
{\mcitedefaultendpunct}{\mcitedefaultseppunct}\relax
\EndOfBibitem
\bibitem[Santra \latin{et~al.}(2013)Santra, Klime\v{s}, Tkatchenko, Alf\`{e},
  Slater, Michaelides, Car, and Scheffler]{santra_on_2013}
Santra,~B.; Klime\v{s},~J.; Tkatchenko,~A.; Alf\`{e},~D.; Slater,~B.;
  Michaelides,~A.; Car,~R.; Scheffler,~M. On the accuracy of van der Waals
  inclusive density-functional theory exchange-correlation functionals for ice
  at ambient and high pressures. \emph{J. Chem. Phys.} \textbf{2013},
  \emph{139}\relax
\mciteBstWouldAddEndPuncttrue
\mciteSetBstMidEndSepPunct{\mcitedefaultmidpunct}
{\mcitedefaultendpunct}{\mcitedefaultseppunct}\relax
\EndOfBibitem
\bibitem[Quigley \latin{et~al.}(2014)Quigley, Alf\`{e}, and
  Slater]{Quigley:Ice_0_i_Ih:jcp2014}
Quigley,~D.; Alf\`{e},~D.; Slater,~B. {Communication: On the stability of ice
  0, ice i, and Ih}. \emph{J. Chem. Phys.} \textbf{2014}, \emph{141},
  161102\relax
\mciteBstWouldAddEndPuncttrue
\mciteSetBstMidEndSepPunct{\mcitedefaultmidpunct}
{\mcitedefaultendpunct}{\mcitedefaultseppunct}\relax
\EndOfBibitem
\bibitem[Lin \latin{et~al.}(2001)Lin, Zong, and
  Ceperley]{Lin:qmctwistavg:pre2001}
Lin,~C.; Zong,~F.~H.; Ceperley,~D.~M. {Twist-averaged boundary conditions in
  continuum quantum Monte Carlo algorithms}. \emph{Phys. Rev. E} \textbf{2001},
  \emph{64}, 016702\relax
\mciteBstWouldAddEndPuncttrue
\mciteSetBstMidEndSepPunct{\mcitedefaultmidpunct}
{\mcitedefaultendpunct}{\mcitedefaultseppunct}\relax
\EndOfBibitem
\bibitem[Chiesa \latin{et~al.}(2006)Chiesa, Ceperley, Martin, and
  Holzmann]{Chiesa:size_effects:prl2006}
Chiesa,~S.; Ceperley,~D.~M.; Martin,~R.~M.; Holzmann,~M. {Finite-Size Error in
  Many-Body Simulations with Long-Range Interactions}. \emph{Phys. Rev. Lett.}
  \textbf{2006}, \emph{97}, 076404\relax
\mciteBstWouldAddEndPuncttrue
\mciteSetBstMidEndSepPunct{\mcitedefaultmidpunct}
{\mcitedefaultendpunct}{\mcitedefaultseppunct}\relax
\EndOfBibitem
\bibitem[Kwee \latin{et~al.}(2008)Kwee, Zhang, and Krakauer]{KZK:prl2008}
Kwee,~H.; Zhang,~S.; Krakauer,~H. {Finite-Size Correction in Many-Body
  Electronic Structure Calculations}. \emph{Phys. Rev. Lett.} \textbf{2008},
  \emph{100}, 126404\relax
\mciteBstWouldAddEndPuncttrue
\mciteSetBstMidEndSepPunct{\mcitedefaultmidpunct}
{\mcitedefaultendpunct}{\mcitedefaultseppunct}\relax
\EndOfBibitem
\bibitem[Drummond \latin{et~al.}(2008)Drummond, Needs, Sorouri, and
  Foulkes]{drummond_finite-size_2008}
Drummond,~N.~D.; Needs,~R.~J.; Sorouri,~A.; Foulkes,~W. M.~C. Finite-size
  errors in continuum quantum {Monte} {Carlo} calculations. \emph{Phys. Rev. B}
  \textbf{2008}, \emph{78}, 125106\relax
\mciteBstWouldAddEndPuncttrue
\mciteSetBstMidEndSepPunct{\mcitedefaultmidpunct}
{\mcitedefaultendpunct}{\mcitedefaultseppunct}\relax
\EndOfBibitem
\bibitem[Dagrada \latin{et~al.}(2016)Dagrada, Karakuzu, Vildosola, Casula, and
  Sorella]{kspecial.Sandro}
Dagrada,~M.; Karakuzu,~S.; Vildosola,~V.~L.; Casula,~M.; Sorella,~S. Exact
  special twist method for quantum Monte Carlo simulations.
  \emph{arXiv:1606.06205} \textbf{2016}, \relax
\mciteBstWouldAddEndPunctfalse
\mciteSetBstMidEndSepPunct{\mcitedefaultmidpunct}
{}{\mcitedefaultseppunct}\relax
\EndOfBibitem
\bibitem[Burke(2012)]{burke:jcp-perspective}
Burke,~K. Perspective on density functional theory. \emph{J. Chem. Phys.}
  \textbf{2012}, \emph{136}, 150901\relax
\mciteBstWouldAddEndPuncttrue
\mciteSetBstMidEndSepPunct{\mcitedefaultmidpunct}
{\mcitedefaultendpunct}{\mcitedefaultseppunct}\relax
\EndOfBibitem
\bibitem[Perdew and Zunger(1981)Perdew, and Zunger]{LDA_PerdewZunger}
Perdew,~J.~P.; Zunger,~A. {Self-interaction correction to density-functional
  approximations for many-electron systems}. \emph{Phys. Rev. B} \textbf{1981},
  \emph{23}, 5048--5079\relax
\mciteBstWouldAddEndPuncttrue
\mciteSetBstMidEndSepPunct{\mcitedefaultmidpunct}
{\mcitedefaultendpunct}{\mcitedefaultseppunct}\relax
\EndOfBibitem
\bibitem[Perdew \latin{et~al.}(1996)Perdew, Burke, and Ernzerhof]{PBE}
Perdew,~J.~P.; Burke,~K.; Ernzerhof,~M. Generalized Gradient Approximation Made
  Simple. \emph{Phys. Rev. Lett.} \textbf{1996}, \emph{77}, 3865\relax
\mciteBstWouldAddEndPuncttrue
\mciteSetBstMidEndSepPunct{\mcitedefaultmidpunct}
{\mcitedefaultendpunct}{\mcitedefaultseppunct}\relax
\EndOfBibitem
\bibitem[Perdew \latin{et~al.}(1997)Perdew, Burke, and Ernzerhof]{PBE_Erratum}
Perdew,~J.~P.; Burke,~K.; Ernzerhof,~M. ERRATA Generalized Gradient
  Approximation Made Simple {[Phys. Rev. Lett. 77, 3865 (1996)]}. \emph{Phys.
  Rev. Lett.} \textbf{1997}, \emph{78}, 1396\relax
\mciteBstWouldAddEndPuncttrue
\mciteSetBstMidEndSepPunct{\mcitedefaultmidpunct}
{\mcitedefaultendpunct}{\mcitedefaultseppunct}\relax
\EndOfBibitem
\bibitem[Hammer \latin{et~al.}(1999)Hammer, Hansen, and N{\o}rskov]{RPBE}
Hammer,~B.; Hansen,~L.~B.; N{\o}rskov,~J.~K. Improved adsorption energetics
  within density-functional theory using revised Perdew-Burke-Ernzerhof
  functionals. \emph{Phys. Rev. B} \textbf{1999}, \emph{59}, 7413\relax
\mciteBstWouldAddEndPuncttrue
\mciteSetBstMidEndSepPunct{\mcitedefaultmidpunct}
{\mcitedefaultendpunct}{\mcitedefaultseppunct}\relax
\EndOfBibitem
\bibitem[Adamo and Barone(1999)Adamo, and Barone]{PBE0}
Adamo,~C.; Barone,~V. Toward reliable density functional methods without
  adjustable parameters: The PBE0 model. \emph{J. Chem. Phys.} \textbf{1999},
  \emph{110}, 6158--6170\relax
\mciteBstWouldAddEndPuncttrue
\mciteSetBstMidEndSepPunct{\mcitedefaultmidpunct}
{\mcitedefaultendpunct}{\mcitedefaultseppunct}\relax
\EndOfBibitem
\bibitem[Vosko \latin{et~al.}(1980)Vosko, Wilk, and Nusair]{Vosko:1980ui}
Vosko,~S.~H.; Wilk,~L.; Nusair,~M. {Accurate spin-dependent electron liquid
  correlation energies for local spin density calculations: a critical
  analysis}. \emph{Can. J. Phys.} \textbf{1980}, \relax
\mciteBstWouldAddEndPunctfalse
\mciteSetBstMidEndSepPunct{\mcitedefaultmidpunct}
{}{\mcitedefaultseppunct}\relax
\EndOfBibitem
\bibitem[Lee \latin{et~al.}(1988)Lee, Yang, and Parr]{LEE:1988ub}
Lee,~C.~T.; Yang,~W.~T.; Parr,~R.~G. {Development of the Colle-Salvetti
  Correlation-Energy Formula Into a Functional of the Electron-Density}.
  \emph{Phys. Rev. B} \textbf{1988}, \emph{37}, 785--789\relax
\mciteBstWouldAddEndPuncttrue
\mciteSetBstMidEndSepPunct{\mcitedefaultmidpunct}
{\mcitedefaultendpunct}{\mcitedefaultseppunct}\relax
\EndOfBibitem
\bibitem[Becke(1993)]{Becke:1993vx}
Becke,~A.~D. {Density-Functional Thermochemistry .3. the Role of Exact
  Exchange}. \emph{J. Chem. Phys.} \textbf{1993}, \emph{98}, 5648--5652\relax
\mciteBstWouldAddEndPuncttrue
\mciteSetBstMidEndSepPunct{\mcitedefaultmidpunct}
{\mcitedefaultendpunct}{\mcitedefaultseppunct}\relax
\EndOfBibitem
\bibitem[Stephens \latin{et~al.}(1994)Stephens, Devlin, Chabalowski, and
  Frisch]{STEPHENS:1994vd}
Stephens,~P.~J.; Devlin,~F.; Chabalowski,~C.~F.; Frisch,~M.~J. {Ab-Initio
  Calculation of Vibrational Absorption and Circular-Dichroism Spectra Using
  Density-Functional Force-Fields}. \emph{J. Phys. Chem.} \textbf{1994},
  \emph{98}, 11623--11627\relax
\mciteBstWouldAddEndPuncttrue
\mciteSetBstMidEndSepPunct{\mcitedefaultmidpunct}
{\mcitedefaultendpunct}{\mcitedefaultseppunct}\relax
\EndOfBibitem
\bibitem[Grimme(2006)]{DFT-D2}
Grimme,~S. Semiempirical {GGA}-type density functional constructed with a
  long-range dispersion correction. \emph{J. Comput. Chem.} \textbf{2006},
  \emph{27}, 1787--1799\relax
\mciteBstWouldAddEndPuncttrue
\mciteSetBstMidEndSepPunct{\mcitedefaultmidpunct}
{\mcitedefaultendpunct}{\mcitedefaultseppunct}\relax
\EndOfBibitem
\bibitem[Grimme \latin{et~al.}(2010)Grimme, Antony, Ehrlich, and Krieg]{DFT-D3}
Grimme,~S.; Antony,~J.; Ehrlich,~S.; Krieg,~H. A consistent and accurate ab
  initio parametrization of density functional dispersion correction
  ({DFT}-{D}) for the 94 elements {H}-{Pu}. \emph{J. Chem. Phys.}
  \textbf{2010}, \emph{132}, 154104\relax
\mciteBstWouldAddEndPuncttrue
\mciteSetBstMidEndSepPunct{\mcitedefaultmidpunct}
{\mcitedefaultendpunct}{\mcitedefaultseppunct}\relax
\EndOfBibitem
\bibitem[Grimme \latin{et~al.}(2011)Grimme, Ehrlich, and
  Goerigk]{Grimme:BJdamping}
Grimme,~S.; Ehrlich,~S.; Goerigk,~L. {Effect of the damping function in
  dispersion corrected density functional theory}. \emph{J. Comput. Chem.}
  \textbf{2011}, \emph{32}, 1456--1465\relax
\mciteBstWouldAddEndPuncttrue
\mciteSetBstMidEndSepPunct{\mcitedefaultmidpunct}
{\mcitedefaultendpunct}{\mcitedefaultseppunct}\relax
\EndOfBibitem
\bibitem[Tkatchenko and Scheffler(2009)Tkatchenko, and Scheffler]{DFT-TS}
Tkatchenko,~A.; Scheffler,~M. Accurate {Molecular} {Van} {Der} {Waals}
  {Interactions} from {Ground}-{State} {Electron} {Density} and {Free}-{Atom}
  {Reference} {Data}. \emph{Phys. Rev. Lett.} \textbf{2009}, \emph{102},
  073005\relax
\mciteBstWouldAddEndPuncttrue
\mciteSetBstMidEndSepPunct{\mcitedefaultmidpunct}
{\mcitedefaultendpunct}{\mcitedefaultseppunct}\relax
\EndOfBibitem
\bibitem[Zhang and Yang(1998)Zhang, and Yang]{revPBE}
Zhang,~Y.; Yang,~W. Comment on ``Generalized Gradient Approximation Made
  Simple''. \emph{Phys. Rev. Lett.} \textbf{1998}, \emph{80}, 890\relax
\mciteBstWouldAddEndPuncttrue
\mciteSetBstMidEndSepPunct{\mcitedefaultmidpunct}
{\mcitedefaultendpunct}{\mcitedefaultseppunct}\relax
\EndOfBibitem
\bibitem[Dion \latin{et~al.}(2004)Dion, Rydberg, Schroder, Langreth, and
  Lundqvist]{vdW-DF}
Dion,~M.; Rydberg,~H.; Schroder,~E.; Langreth,~D.~C.; Lundqvist,~B.~I.
  \emph{Phys. Rev. Lett.} \textbf{2004}, \emph{92}, 246401\relax
\mciteBstWouldAddEndPuncttrue
\mciteSetBstMidEndSepPunct{\mcitedefaultmidpunct}
{\mcitedefaultendpunct}{\mcitedefaultseppunct}\relax
\EndOfBibitem
\bibitem[Lee \latin{et~al.}(2010)Lee, Murray, Kong, Lundqvist, and
  Langreth]{vdW-DF2}
Lee,~K.; Murray,~E.~D.; Kong,~L.; Lundqvist,~B.~I.; Langreth,~D.~C. \emph{Phys.
  Rev. B} \textbf{2010}, \emph{82}, 081101\relax
\mciteBstWouldAddEndPuncttrue
\mciteSetBstMidEndSepPunct{\mcitedefaultmidpunct}
{\mcitedefaultendpunct}{\mcitedefaultseppunct}\relax
\EndOfBibitem
\bibitem[Klime\v{s} \latin{et~al.}(2010)Klime\v{s}, Bowler, and
  Michaelides]{klimes-vdW-DF}
Klime\v{s},~J.; Bowler,~D.~R.; Michaelides,~A. \emph{J. Phys.: Cond. Mat.}
  \textbf{2010}, \emph{22}, 022201\relax
\mciteBstWouldAddEndPuncttrue
\mciteSetBstMidEndSepPunct{\mcitedefaultmidpunct}
{\mcitedefaultendpunct}{\mcitedefaultseppunct}\relax
\EndOfBibitem
\bibitem[Klime\v{s} \latin{et~al.}(2011)Klime\v{s}, Bowler, and
  Michaelides]{optB86b-vdW}
Klime\v{s},~J.; Bowler,~D.~R.; Michaelides,~A. \emph{Phys. Rev. B}
  \textbf{2011}, \emph{83}, 195131\relax
\mciteBstWouldAddEndPuncttrue
\mciteSetBstMidEndSepPunct{\mcitedefaultmidpunct}
{\mcitedefaultendpunct}{\mcitedefaultseppunct}\relax
\EndOfBibitem
\bibitem[Cygan \latin{et~al.}(2004)Cygan, Liang, and Kalinichev]{cygan:clayff}
Cygan,~R.~T.; Liang,~J.~J.; Kalinichev,~A.~G. Molecular models of hydroxide,
  oxyhydroxide, and clay phases and the development of a general force field.
  \emph{J. Phys. Chem. B} \textbf{2004}, \emph{108}, 1255\relax
\mciteBstWouldAddEndPuncttrue
\mciteSetBstMidEndSepPunct{\mcitedefaultmidpunct}
{\mcitedefaultendpunct}{\mcitedefaultseppunct}\relax
\EndOfBibitem
\bibitem[Hess \latin{et~al.}(1997)Hess, Bekker, Berendsen, and
  Fraaije]{hess:lincs}
Hess,~B.; Bekker,~H.; Berendsen,~H. J.~C.; Fraaije,~J. G. E.~M. \texttt{LINCS}:
  a linear constraint solver for molecular simulations. \emph{J. Comput. Chem.}
  \textbf{1997}, \emph{18}, 1463--1472\relax
\mciteBstWouldAddEndPuncttrue
\mciteSetBstMidEndSepPunct{\mcitedefaultmidpunct}
{\mcitedefaultendpunct}{\mcitedefaultseppunct}\relax
\EndOfBibitem
\bibitem[Abascal and Vega(2005)Abascal, and Vega]{vega:tip4p-2005}
Abascal,~J. L.~F.; Vega,~C. A general purpose model for the condensed phases of
  water: {TIP4P/2005}. \emph{J. Chem. Phys.} \textbf{2005}, \emph{123},
  234505\relax
\mciteBstWouldAddEndPuncttrue
\mciteSetBstMidEndSepPunct{\mcitedefaultmidpunct}
{\mcitedefaultendpunct}{\mcitedefaultseppunct}\relax
\EndOfBibitem
\bibitem[Jorgensen(1986)]{jorgensen1986optimized}
Jorgensen,~W.~L. Optimized intermolecular potential functions for liquid
  alcohols. \emph{J. Phys. Chem.} \textbf{1986}, \emph{90}, 1276--1284\relax
\mciteBstWouldAddEndPuncttrue
\mciteSetBstMidEndSepPunct{\mcitedefaultmidpunct}
{\mcitedefaultendpunct}{\mcitedefaultseppunct}\relax
\EndOfBibitem
\bibitem[Hess \latin{et~al.}(2008)Hess, Kutzner, van~der Spoel, and
  Lindahl]{gromacs4}
Hess,~B.; Kutzner,~C.; van~der Spoel,~D.; Lindahl,~E. \texttt{GROMACS 4}:
  Algorithms for Highly Efficient, Load-Balanced, and Scalable Molecular
  Simulation. \emph{J. Chem. Theor. Comput.} \textbf{2008}, \emph{4}, 435\relax
\mciteBstWouldAddEndPuncttrue
\mciteSetBstMidEndSepPunct{\mcitedefaultmidpunct}
{\mcitedefaultendpunct}{\mcitedefaultseppunct}\relax
\EndOfBibitem
\bibitem[Darden \latin{et~al.}(1993)Darden, York, and Pedersen]{pme1}
Darden,~T.; York,~D.; Pedersen,~L. Particle mesh {Ewald}: An {$N\cdot\log(N)$}
  method for {Ewald} sums in large systems. \emph{J. Chem. Phys.}
  \textbf{1993}, \emph{98}, 10089--10092\relax
\mciteBstWouldAddEndPuncttrue
\mciteSetBstMidEndSepPunct{\mcitedefaultmidpunct}
{\mcitedefaultendpunct}{\mcitedefaultseppunct}\relax
\EndOfBibitem
\bibitem[Essmann \latin{et~al.}(1995)Essmann, Perera, Berkowitz, Darden, Lee,
  and Pedersen]{pme2}
Essmann,~U.; Perera,~L.; Berkowitz,~M.~L.; Darden,~T.; Lee,~H.; Pedersen,~L.~G.
  A smooth particle mesh {Ewald} method. \emph{J. Chem. Phys.} \textbf{1995},
  \emph{103}, 8577--8593\relax
\mciteBstWouldAddEndPuncttrue
\mciteSetBstMidEndSepPunct{\mcitedefaultmidpunct}
{\mcitedefaultendpunct}{\mcitedefaultseppunct}\relax
\EndOfBibitem
\bibitem[Yeh and Berkowitz(1999)Yeh, and Berkowitz]{yeh:2dEwald}
Yeh,~I.~C.; Berkowitz,~M.~L. Ewald summation for systems with slab geometry.
  \emph{J. Chem. Phys.} \textbf{1999}, \emph{111}, 3155--3162\relax
\mciteBstWouldAddEndPuncttrue
\mciteSetBstMidEndSepPunct{\mcitedefaultmidpunct}
{\mcitedefaultendpunct}{\mcitedefaultseppunct}\relax
\EndOfBibitem
\bibitem[Costanzo \latin{et~al.}(1984)Costanzo, Giesse, and Lipsicas]{Costanzo}
Costanzo,~P.~M.; Giesse,~F.~R.; Lipsicas,~M. Static and Dynamic Structure of
  Water in Hydrated Kaolinites. I. The Static Structure. \emph{Clays Clay
  Miner.} \textbf{1984}, \emph{32}, 419--428\relax
\mciteBstWouldAddEndPuncttrue
\mciteSetBstMidEndSepPunct{\mcitedefaultmidpunct}
{\mcitedefaultendpunct}{\mcitedefaultseppunct}\relax
\EndOfBibitem
\bibitem[Lipsicas \latin{et~al.}(1984)Lipsicas, Straley, Costanzo, and
  Giesse]{Lipsicas}
Lipsicas,~M.; Straley,~C.; Costanzo,~P.~M.; Giesse,~F.~R. Static and dynamic
  structure of water in hydrated kaolinites. II. The dynamic structure.
  \emph{J. Colloid Interface Sci.} \textbf{1984}, \emph{107}, 221--230\relax
\mciteBstWouldAddEndPuncttrue
\mciteSetBstMidEndSepPunct{\mcitedefaultmidpunct}
{\mcitedefaultendpunct}{\mcitedefaultseppunct}\relax
\EndOfBibitem
\bibitem[Khalfi and Banchart(1999)Khalfi, and Banchart]{Khalfi}
Khalfi,~A.; Banchart,~P. Desorption of water during the drying of clay
  minerals. Enthalpy and entropy variation. \emph{Ceram. Int.} \textbf{1999},
  \emph{25}, 409--414\relax
\mciteBstWouldAddEndPuncttrue
\mciteSetBstMidEndSepPunct{\mcitedefaultmidpunct}
{\mcitedefaultendpunct}{\mcitedefaultseppunct}\relax
\EndOfBibitem
\bibitem[Hu and Michaelides(2007)Hu, and Michaelides]{Kaolinite_Hu0}
Hu,~X.~L.; Michaelides,~A. Ice formation on kaolinite: Lattice match or
  amphoterism? \emph{Surf. Sci.} \textbf{2007}, \emph{601}, 5378--5381\relax
\mciteBstWouldAddEndPuncttrue
\mciteSetBstMidEndSepPunct{\mcitedefaultmidpunct}
{\mcitedefaultendpunct}{\mcitedefaultseppunct}\relax
\EndOfBibitem
\bibitem[Hu and Michaelides(2008)Hu, and Michaelides]{Kaolinite_Hu1}
Hu,~X.~L.; Michaelides,~A. Water on the hydroxylated (0 0 1) surface of
  kaolinite: {From} monomer adsorption to a flat 2D wetting layer. \emph{Surf.
  Sci.} \textbf{2008}, \emph{602}, 960--974\relax
\mciteBstWouldAddEndPuncttrue
\mciteSetBstMidEndSepPunct{\mcitedefaultmidpunct}
{\mcitedefaultendpunct}{\mcitedefaultseppunct}\relax
\EndOfBibitem
\bibitem[Hu and Michaelides(2010)Hu, and Michaelides]{Kaolinite_Hu2}
Hu,~X.~L.; Michaelides,~A. The kaolinite (0 0 1) polar basal plane.
  \emph{Surf. Sci.} \textbf{2010}, \emph{604}, 111--117\relax
\mciteBstWouldAddEndPuncttrue
\mciteSetBstMidEndSepPunct{\mcitedefaultmidpunct}
{\mcitedefaultendpunct}{\mcitedefaultseppunct}\relax
\EndOfBibitem
\bibitem[Tunega \latin{et~al.}(2002)Tunega, Benco, Haberhauer, Gerzabek, and
  Lischka]{Tunega_1}
Tunega,~D.; Benco,~L.; Haberhauer,~G.; Gerzabek,~M.~H.; Lischka,~H. Ab Initio
  Molecular Dynamics Study of Adsorption Sites on the {(001)} Surfaces of {1:1}
  Dioctahedral Clay Minerals. \emph{J. Phys. Chem. B} \textbf{2002},
  \emph{106}, 11515--11525\relax
\mciteBstWouldAddEndPuncttrue
\mciteSetBstMidEndSepPunct{\mcitedefaultmidpunct}
{\mcitedefaultendpunct}{\mcitedefaultseppunct}\relax
\EndOfBibitem
\bibitem[Tunega \latin{et~al.}(2004)Tunega, Gerzabek, and Lischka]{Tunega_2}
Tunega,~D.; Gerzabek,~M.~H.; Lischka,~H. Ab Initio Molecular Dynamics Study of
  a Monomolecular Water Layer on Octahedral and Tetrahedral Kaolinite Surfaces.
  \emph{J. Phys. Chem. B} \textbf{2004}, \emph{108}, 5930--5936\relax
\mciteBstWouldAddEndPuncttrue
\mciteSetBstMidEndSepPunct{\mcitedefaultmidpunct}
{\mcitedefaultendpunct}{\mcitedefaultseppunct}\relax
\EndOfBibitem
\bibitem[Tunega \latin{et~al.}(2002)Tunega, Haberhauer, Gerzabek, and
  Lischka]{Tunega_3}
Tunega,~D.; Haberhauer,~G.; Gerzabek,~M.~H.; Lischka,~H. Theoretical Study of
  Adsorption Sites on the (001) Surfaces of 1:1 Clay Minerals. \emph{Langmuir}
  \textbf{2002}, \emph{18}, 139--147\relax
\mciteBstWouldAddEndPuncttrue
\mciteSetBstMidEndSepPunct{\mcitedefaultmidpunct}
{\mcitedefaultendpunct}{\mcitedefaultseppunct}\relax
\EndOfBibitem
\bibitem[Croteau \latin{et~al.}(2009)Croteau, Bertram, and
  Patey]{Patey_FFsim_2009}
Croteau,~T.; Bertram,~A.~K.; Patey,~G.~N. Simulation of {Water} {Adsorption} on
  {Kaolinite} under {Atmospheric} {Conditions}. \emph{J. Phys. Chem. A}
  \textbf{2009}, \emph{113}, 7826--7833\relax
\mciteBstWouldAddEndPuncttrue
\mciteSetBstMidEndSepPunct{\mcitedefaultmidpunct}
{\mcitedefaultendpunct}{\mcitedefaultseppunct}\relax
\EndOfBibitem
\bibitem[Carrasco \latin{et~al.}(2011)Carrasco, Santra, Klime\v{s}, and
  Michaelides]{javi:wetting}
Carrasco,~J.; Santra,~B.; Klime\v{s},~J.; Michaelides,~A. To Wet or Not to Wet?
  Dispersion Forces Tip the Balance for Water Ice on Metals. \emph{Phys. Rev.
  Lett.} \textbf{2011}, \emph{106}, 026101\relax
\mciteBstWouldAddEndPuncttrue
\mciteSetBstMidEndSepPunct{\mcitedefaultmidpunct}
{\mcitedefaultendpunct}{\mcitedefaultseppunct}\relax
\EndOfBibitem
\bibitem[Carrasco \latin{et~al.}(2013)Carrasco, Klime\v{s}, and
  Michaelides]{JCP:Javier}
Carrasco,~J.; Klime\v{s},~J.; Michaelides,~A. The role of van der Waals forces
  in water adsorption on metals. \emph{J. Chem. Phys.} \textbf{2013},
  \emph{138}, 024708\relax
\mciteBstWouldAddEndPuncttrue
\mciteSetBstMidEndSepPunct{\mcitedefaultmidpunct}
{\mcitedefaultendpunct}{\mcitedefaultseppunct}\relax
\EndOfBibitem
\bibitem[Tonigold and Gross(2012)Tonigold, and Gross]{JCC:Gross}
Tonigold,~K.; Gross,~A. Dispersive Interactions in Water Bilayers at Metallic
  Surfaces: A Comparison of the PBE and RPBE Functional Including Semiempirical
  Dispersion Corrections. \emph{J. Comput. Chem.} \textbf{2012}, \emph{33},
  695\relax
\mciteBstWouldAddEndPuncttrue
\mciteSetBstMidEndSepPunct{\mcitedefaultmidpunct}
{\mcitedefaultendpunct}{\mcitedefaultseppunct}\relax
\EndOfBibitem
\bibitem[Hamada(2014)]{PhysRevB.89.121103}
Hamada,~I. van der Waals density functional made accurate. \emph{Phys. Rev. B}
  \textbf{2014}, \emph{89}, 121103\relax
\mciteBstWouldAddEndPuncttrue
\mciteSetBstMidEndSepPunct{\mcitedefaultmidpunct}
{\mcitedefaultendpunct}{\mcitedefaultseppunct}\relax
\EndOfBibitem
\bibitem[Tkatchenko \latin{et~al.}(2012)Tkatchenko, DiStasio, Car, and
  Scheffler]{PhysRevLett.108.236402}
Tkatchenko,~A.; DiStasio,~R.~A.; Car,~R.; Scheffler,~M. Accurate and Efficient
  Method for Many-Body van der Waals Interactions. \emph{Phys. Rev. Lett.}
  \textbf{2012}, \emph{108}, 236402\relax
\mciteBstWouldAddEndPuncttrue
\mciteSetBstMidEndSepPunct{\mcitedefaultmidpunct}
{\mcitedefaultendpunct}{\mcitedefaultseppunct}\relax
\EndOfBibitem
\bibitem[Becke(1988)]{PhysRevA.38.3098}
Becke,~A.~D. Density-functional exchange-energy approximation with correct
  asymptotic behavior. \emph{Phys. Rev. A} \textbf{1988}, \emph{38},
  3098--3100\relax
\mciteBstWouldAddEndPuncttrue
\mciteSetBstMidEndSepPunct{\mcitedefaultmidpunct}
{\mcitedefaultendpunct}{\mcitedefaultseppunct}\relax
\EndOfBibitem
\bibitem[Lee \latin{et~al.}(1988)Lee, Yang, and Parr]{PhysRevB.37.785}
Lee,~C.; Yang,~W.; Parr,~R.~G. Development of the Colle-Salvetti
  correlation-energy formula into a functional of the electron density.
  \emph{Phys. Rev. B} \textbf{1988}, \emph{37}, 785--789\relax
\mciteBstWouldAddEndPuncttrue
\mciteSetBstMidEndSepPunct{\mcitedefaultmidpunct}
{\mcitedefaultendpunct}{\mcitedefaultseppunct}\relax
\EndOfBibitem
\bibitem[Sun \latin{et~al.}(2016)Sun, Remsing, Zhang, Sun, Ruzsinszky, Peng,
  Yang, Paul, Waghmare, Wu, Klein, and Perdew]{sun_accurate_2016}
Sun,~J.; Remsing,~R.~C.; Zhang,~Y.; Sun,~Z.; Ruzsinszky,~A.; Peng,~H.;
  Yang,~Z.; Paul,~A.; Waghmare,~U.; Wu,~X. \latin{et~al.}  Accurate
  first-principles structures and energies of diversely bonded systems from an
  efficient density functional. \emph{Nature Chemistry} \textbf{2016},
  \emph{8}, 831--836\relax
\mciteBstWouldAddEndPuncttrue
\mciteSetBstMidEndSepPunct{\mcitedefaultmidpunct}
{\mcitedefaultendpunct}{\mcitedefaultseppunct}\relax
\EndOfBibitem
\bibitem[Krukau \latin{et~al.}(2006)Krukau, Vydrov, Izmaylov, and
  Scuseria]{HSE06}
Krukau,~A.~V.; Vydrov,~O.~A.; Izmaylov,~A.~F.; Scuseria,~G.~E. {Influence of
  the exchange screening parameter on the performance of screened hybrid
  functionals}. \emph{J. Chem. Phys.} \textbf{2006}, \emph{125}, 224106\relax
\mciteBstWouldAddEndPuncttrue
\mciteSetBstMidEndSepPunct{\mcitedefaultmidpunct}
{\mcitedefaultendpunct}{\mcitedefaultseppunct}\relax
\EndOfBibitem
\bibitem[Grimme(2006)]{grimme_D2}
Grimme,~S. Semiempirical GGA-type density functional constructed with a
  long-range dispersion correction. \emph{J. Comput. Chem.} \textbf{2006},
  \emph{27}, 1787--1799\relax
\mciteBstWouldAddEndPuncttrue
\mciteSetBstMidEndSepPunct{\mcitedefaultmidpunct}
{\mcitedefaultendpunct}{\mcitedefaultseppunct}\relax
\EndOfBibitem
\bibitem[Grimme \latin{et~al.}(2010)Grimme, Antony, Ehrlich, and
  Krieg]{grimme_D3}
Grimme,~S.; Antony,~J.; Ehrlich,~S.; Krieg,~H. A consistent and accurate ab
  initio parametrization of density functional dispersion correction (DFT-D)
  for the 94 elements H-Pu. \emph{J. Chem. Phys.} \textbf{2010}, \emph{132},
  154104\relax
\mciteBstWouldAddEndPuncttrue
\mciteSetBstMidEndSepPunct{\mcitedefaultmidpunct}
{\mcitedefaultendpunct}{\mcitedefaultseppunct}\relax
\EndOfBibitem
\bibitem[Tkatchenko and Scheffler(2009)Tkatchenko, and
  Scheffler]{PRL09:Scheffler}
Tkatchenko,~A.; Scheffler,~M. Accurate {Molecular} {Van} {Der} {Waals}
  {Interactions} from {Ground}-{State} {Electron} {Density} and {Free}-{Atom}
  {Reference} {Data}. \emph{Phys. Rev. Lett.} \textbf{2009}, \emph{102},
  073005\relax
\mciteBstWouldAddEndPuncttrue
\mciteSetBstMidEndSepPunct{\mcitedefaultmidpunct}
{\mcitedefaultendpunct}{\mcitedefaultseppunct}\relax
\EndOfBibitem
\bibitem[Needs \latin{et~al.}(2010)Needs, Towler, Drummond, and {L\'opez
  R\'ios}]{casino}
Needs,~R.~J.; Towler,~M.~D.; Drummond,~N.~D.; {L\'opez R\'ios},~P. {Continuum
  variational and diffusion quantum Monte Carlo calculations}. \emph{Journal of
  Physics: Condensed Matter} \textbf{2010}, \emph{22}, 023201\relax
\mciteBstWouldAddEndPuncttrue
\mciteSetBstMidEndSepPunct{\mcitedefaultmidpunct}
{\mcitedefaultendpunct}{\mcitedefaultseppunct}\relax
\EndOfBibitem
\bibitem[Needs \latin{et~al.}(2009)Needs, Towler, Drummond, and {L\'opez
  R\'ios}]{CASINO2.3}
Needs,~R.~J.; Towler,~M.~D.; Drummond,~N.~D.; {L\'opez R\'ios},~P. CASINO
  Version 2.3 User Manual, University of Cambridge, Cambridge, UK. 2009\relax
\mciteBstWouldAddEndPuncttrue
\mciteSetBstMidEndSepPunct{\mcitedefaultmidpunct}
{\mcitedefaultendpunct}{\mcitedefaultseppunct}\relax
\EndOfBibitem
\bibitem[PPu()]{PPurl}
\url{www.tcm.phy.cam.ac.uk/~mdt26/casino2_pseudopotentials.html}, access date:
  March 2015\relax
\mciteBstWouldAddEndPuncttrue
\mciteSetBstMidEndSepPunct{\mcitedefaultmidpunct}
{\mcitedefaultendpunct}{\mcitedefaultseppunct}\relax
\EndOfBibitem
\bibitem[Trail and Needs(2005)Trail, and Needs]{trail05_NCHF}
Trail,~J.~R.; Needs,~R.~J. {Norm-conserving Hartree-Fock pseudopotentials and
  their asymptotic behavior}. \emph{J. Chem. Phys.} \textbf{2005}, \emph{122},
  014112\relax
\mciteBstWouldAddEndPuncttrue
\mciteSetBstMidEndSepPunct{\mcitedefaultmidpunct}
{\mcitedefaultendpunct}{\mcitedefaultseppunct}\relax
\EndOfBibitem
\bibitem[Trail and Needs(2005)Trail, and Needs]{trail05_SRHF}
Trail,~J.~R.; Needs,~R.~J. {Smooth relativistic Hartree-Fock pseudopotentials
  for H to Ba and Lu to Hg}. \emph{J. Chem. Phys.} \textbf{2005}, \emph{122},
  174109\relax
\mciteBstWouldAddEndPuncttrue
\mciteSetBstMidEndSepPunct{\mcitedefaultmidpunct}
{\mcitedefaultendpunct}{\mcitedefaultseppunct}\relax
\EndOfBibitem
\bibitem[pws()]{pwscf}
S. Baroni, A. Dal Corso, S. de Gironcoli, and P. Giannozzi,
  \url{http://www.pwscf.org}, access date: June 2015\relax
\mciteBstWouldAddEndPuncttrue
\mciteSetBstMidEndSepPunct{\mcitedefaultmidpunct}
{\mcitedefaultendpunct}{\mcitedefaultseppunct}\relax
\EndOfBibitem
\bibitem[Alf\`{e} and Gillan(2004)Alf\`{e}, and Gillan]{splinesQMC}
Alf\`{e},~D.; Gillan,~M.~J. Efficient localized basis set for quantum {Monte}
  {Carlo} calculations on condensed matter. \emph{Phys. Rev. B} \textbf{2004},
  \emph{70}, 161101\relax
\mciteBstWouldAddEndPuncttrue
\mciteSetBstMidEndSepPunct{\mcitedefaultmidpunct}
{\mcitedefaultendpunct}{\mcitedefaultseppunct}\relax
\EndOfBibitem
\bibitem[Umrigar \latin{et~al.}(1993)Umrigar, Nightingale, and
  Runge]{UMRIGAR:1993}
Umrigar,~C.~J.; Nightingale,~M.~P.; Runge,~K.~J. {A diffusion Monte Carlo
  algorithm with very small time-step errors}. \emph{J. Chem. Phys.}
  \textbf{1993}, \emph{99}, 2865--2890\relax
\mciteBstWouldAddEndPuncttrue
\mciteSetBstMidEndSepPunct{\mcitedefaultmidpunct}
{\mcitedefaultendpunct}{\mcitedefaultseppunct}\relax
\EndOfBibitem
\bibitem[Flyvbjerg and Petersen(1989)Flyvbjerg, and Petersen]{blockingmethod}
Flyvbjerg,~H.; Petersen,~H. {Error-Estimates on Averages of Correlated Data}.
  \emph{J. Chem. Phys.} \textbf{1989}, \emph{91}, 461--466\relax
\mciteBstWouldAddEndPuncttrue
\mciteSetBstMidEndSepPunct{\mcitedefaultmidpunct}
{\mcitedefaultendpunct}{\mcitedefaultseppunct}\relax
\EndOfBibitem
\bibitem[Tur()]{TurboRVB}
The {\sc TurboRVB} Quantum Monte Carlo package includes a complete suite for
  variational, diffusion and Green function quantum Monte Carlo calculations on
  molecules and solids, for wave function and geometry optimization, and for
  quantum Monte Carlo based molecular dynamics simulations. It is developed by
  S. Sorella and coworkers.
  \url{http://people.sissa.it/~sorella/web/index.html}, access date: June
  2015\relax
\mciteBstWouldAddEndPuncttrue
\mciteSetBstMidEndSepPunct{\mcitedefaultmidpunct}
{\mcitedefaultendpunct}{\mcitedefaultseppunct}\relax
\EndOfBibitem
\bibitem[Zen \latin{et~al.}(2013)Zen, Luo, Sorella, and Guidoni]{Zen:2013is}
Zen,~A.; Luo,~Y.; Sorella,~S.; Guidoni,~L. {Molecular Properties by Quantum
  Monte Carlo: An Investigation on the Role of the Wave Function Ansatz and the
  Basis Set in the Water Molecule}. \emph{J. Chem. Theory Comput.}
  \textbf{2013}, \emph{9}, 4332--4350\relax
\mciteBstWouldAddEndPuncttrue
\mciteSetBstMidEndSepPunct{\mcitedefaultmidpunct}
{\mcitedefaultendpunct}{\mcitedefaultseppunct}\relax
\EndOfBibitem
\bibitem[Burkatzki \latin{et~al.}(2007)Burkatzki, Filippi, and
  Dolg]{Burkatzki:2007p25447}
Burkatzki,~M.; Filippi,~C.; Dolg,~M. {Energy-consistent pseudopotentials for
  quantum monte carlo calculations}. \emph{J. Chem. Phys.} \textbf{2007},
  \emph{126}, 234105\relax
\mciteBstWouldAddEndPuncttrue
\mciteSetBstMidEndSepPunct{\mcitedefaultmidpunct}
{\mcitedefaultendpunct}{\mcitedefaultseppunct}\relax
\EndOfBibitem
\bibitem[fil()]{filippi-basis}
\url{http://burkatzki.com/pseudos/index.2.html}, access date: March 2015\relax
\mciteBstWouldAddEndPuncttrue
\mciteSetBstMidEndSepPunct{\mcitedefaultmidpunct}
{\mcitedefaultendpunct}{\mcitedefaultseppunct}\relax
\EndOfBibitem
\bibitem[Azadi \latin{et~al.}(2010)Azadi, Cavazzoni, and Sorella]{TurboPREP}
Azadi,~S.; Cavazzoni,~C.; Sorella,~S. {Systematically convergent method for
  accurate total energy calculations with localized atomic orbitals}.
  \emph{Phys. Rev. B} \textbf{2010}, \emph{82}, 125112\relax
\mciteBstWouldAddEndPuncttrue
\mciteSetBstMidEndSepPunct{\mcitedefaultmidpunct}
{\mcitedefaultendpunct}{\mcitedefaultseppunct}\relax
\EndOfBibitem
\bibitem[Binnie \latin{et~al.}(2009)Binnie, Sola, Alfe, and
  Gillan]{Binnie:2009ba}
Binnie,~S.~J.; Sola,~E.; Alfe,~D.; Gillan,~M.~J. {Benchmarking DFT surface
  energies with quantum Monte Carlo}. \emph{Mol. Simul.} \textbf{2009},
  \emph{35}, 609--612\relax
\mciteBstWouldAddEndPuncttrue
\mciteSetBstMidEndSepPunct{\mcitedefaultmidpunct}
{\mcitedefaultendpunct}{\mcitedefaultseppunct}\relax
\EndOfBibitem
\bibitem[Kresse and Hafner(1993)Kresse, and Hafner]{VASP1}
Kresse,~G.; Hafner,~J. {\em Ab initio} molecular dynamics for liquid metals.
  \emph{Phys. Rev. B} \textbf{1993}, \emph{558}, 47\relax
\mciteBstWouldAddEndPuncttrue
\mciteSetBstMidEndSepPunct{\mcitedefaultmidpunct}
{\mcitedefaultendpunct}{\mcitedefaultseppunct}\relax
\EndOfBibitem
\bibitem[Kresse and Hafner(1994)Kresse, and Hafner]{VASP2}
Kresse,~G.; Hafner,~J. {\em Ab initio} molecular-dynamics simulation of the
  liquid-metal-amorphous-semiconductor transition in germanium. \emph{Phys.
  Rev. B} \textbf{1994}, \emph{49}, 14251\relax
\mciteBstWouldAddEndPuncttrue
\mciteSetBstMidEndSepPunct{\mcitedefaultmidpunct}
{\mcitedefaultendpunct}{\mcitedefaultseppunct}\relax
\EndOfBibitem
\bibitem[Kresse and Furthmuller(1996)Kresse, and Furthmuller]{VASP3}
Kresse,~G.; Furthmuller,~J. Efficiency of ab-initio total energy calculations
  for metals and semiconductors using a plane-wave basis set. \emph{Comput.
  Mat. Sci.} \textbf{1996}, \emph{6}, 15--50\relax
\mciteBstWouldAddEndPuncttrue
\mciteSetBstMidEndSepPunct{\mcitedefaultmidpunct}
{\mcitedefaultendpunct}{\mcitedefaultseppunct}\relax
\EndOfBibitem
\bibitem[Kresse and Furthmuller(1996)Kresse, and Furthmuller]{VASP4}
Kresse,~G.; Furthmuller,~J. Efficient iterative schemes for {\em ab initio}
  total-energy calculations using a plane-wave basis set. \emph{Phys. Rev. B}
  \textbf{1996}, \emph{54}, 11169\relax
\mciteBstWouldAddEndPuncttrue
\mciteSetBstMidEndSepPunct{\mcitedefaultmidpunct}
{\mcitedefaultendpunct}{\mcitedefaultseppunct}\relax
\EndOfBibitem
\bibitem[Rom\'an-P\'erez and Soler(2009)Rom\'an-P\'erez, and
  Soler]{perez&soler:implementation}
Rom\'an-P\'erez,~G.; Soler,~J.~M. Efficient Implementation of a van der Waals
  Density Functional: Application to Double-Wall Carbon Nanotubes. \emph{Phys.
  Rev. Lett.} \textbf{2009}, \emph{103}, 096102\relax
\mciteBstWouldAddEndPuncttrue
\mciteSetBstMidEndSepPunct{\mcitedefaultmidpunct}
{\mcitedefaultendpunct}{\mcitedefaultseppunct}\relax
\EndOfBibitem
\bibitem[Bl\"{o}chl(1994)]{PAW1}
Bl\"{o}chl,~P.~E. Projector augmented-wave method. \emph{Phys. Rev. B}
  \textbf{1994}, \emph{50}, 17953\relax
\mciteBstWouldAddEndPuncttrue
\mciteSetBstMidEndSepPunct{\mcitedefaultmidpunct}
{\mcitedefaultendpunct}{\mcitedefaultseppunct}\relax
\EndOfBibitem
\bibitem[Kresse and Joubert(1999)Kresse, and Joubert]{PAW2}
Kresse,~G.; Joubert,~D. From ultrasoft pseudopotentials to the projector
  augmented-wave method. \emph{Phys. Rev. B} \textbf{1999}, \emph{59},
  1758\relax
\mciteBstWouldAddEndPuncttrue
\mciteSetBstMidEndSepPunct{\mcitedefaultmidpunct}
{\mcitedefaultendpunct}{\mcitedefaultseppunct}\relax
\EndOfBibitem
\bibitem[Graziano \latin{et~al.}(2012)Graziano, Klime\v{s}, Fernandez-Alonso,
  and Michaelides]{JPCM:Graziano}
Graziano,~G.; Klime\v{s},~J.; Fernandez-Alonso,~F.; Michaelides,~A. Improved
  description of soft layered materials with van der Waals density functional
  theory. \emph{J. Phys.: Condens. Matter} \textbf{2012}, \emph{24},
  424216\relax
\mciteBstWouldAddEndPuncttrue
\mciteSetBstMidEndSepPunct{\mcitedefaultmidpunct}
{\mcitedefaultendpunct}{\mcitedefaultseppunct}\relax
\EndOfBibitem
\bibitem[Makov and Payne(1995)Makov, and Payne]{Dipole1}
Makov,~G.; Payne,~M.~C. Periodic boundary conditions in {\em ab initio}
  calculations. \emph{Phys. Rev. B} \textbf{1995}, \emph{51}, 4014\relax
\mciteBstWouldAddEndPuncttrue
\mciteSetBstMidEndSepPunct{\mcitedefaultmidpunct}
{\mcitedefaultendpunct}{\mcitedefaultseppunct}\relax
\EndOfBibitem
\bibitem[Neugebauer and Scheffler(1992)Neugebauer, and Scheffler]{Dipole2}
Neugebauer,~J.; Scheffler,~M. Adsorbate-substrate and adsorbate-adsorbate
  interactions of Na and K adlayers on {Al(111)}. \emph{Phys. Rev. B}
  \textbf{1992}, \emph{46}, 16067\relax
\mciteBstWouldAddEndPuncttrue
\mciteSetBstMidEndSepPunct{\mcitedefaultmidpunct}
{\mcitedefaultendpunct}{\mcitedefaultseppunct}\relax
\EndOfBibitem
\end{mcitethebibliography}
\providecommand{\latin}[1]{#1}
\providecommand*\mcitethebibliography{\thebibliography}
\csname @ifundefined\endcsname{endmcitethebibliography}
  {\let\endmcitethebibliography\endthebibliography}{}

\end{document}